\documentclass[aps,prc,preprint,groupedaddress,tightenlines,showpacs,floatfix]{revtex4}
\usepackage{epsfig}
\usepackage{bm}
\def\NEG#1{{\rlap/#1}}
\begin{document}

\preprint{NT@UW-2001-28}
\title{Deuteron binding energies and form factors from light front field
theory}
\author{Jason R.~Cooke}
\author{Gerald A.~Miller}
\affiliation{
Department of Physics \\
University of Washington \\
Box 351560 \\
Seattle WA 98195-1560, USA}
\date{\today{}}

\begin{abstract}
  The objective of this paper is to investigate how the breaking of
manifest rotational invariance in light-front dynamics affects the binding
energy and the form factors of the deuteron. To do this, we derive new
light-front nucleon-nucleon one- and two-meson-exchange potentials, and use the
potentials to solve for the deuteron wave function and binding energy. We find
that including two-meson-exchange (TME) potentials partially repairs the broken
rotational invariance of the one-meson-exchange (OME) potential. This is shown
by a decrease in binding energy difference of different $m$ states of the
deuteron. We calculate the matrix elements of the electromagnetic current using
the deuteron wave functions obtained from the OME and OME+TME potentials.
Rotational invariance requires that the matrix elements satisfy an angular
condition, but in light-front dynamics that condition is only partially
satisfied. The current matrix elements from the OME calculation satisfy the
angular condition better than
the ones from the OME+TME calculation. The matrix elements of the axial current
satisfy the angular condition to the same extent regardless of which
wave functions are used for the calculation. Finally, we find that at momentum
transfers greater than about 2~GeV$^2$, the breaking of rotational invariance
causes less uncertainty in the computed deuteron form factors than do the
uncertainties in the nucleon form factors.
\end{abstract}

\pacs{
21.45.+v, % Few-body systems
03.65.Ge, % Solutions of wave equations: bound states
03.65.Pm, % Relativistic wave equations
11.10.Ef  % Lagrangian and Hamiltonian approach
}

\maketitle

\section{Introduction} \label{ch:int}

Recent experiments at Thomas Jefferson National Accelerator Facility
have measured the $A(Q^2)$ structure function of the deuteron for
momentum transfers up to 6 (GeV/c)$^2$ \cite{Alexa:1999fe}, and
measurements for $B(Q^2)$ are planned. Eventually, even higher momentum
transfers will be achieved. At such large momentum
transfers, a relativistic description of the deuteron is required.  Even
at lower momentum transfers, a relativistic description is important to
understand the details of the form factors. In addition, incorporating
relativity is important for the deuteron wave function to transform
correctly under boosts to large momentum, which is important for
calculating form factors.

One approach that gives a relativistic description of the deuteron is
light-front dynamics. The subject of this work is to investigate the
consequences of combining light front dynamics with various nuclear
models to calculate bound state wave functions, and use them to
calculate the deuteron form factors.

The utility of the light-front dynamics was first discussed by Dirac
\cite{Dirac:1949cp}. Light-front dynamics makes use of the light-front
coordinate system, where a four-vector $x^\mu$ is expressed as
$x^\mu=(x^+,x^-,x^1,x^2)$, with $x^\pm=x^0\pm x^3$.  This is simply a
change of variables, but an especially convenient one.  Using this
coordinate system and defining the commutation relations at equal
light-front time ($x^+=t_{\text{LF}}$), we obtain a light-front
Hamiltonian \cite{Brodsky:1998de,Harindranath:1996hq,Heinzl:1998kz}.
We use Hamiltonian in the light-front Schr\"odinger equation to solve
for bound states. There are many desirable features of the light-front
dynamics and the use of light-front coordinates.

First of all, high-energy experiments are naturally described using
light-front coordinates.  The wave front of a beam of high-energy
particles traveling in the (negative) three-direction is defined by a
surface where $x^+$ is (approximately) constant.  Such a beam can probe
the wave function of a target described in terms of light-front
variables \cite{Brodsky:1998de,Miller:1997cr,Miller:2000kv}: the Bjorken
$x$ variable
used to describe high-energy experiments is simply the ratio of the plus
momentum of the struck constituent particle to the total plus momentum 
($p^+$) of the bound state.

Secondly, the vacuum for a theory with massive particles can be very
simple on the light front.  This is because all massive particles and
anti-particles have positive plus momentum, and the total plus momentum
is a conserved quantity.  Thus, the na\"{\i}ve vacuum (with $p^+=0$) is
empty, and diagrams that couple to this vacuum are zero.  This greatly
reduces the number of non-trivial light-front time-ordered diagrams.

Thirdly, the generators of boosts in the one, two, and plus directions
are kinematic, meaning they are  independent of the interaction.
Thus, even when the Hamiltonian is truncated, the wave functions will
transform correctly under boosts.  Thus, light-front dynamics is useful
for describing form factors at high momentum transfers.

Finally, it is easy to perform relativistic calculations using
light-front dynamics. This is partly due to the simplicity of the
vacuum, and partly due the the fact that, with light-front variables,
center-of-momentum variables can be cleanly separated from the relative
momentum variables. This allows us to write relativistic equations which
have the simple form of a non-relativistic Schr\"odinger equation.

One serious drawback of light-front dynamics is that rotational
invariance is not manifest in any light-front Hamiltonian
\cite{breakpoin}. This is a result of selecting a particular direction
in space for the orientation of the light-front.

An untruncated light-front Hamiltonian will commute with the total
relative angular momentum operator, since the total momentum commutes
with the relative momentum.  Thus, eigenstates of the full Hamiltonian
will also be eigenstates of the angular momentum.  However, as mentioned
earlier, a Fock-space truncation of the light-front  Hamiltonian results
in the momentum operator four-vector losing covariance under rotations.
Hence $J^2$ and the truncated Hamiltonian do not commute and this
implies that the eigenstates of the truncated Hamiltonian will not be
eigenstates of the angular momentum.

How will this violation of rotational invariance affect physical
observables? One way to observe this violation is to note that on the
light front, rotational invariance about the $z$-axis is maintained.
This allows us to classify states as eigenstates of $J_3$ with
eigenvalues $m$. We compare the binding energy of deuteron states (which
have $j=1$) with different $m$ values. If the Hamiltonian were
rotationally invariant, the energies should be the same; the breaking of
rotational invariance causes the energies to be different
\cite{Trittmann:1997ga}.

Another symptom of the breaking of rotational invariance is that the
angular condition (a relation between the matrix elements of the current, defined
in section~\ref{sec:rotinv}) for the deuteron current is not exactly
satisfied \cite{Frankfurt:1993ut,Frederico:1991vb,%
Kondratyuk:1984kq,Cardarelli:1995yq,Arndt:1999wx}.
This means that different prescriptions for calculating the
deuteron form factors from the deuteron current will in principle give
different results when light-front dynamics is used. This dependence on
the prescription used has caused concern about the validity of applying
light-front dynamics to calculate form factors. 
We address both the breaking of binding-energy degeneracy and the angular
condition in this paper.

One notable feature of this calculation is that it is done entirely with
light-front dynamics. The covariant Lagrangian generates light-front
potentials, which generate light-front wave functions, which are used in
a light-front calculation of the deuteron current and form factors. This
is different from other approaches which use deuteron wave functions
calculated from equal-time dynamics, then transformed to the light front
\cite{Carbonell:1994uy,Carbonell:1995yi,Chung:1988my,Frankfurt:1993ut,%
Frederico:1991vb,Kondratyuk:1984kq,Cardarelli:1995yq,Arndt:1999wx}.

We start by introducing a model Lagrangian for nuclear physics which
includes chiral symmetry \cite{Miller:1997cr} in
section~\ref{ch:pionly}. The methods introduced in
Refs.~\cite{Cooke:1999yi,Cooke:2000ef} are generalized for use with
this nuclear model. The Hamiltonian is derived and used to calculate new
light-front nucleon-nucleon one-meson-exchange (OME) and
two-meson-exchange (TME) potentials. The notation and conventions
defined in the appendix are used extensively in this section.
We have some freedom in how to
choose the TME potentials, and we consider several different choices. 

In section~\ref{nnresults}, the OME and TME potentials are used to
calculate the deuteron wave functions and binding energies. We find that
by including the TME potentials for the calculation of the deuteron, the
binding energy degeneracy is broken by a smaller amount. 

The wave functions obtained in section~\ref{nnresults}, are used in
section~\ref{ch:ffdeut} to calculate the electromagnetic and axial form
factors of the
deuteron. Although rotational invariance demands that there be only
three independent components of the deuteron current, the light-front
calculation of the deuteron current results in four independent
components. This is a result of the lack of manifest rotational
invariance on the light front. There are several prescriptions for
choosing which deuteron current component should be eliminated, and in
principle this choice will affect the form factors. We attempt to find
currents that transform correctly (or well enough) under rotations so
that the choice of ``bad'' component does not matter too much.

We discuss the results of that search in section~\ref{ch:ff:results}. We
find that for most of the currents, the angular condition does not
depend strongly on which potential is used to calculate the wave
function. The only exception to this is that part of the electromagnetic
current which is multiplied by the isoscalar $F_1$ nucleon form factor
satisfies the angular condition much better when using the wave function
calculated with the OME potential than with wave function with other
potentials. We also find that the major uncertainty in the calculated
deuteron form factors at momentum transfers greater than 2~GeV$^2$ is
due not to the prescription used to determine the form factors from the
current, but instead is from the uncertainties in the nucleon form
factors.

\section{Realistic Nuclear Model} \label{ch:pionly}

We consider a nuclear Lagrangian which uses ($\pi$, $\sigma$, $\rho$,
$\omega$, $\eta$, and $\delta$) for the nucleon-nucleon interaction. This
is used to calculate a new light-front nucleon-nucleon potential (LFNN).
This model is an extension of the light-front
model used by Miller and Machleidt \cite{Miller:1999ap}. A new feature
of this model is that light-front energy dependent denominators
are used in the potentials; the denominators used in
Ref.~\cite{Miller:1999ap} are energy independent.

Note that most of the material covered here can be considered an
extension of the work done with the Wick-Cutkosky model in the
Refs~\cite{Cooke:1999yi,Cooke:2000ef}. This allows us to build upon the
previous results.

\subsection{Model and Formalism}

Our starting point is a nuclear Lagrangian \cite{Miller:1997cr}
which incorporates a non-linear chiral model for the pions. The
Lagrangian is based on the linear representations of chiral symmetry
used by Gursey \cite{Gursey:1960yy}. It is invariant (in the limit
where $m_\pi\rightarrow0$) under chiral transformations. 

The model prescribes the use of nucleons $\psi$ (or $\psi'$) and six
mesons: the $\pi$, $\delta$ (also known as the $a_0(980)$), $\sigma$
(also known as the $f_0(400-1200)$), $\eta$, $\rho$, and $\omega$
mesons. The coupling of each meson to the nucleon is governed by the
combination of the meson's spin and isospin. The $\pi$ and $\eta$ are
pseudoscalars, the
$\rho$ and $\omega$ are vectors, and the $\delta$ and $\sigma$ are
scalars. Under isospin transformations, the $\pi$, $\rho$, and
$\delta$ are isovector particles while the $\eta$, $\omega$, and
$\sigma$ are isoscalar particles.

The use of scalar mesons is meant as a simple representation of part of
the two-pion-exchange potential which causes much of the medium
range attraction between nucleons \cite{Machleidt:1987hj,Machleidt:1989}.
It can also be interpreted as the effect of fundamental scalar mesons
\cite{Black:1998wt,Black:1999dx,Black:2000qq}.

The Lagrangian ${\mathcal L}$ is based on the one used in
Refs.~\cite{Miller:1997xh,Miller:1997cr,Miller:1999ap}. It is given by
\begin{eqnarray}
{\mathcal L} &=&
-\frac{1}{4} \bm{\rho}^{\mu\nu} \cdot \bm{\rho}_{\mu\nu} 
+ \frac{m_\rho^2}{2}\bm{\rho}^\mu \cdot \bm{\rho}_\mu
-\frac{1}{4} \omega^{\mu\nu}\omega_{\mu\nu} +
\frac{m_\omega^2}{2}\omega^\mu \omega _\mu
\nonumber \\ & &
+ \frac{1}{4}f^2 \mbox{Tr} (\partial_\mu U \, \partial^\mu U^\dagger)
+\frac{1}{4}m_\pi^2f^2 \, \mbox{Tr}(U +U^\dagger-2)
\nonumber \\ & &
+\frac{1}{2} (\partial_\mu \sigma \partial^\mu \sigma -m_\sigma^2 \sigma^2) 
+\frac{1}{2} (\partial_\mu \bm{\delta} \cdot 
              \partial^\mu \bm{\delta}       -m_\delta^2 \bm{\delta}^2) 
+\frac{1}{2} (\partial_\mu \eta   \partial^\mu \eta   -m_\eta^2   \eta^2  ) 
\nonumber\\&&
+\overline{\psi}' \Big[
\gamma^\mu
(i\partial_\mu
-g_\rho   \bm{\rho  }_\mu\cdot\bm{\tau}
-g_\omega \omega_\mu
) - U (M
+g_\sigma\sigma
+g_\delta\bm{\delta}\cdot\bm{\tau}
+ig_\eta\gamma_5\eta)
\Big] \psi' \label{eq:mainNNlag},
\end{eqnarray}
where the bare masses of the nucleon and the mesons are given by $M$ and
$m_\alpha$ where $\alpha=\pi,\eta,\sigma,\delta,\rho,\omega$. We have
defined $V^{\mu\nu}\equiv\partial^\mu{}V^\nu-\partial^\nu{}V^\mu$ for
$V=\rho,\omega$. The unitary 
matrix $U$ can be chosen to have one of the three forms $U_i$:
\begin{eqnarray}
U_1\equiv e^{i  \gamma_5 \bm{\tau\cdot\pi}/f},\quad
U_2\equiv \frac
{1+i\gamma_5\bm{\tau}\cdot\bm{\pi}/2f}
{1-i\gamma_5\bm{\tau}\cdot\bm{\pi}/2f}, \quad
U_3\equiv \sqrt{1-\pi^2/f^2}+i\gamma_5\bm{\tau\cdot\pi}/f, \label{us}
\end{eqnarray}
which correspond to different definitions of the fields. Note that
each of these definitions can be expanded to give
\begin{eqnarray}
U &=& 1 + i \gamma_5 \frac{\bm{\tau}\cdot\bm{\pi}}{f}
- \frac{\pi^2}{2f^2} + {\mathcal O}
\left(\frac{\pi^3}{f^3}\right) \label{pi:pertU}.
\end{eqnarray}
In this work, we consider at most two meson exchange potentials, so we
consider $U$ to be defined by Eq.~(\ref{pi:pertU}).

In the limit where $m_\pi\rightarrow0$, this Lagrangian, is invariant
under the chiral transformation 
\begin{eqnarray}
\psi^\prime\to e^{i \gamma_5 \bm{\tau}\cdot\bm{a}}\psi^\prime,\qquad
U\to e^{-i \gamma_5 \bm{\tau}\cdot\bm{ a}} \;U\; 
e^{-i \gamma_5 \bm{\tau}\cdot \bm{a}}.
\label{chiral}\end{eqnarray}
In this model the other mesons are not affected by the transformation
because they are not chiral partners of the pion. This is in contrast to
the Lagrangian given in Refs.~\cite{Miller:1997xh,Miller:1997cr}, where
the mass and scalar interaction terms for the nucleon were written as
$MU+g_s\phi$ instead of $U(M+g_s\phi)$.

\subsection{Non-interacting Nucleon-Nucleon Theory}

The light-front Hamiltonian is derived from this Lagrangian using the
same approach used in Refs.~\cite{Cooke:1999yi,Cooke:2000ef},
the approach used by Miller \cite{Miller:1997cr} and others
\cite{Chang:1973xt,Chang:1973qi,Yan:1973qf,Yan:1973qg}. The basic idea
is to write the light-front Hamiltonian ($P^-$) as the sum of a free,
non-interacting part and a term containing the interactions. We consider
the free part first.

\subsubsection{Free Field Expansions}

The solutions for the free fields are similar to those obtained by using
equal-time dynamics. In fact, the solutions are formally related by a
change of variable, and so the most obvious difference between the two is
due to the Jacobian. The field equations have the general form
(when Lorentz, spinor, and isospin indices are suppressed) of
\begin{eqnarray}
\alpha(x) &=& \int \frac{d^2k_\perp dk^+ \theta(k^+)}{(2\pi)^{3/2} \sqrt{2
k^+}} \left[
a_\alpha        (\bm{k}) e^{-i k^\mu x_\mu} +
a_\alpha^\dagger(\bm{k}) e^{+i k^\mu x_\mu} \right],
\end{eqnarray}
where $\alpha=\pi,\eta,\sigma,\delta,\rho,\omega,\psi$. Note that in the
exponentials, 
\begin{eqnarray}
k^\mu x_\mu &=& \frac{1}{2} \left( k^+ x^- + k^- x^+ \right) -
\bm{k}_\perp \cdot \bm{x}_\perp.
\end{eqnarray}
In particular, the solutions for all the mesons and the nucleon field
are
\begin{eqnarray}
\bm{\pi}(x) &=& \int \frac{d^2k_\perp dk^+ \theta(k^+)}{(2\pi)^{3/2} \sqrt{2
k^+}} \left[
\bm{a}_\pi        (\bm{k}) e^{-i k^\mu x_\mu} +
\bm{a}_\pi^\dagger(\bm{k}) e^{+i k^\mu x_\mu} \right], \\
\eta(x) &=& \int \frac{d^2k_\perp dk^+ \theta(k^+)}{(2\pi)^{3/2} \sqrt{2
k^+}} \left[
a_\eta        (\bm{k}) e^{-i k^\mu x_\mu} +
a_\eta^\dagger(\bm{k}) e^{+i k^\mu x_\mu} \right], \\
\bm{\delta}(x) &=& \int \frac{d^2k_\perp dk^+
\theta(k^+)}{(2\pi)^{3/2} \sqrt{2 k^+}} \left[
\bm{a}_\delta        (\bm{k}) e^{-i k^\mu x_\mu} +
\bm{a}_\delta^\dagger(\bm{k}) e^{+i k^\mu x_\mu} \right], \\
\sigma(x) &=& \int \frac{d^2k_\perp dk^+ \theta(k^+)}{(2\pi)^{3/2} \sqrt{2
k^+}} \left[
a_\sigma        (\bm{k}) e^{-i k^\mu x_\mu} +
a_\sigma^\dagger(\bm{k}) e^{+i k^\mu x_\mu} \right], \\
\bm{\rho}^\mu(x) &=& \int \frac{d^2k_\perp dk^+
\theta(k^+)}{(2\pi)^{3/2} \sqrt{2 k^+}}
\sum_{s=1,3} \epsilon^\mu(\bm{k},s) \left[
\bm{a}_\rho        (\bm{k},s) e^{-i k^\mu x_\mu} +
\bm{a}_\rho^\dagger(\bm{k},s) e^{+i k^\mu x_\mu} \right], \\
\omega^\mu(x) &=& \int \frac{d^2k_\perp dk^+ \theta(k^+)}{(2\pi)^{3/2}
\sqrt{2 k^+}} \sum_{s=1,3} \epsilon^\mu(\bm{k},s) \left[
a_\omega        (\bm{k},s) e^{-i k^\mu x_\mu} +
a_\omega^\dagger(\bm{k},s) e^{+i k^\mu x_\mu} \right], \\
\psi(x) &=& \sqrt{2M}
\int \frac{d^2k_\perp dk^+ \theta(k^+)}{(2\pi)^{3/2} \sqrt{2
k^+}} \nonumber \\
&& \qquad \qquad \times 
\sum_{\lambda=+,-}
\sum_{t_3=+,-}
 \!\! \left[
u(\bm{k},\lambda) b        (\bm{k}) e^{-i k^\mu x_\mu} +
v(\bm{k},\lambda) d^\dagger(\bm{k}) e^{+i k^\mu x_\mu} \right]
\chi_{t_3}.
\end{eqnarray}
The polarization vectors are the usual ones. The most general of the
commutation relations is
\begin{eqnarray}
\left[ a_{\alpha,i}(\bm{k},s),a_{\beta,j}^\dagger(\bm{k}',s')
\right] &=&
\delta_{\alpha,\beta} \delta_{i,j} \delta_{s,s'}
\delta^{(2,+)}(\bm{k}-\bm{k}'),
\end{eqnarray}
where $\alpha$, $i$, and $s$ denote the meson type, isospin, and
spin. The anti-commutation relations are
\begin{eqnarray}
\left\{ b(\bm{k},\lambda),b^\dagger(\bm{k}',\lambda') \right\}  =
\left\{ d(\bm{k},\lambda),d^\dagger(\bm{k}',\lambda') \right\} &=&
\delta_{\lambda,\lambda'} \delta^{(2,+)}(\bm{k}-\bm{k}').
\end{eqnarray}
All other (anti-)commutation relations vanish. The spinors are
normalized so that
$\overline{u}(\bm{p},\lambda')u(\bm{p},\lambda)=\delta_{\lambda'\lambda}$.
For more information on the definition of the spinors, see the
Appendix.

\subsubsection{Non-interacting Hamiltonians}

The general form of the non-interacting Hamiltonian for each meson is 
\begin{eqnarray}
P^-_0(\alpha) &=& \int d^2k_\perp dk^+ \theta(k^+)
a_\alpha^\dagger(\bm{k}) a_\alpha(\bm{k})
\frac{k_\perp^2 + m_\alpha^2}{k^+}.
\end{eqnarray}
For the vector mesons ($\rho$ and $\omega$), there is an implicit sum
over the meson spins. Explicitly, this means that for vector mesons
$a_V^\dagger(\bm{k})a_V(\bm{k})\rightarrow\sum_{s=1,3}a_V^\dagger(\bm{k},s)a_V(\bm{k},s)$.
Likewise, the sum over the isospin of the isovector mesons ($\pi$,
$\delta$, and $\rho$) is implicit. The sum over isospin can be made
explicit by writing
$a_I^\dagger(\bm{k})a_I(\bm{k})\rightarrow\sum_{i=1,3}a_{I,i}^\dagger(\bm{k})a_{I,i}(\bm{k})$.
The non-interacting Hamiltonian for the nucleons has a similar form as
well, 
\begin{eqnarray}
P^-_0(\psi) &=& \int d^2k_\perp dk^+ \theta(k^+)
\left[ \sum_{\lambda=+,-}
b^\dagger(\bm{k},\lambda) b(\bm{k},\lambda) +
d^\dagger(\bm{k},\lambda) d(\bm{k},\lambda) \right]
\frac{k_\perp^2 + M^2}{k^+}.
\end{eqnarray}
These equations are what one expects, since a free particle with momenta
$k_\perp$ and $k^+$ has light-front energy
$k^-=\frac{k_\perp^2+m^2}{k^+}$.

\subsection{Interacting Nucleon-Nucleon Theory}

The interaction Hamiltonians are derived from the Lagrangian in
Eq.~(\ref{eq:mainNNlag}) using the techniques presented in
Refs.~\cite{Cooke:1999yi,Cooke:2000ef}. However, there are some additional
complications due to the structure of the interactions.

One complication 
is that the chiral coupling of the pion field to the nucleons through
the $U$ matrix generates vertices with any number of pions. This is
addressed simply by expanding the $U$ matrix in powers of $\frac{1}{f}$,
and considering the interaction Hamiltonians order by order.

Another complication is due to the fact that both the vector mesons and
the fermions have components which depend on other components of the
field
\cite{Miller:1997cr,Soper:1971sr,Kogut:1970xa,Bjorken:1971ah,Soper:1971wn}.
Vector meson fields have four components, but only three degrees of
freedom. Likewise, fermion fields have four spinor components, but only
two degrees of freedom. When the dependent components are expressed
explicitly in terms of the independent components, we obtain new
(effective) interaction Hamiltonians for instantaneous vector mesons and
fermions. A complete derivation is given by Miller in
Ref.~\cite{Miller:1997cr} and earlier workers cited therein. We
illustrate only the main points of the derivation here. 

\subsubsection{Expanding the Pion Interaction}

We start by Taylor-expanding $U$ in powers of $\frac{1}{f}$,
after which the derivation of the
one-meson-interaction Hamiltonian ${P'}^-_{I,1}$ is 
straightforward. (The prime indicates that it is in terms of $\psi'$,
not $\psi$. We derive the expressions for $P^-_{I,1}$ in the
section~\ref{sec:elemferm}.) The result is
\begin{eqnarray}
{P'}^-_{I,1} &=&
\int d^2x_\perp dx^- \overline{\psi}'(x)
\Big[
  g_\rho   \gamma^\mu   \rho_{\mu,i}(x) \tau_i
+ g_\omega \gamma^\mu \omega_\mu    (x)
+ g_\delta            \delta_i      (x) \tau_i
+ g_\sigma            \sigma        (x)
\nonumber\\&&
\phantom{\int d^2x_\perp dx^- \overline{\psi}(x)\Big[}\,\,
+ g_\pi  (i\gamma_5) \tau_i \pi_i(x)
+ g_\chi (i\gamma_5)        \chi (x)
\Big] \psi'(x). \label{eq:pm1Primespecific}
\end{eqnarray}
We have defined a dimensionless coupling constant
$g_\pi\equiv\frac{M}{f}$. To save space and to generalize, we define
\begin{eqnarray}
\Gamma_\alpha &=& \left\{\begin{array}{cl}
i\gamma^5  & \mbox{ if $\alpha$ is a pseudoscalar meson ($\pi$, $\eta$)} \\
1          & \mbox{ if $\alpha$ is a scalar meson ($\delta$, $\sigma$)} \\
\gamma^\mu & \mbox{ if $\alpha$ is a vector meson ($\rho$, $\omega$)} \\
\end{array} \right. \\
T_\alpha &=& \left\{\begin{array}{cl}
\tau_i & \mbox{ if $\alpha$ is an isovector meson ($\pi$, $\delta$, $\rho$)} \\
1      & \mbox{ if $\alpha$ is an isoscalar meson ($\eta$, $\sigma$, $\omega$)}
\\
\end{array} \right.
\end{eqnarray}
and denote the meson fields by $\Phi_\alpha$. This allows us to write
\begin{eqnarray}
P^-_{I,1} &=&
\sum_{\alpha=\pi,\eta,\sigma,\delta,\rho,\omega}
\int d^2x_\perp dx^- \overline{\psi}'(x)
g_\alpha \Gamma_\alpha T_\alpha \Phi_\alpha(x)
\psi'(x) \label{eq:pm1Prime},
\end{eqnarray}
where the appropriate sums over the meson indices are implicit.

The next step is to derive the two-meson-interaction Hamiltonian which
arises from chiral symmetry, ${P'}^-_{I,2c}$:
\begin{eqnarray}
{P'}^-_{I,2c} &=&
\int d^2x_\perp dx^- \overline{\psi}'(x)\Big[
- \frac{g^2_\pi}{2M}                 \tau_i \tau_j \pi_i(x) \pi_j(x)
\nonumber\\&&\phantom{\int d^2x_\perp dx^- \overline{\psi}(x)\Big[}\,\,
+ \frac{g_\pi g_\phi}{M} (i\gamma^5) \tau_i        \pi_i(x) \phi (x)
\nonumber\\&&\phantom{\int d^2x_\perp dx^- \overline{\psi}(x)\Big[}\,\,
- \frac{g_\pi g_\chi}{M}             \tau_i        \pi_i(x) \chi (x)
\Big] \psi'(x) \\
&=&
\sum_{\alpha=\pi,\eta,\sigma,\delta} \frac{g_\pi g_\alpha}{M} s_\alpha
\int d^2x_\perp dx^- \overline{\psi}'(x) 
\left[\Gamma_\pi    T_\pi    \Phi_\pi   (x) \right]
\left[\Gamma_\alpha T_\alpha \Phi_\alpha(x) \right]
\psi'(x) \label{eq:rawtme2},
\end{eqnarray}
where $s_\alpha$ is a symmetry factor, equal to $\frac{1}{2}$ when
$\alpha=\pi$, and $1$ otherwise. When the contact interaction is used to
calculate diagrams, an additional factor is picked up for the $\pi\pi$
contact term (due to indistinguishability) which cancels the
symmetry factor $s_\pi$. Note that this contact interaction
involves only scalar and pseudoscalar mesons.

\subsubsection{Elimination of Dependent Fermion Components}
\label{sec:elemferm}

We are now ready to express the dependent components of $\psi'$ in terms
of the independent components, and address the problem of instantaneous
nucleons. The generation of instantaneous interactions is a general
feature of theories with interacting fermions in light-front
dynamics. This allows us to use a simplified model to demonstrate how
these instantaneous nucleons arise. In particular, we want to postpone
the discussion of the complication introduced by the vector mesons
until the next section. To this end, we choose to remove all mesons
except the $\sigma$ from the Lagrangian given in
Eq.~(\ref{eq:mainNNlag}). (The $\sigma$ is chosen since it has the simplest
coupling to the nucleon.) The equation of motion for the nucleons is then
\begin{eqnarray}
i \NEG\partial \psi' &=& ( M + g_\sigma\sigma ) \psi'.
\label{eq:eomNucleon} 
\end{eqnarray}
Applying the projection operators $\Lambda_\pm=\frac{1}{2}\gamma^0\gamma^\pm$
(defined in the Appendix) to Eq.~(\ref{eq:eomNucleon})
splits it into two equations,
\begin{eqnarray}
i \partial^- \psi_+' &=& [ \bm{\alpha}_\perp \cdot \bm{p}_\perp +
\beta (M + g_\sigma\sigma ) ] \psi_-' \label{eomNucleonP}, \\
i \partial^+ \psi_-' &=& [ \bm{\alpha}_\perp \cdot \bm{p}_\perp +
\beta (M + g_\sigma\sigma ) ] \psi_+' \label{eomNucleonM},
\end{eqnarray}
where $\psi_\pm'=\Lambda_\pm\psi'$. This split is useful because in
Eq.~(\ref{eq:eomNucleon}), all four components of the nucleon field are
interrelated, while in Eqs.~(\ref{eomNucleonP}) and (\ref{eomNucleonM}),
the two components of $\psi'_+$ are related to the two components of
$\psi'_-$, and vice versa.

First, notice that Eq.~(\ref{eomNucleonP}) involves $\partial^-$, a
dynamic operator in light-front dynamics. Dynamic operators should be
avoided since they involve the interaction, and are therefore
complicated. We use Eq.~(\ref{eomNucleonP}) to avoid that complication and
relate the components of $\psi$. Secondly, to
keep the relation as simple as possible, we do not attempt to invert the
spinor matrix on the right side of Eq.~(\ref{eomNucleonP}). Requiring that
the equation for the dependent components be both a kinematic equation
and simple equation forces us to choose $\psi_+'$ as the independent
components. The dependent components, $\psi_-'$, are defined by
\begin{eqnarray}
\psi_-' &=& \frac{1}{p^+} \left[ \bm{\alpha}_\perp \cdot
\bm{p}_\perp + \beta (M + g_\sigma \sigma ) \right] \psi_+'.
\end{eqnarray}
Notice that the dependent components consist of a non-interacting part
and an interacting part. We separate these parts by defining $\psi$
(without a prime) to be the free nucleon field, and $\xi$ to be the part
of $\psi'$ that is due to interactions. So
\begin{eqnarray}
\psi' &=& \psi + \xi \label{eq:psippsi},
\end{eqnarray}
where
\begin{eqnarray}
\xi = \xi_- 
&=& \frac{1}{p^+} \beta ( g_\sigma \sigma ) \psi_+
 = \frac{\gamma^+}{2p^+} ( g_\sigma \sigma ) \psi_+.
\end{eqnarray}
This allows us to write
\begin{eqnarray}
\psi'
&=& \psi + \frac{\gamma^+}{2p^+} ( g_\sigma \sigma ) \psi_+ \\
&=& \psi + \frac{\gamma^+}{2p^+} ( g_\sigma \sigma ) \psi'
\label{eq:psipSigma}
\end{eqnarray}
The last equation follows is obtained from noting that $(\gamma^+)^2=0$,
which implies that $\gamma^+\psi'=\gamma^+\psi_+$. 

Plugging Eq.~(\ref{eq:psipSigma}) into Eq.~(\ref{eq:pm1Primespecific})
and removing all mesons except the $\sigma$ meson, we obtain
\begin{eqnarray}
{P'}^-_{I,1} &=& P^-_{I,1} + P^-_{I,2}, \\
P^-_{I,1} &=&
\int d^2x_\perp dx^- \overline{\psi}(x) g_\sigma \sigma(x) \psi(x),
\\
P^-_{I,2} &=&
\int d^2x_\perp dx^-
\overline{\psi}(x) \left[
g_\sigma \sigma(x)
\frac{\gamma^+}{2p^+}
g_\sigma \sigma(x)
\right] \psi(x)
\Big] \label{eq:sigmaHamWinst}.
\end{eqnarray}

We interpret the $\frac{\gamma^+}{2p^+}$ factor as a
type of nucleon propagator that joins any two meson interactions
(although having two of these propagators adjacent to each other causes
the interaction to vanish since $(\gamma^+)^2$). Because this propagator
does not allow for an
energy denominator (as it is already between two potentials),
$\frac{\gamma^+}{2p^+}$ is called an {\it instantaneous} propagator.

Thus, when constructing the diagrams for the light-front potentials,
we must also include instantaneous propagators for the nucleons in
addition to the usual propagators.

\subsubsection{Elimination of Dependent Vector Meson Components}

Like the nucleons, the vector mesons have a dependent component that
contains interactions and must be eliminated. This process is
complicated somewhat by the fact that the dependent nucleon components
must be eliminated at the same time. The salient points of the combined
elimination of the dependent nucleon and vector meson components are
discussed in detail by Miller \cite{Miller:1997cr}. 

The result is that the vector meson field must be redefined and 
an instantaneous vector meson propagator is generated
in addition to an instantaneous nucleon propagator. However, when the
nucleon-nucleon potential is calculated, the redefinition of the vector
meson field exactly cancels the contribution of the instantaneous
vector meson. The result is that the potentials can formally be
calculated using the original vector meson field.

In this work, we use that result to simplify our derivations of
nucleon-nucleon potentials by formally using the na\"{\i}ive form of the
vector meson field. Thus, we find an interaction Hamiltonian similar
to the one shown in Eq.~(\ref{eq:sigmaHamWinst}),
\begin{eqnarray}
{P'}^-_{I,1} &=& P^-_{I,1} + P^-_{I,2}, \\
P^-_{I,1} &=& 
\sum_{\alpha=\pi,\eta,\sigma,\delta,\rho,\omega}
\int d^2x_\perp dx^- \overline{\psi}(x)
g_\alpha \Gamma_\alpha T_\alpha \Phi_\alpha(x)
\psi(x), \label{eq:rawome1} \\
P^-_{I,2} &=&
\sum_{\alpha_1,\alpha_2=\pi,\eta,\sigma,\delta,\rho,\omega}
\int d^2x_\perp dx^-
\overline{\psi}(x)
\left[g_{\alpha_1} \Gamma_{\alpha_1} T_{\alpha_1} \Phi_{\alpha_1}(x)\right]
\frac{\gamma^+}{2p^+}
\nonumber\\&&\qquad\qquad\qquad\qquad\qquad\qquad\qquad
\left[g_{\alpha_2} \Gamma_{\alpha_2} T_{\alpha_2} \Phi_{\alpha_2}(x)\right]
\psi(x) \label{eq:rawtme1}.
\end{eqnarray}
We may continue to interpret $\frac{\gamma^+}{2p^+}$ as an instantaneous
nucleon propagator, since in the derivation in Ref.~\cite{Miller:1997cr}
it is clear that the potential vanishes when there are two adjacent 
instantaneous propagators. In Refs.~\cite{Miller:1997cr,Miller:1999ap},
sign of the coupling of the vector mesons in the equations equivalent to
Eq.~(\ref{eq:rawtme1}) has the wrong sign; the coupling of the mesons in
Eq.~(\ref{eq:rawtme1}) must be the same as in Eq.~(\ref{eq:rawome1}).

Also note that in principle the same prescription has to be applied to
the contact interaction, although to the order of two mesons, this
simply has the effect of removing the primes. We find that
\begin{eqnarray}
P^-_{I,2c} &=&
\sum_{\alpha=\pi,\eta,\sigma,\delta} \frac{s_\alpha}{M}
\int d^2x_\perp dx^- \overline{\psi}(x) 
\left[g_\pi    \Gamma_\pi    T_\pi    \Phi_\pi   (x) \right]
\left[g_\alpha \Gamma_\alpha T_\alpha \Phi_\alpha(x) \right]
\psi(x) \label{eq:mediumtme2}.
\end{eqnarray}
Note that $P^-_{I,2c}$ and $P^-_{I,2}$ have forms that are very
similar. In fact, we can obtain $P^-_{I,2c}$ from $P^-_{I,2}$ by making
the following changes:
\begin{enumerate}
\item Replace $\frac{\gamma^+}{2p^+}$ with $\frac{s_\alpha}{M}$.
\item Replace $\alpha_1$ with $\pi$.
\item Restrict the sum on $\alpha_2$ to $\pi,\eta,\sigma,\delta$.
\end{enumerate}

\subsubsection{Interaction Hamiltonians in Momentum Space}

We look at the matrix element of the
interaction Hamiltonians between initial and final states given by
$|k_i,\lambda_i,\tau_i\rangle=b^\dagger(k_i,\lambda_i)\chi_{\tau_i}|0\rangle$
and 
$\langle k_f,\lambda_f,\tau_f|=\langle0|b(k_f,\lambda_f)\chi^\dagger_{\tau_f}$
where both the bra and the ket have units of $[M^{-3/2}]$.
We find that
\begin{eqnarray}
P^-_{I,1}(f,i)
&=&
\sum_{\alpha=\pi,\eta,\sigma,\delta,\rho,\omega}
\frac{g_\alpha 2M}{2(2\pi)^3 \sqrt{k_f^+ k_i^+}} 
\int \frac{d^2q_\perp dq^+ \theta(q^+)}{ \sqrt{2(2\pi)^3} \sqrt{q^+}}
\nonumber\\&&
\overline{u}(\bm{k}_f,\lambda_f) \Gamma_\alpha u(\bm{k}_i,\lambda_i)
\chi^\dagger_{\tau_f} T_\alpha \chi_{\tau_i}
\nonumber\\&&
\int d^2x_\perp dx^- 
e^{+i k_f^\mu x_\mu}e^{-i k_i^\mu x_\mu} \left[
a_\alpha        (\bm{q}) e^{-i q^\mu x_\mu} +
a_\alpha^\dagger(\bm{q}) e^{+i q^\mu x_\mu} \right] F_\alpha(\bm{q}),
\end{eqnarray}
where $\bm{q}=(q^+,\bm{q}_\perp)$
and there are implicit
$\theta$-functions on $q^+$ for each particle to ensure that the
light-front energy is positive. These are occasionally suppressed to
simplify the equations. We include a meson-nucleon form factor
$F_\alpha$ to phenomenologically account for the fact that the mesons
and nucleons are composite objects. 

Now, we evaluate $P^-$ at $x^+=0$ and use
\begin{eqnarray}
\int d^2x_\perp dx^- e^{i (k_f-k_i-q)^\mu x_\mu}
&=& 2(2\pi)^3 \delta^{(2,+)}(k_f-k_i-q),
\end{eqnarray}
to write
\begin{eqnarray}
P^-_{I,1}(f,i)
&=& 
\sum_{\alpha=\pi,\eta,\sigma,\delta,\rho,\omega}
g_\alpha 2M \frac{
\overline{u}(\bm{k}_f,\lambda_f) \Gamma_\alpha u(\bm{k}_i,\lambda_i)
}{\sqrt{2(2\pi)^3} \sqrt{k_f^+ k_i^+ q^+}} 
\chi^\dagger_{\tau_f} T_\alpha \chi_{\tau_i}
\nonumber\\&&
\left[
a_\alpha        (\bm{q}) \theta(  k_f^+-k_i^+ ) +
a_\alpha^\dagger(\bm{q}) \theta(-(k_f^+-k_i^+)) \right] F_\alpha(\bm{q}).
\end{eqnarray}
We define $\bm{q} = \mbox{sign}(k_f^+-k_i^+) (\bm{k}_f-\bm{k}_i)$, which
ensures that the $q^+$ of the meson is positive.

Next we consider the two-meson interactions. First take the interaction
with the instantaneous propagator given by Eq.~(\ref{eq:rawtme1}). Plugging
in the field definitions gives
\begin{eqnarray}
P^-_{I,2}(f,i)
&=& 
\sum_{\alpha_1,\alpha_2=\pi,\eta,\sigma,\delta,\rho,\omega}
\frac{g_{\alpha_1} g_{\alpha_2}  2M}{2(2\pi)^3 \sqrt{k_f^+ k_i^+}} 
\int \frac{d^2q_{1\perp} dq_1^+ \theta(q_1^+)}{(2\pi)^{3/2} \sqrt{2 q_1^+}}
\int \frac{d^2q_{2\perp} dq_2^+ \theta(q_2^+)}{(2\pi)^{3/2} \sqrt{2 q_2^+}}
\nonumber\\&&
\frac{\theta(k_m^+)}{2k_m^+}
\overline{u}(\bm{k}_f,\lambda_f) 
\Gamma_{\alpha_2}
\gamma^+
\Gamma_{\alpha_1}
          u (\bm{k}_i,\lambda_i)
\chi^\dagger_{\tau_f} T_{\alpha_2} T_{\alpha_1} \chi_{\tau_i}
\nonumber\\&&
\int d^2x_\perp dx^- 
e^{+i k_f^\mu x_\mu}e^{-i k_i^\mu x_\mu} \left[
a_{\alpha_2}        (\bm{q}_2) e^{-i q_2^\mu x_\mu} +
a_{\alpha_2}^\dagger(\bm{q}_2) e^{+i q_2^\mu x_\mu} \right]
F_{\alpha_2} (\bm{q}_2)
\nonumber\\&&
\left[
a_{\alpha_1}        (\bm{q}_1) e^{-i q_1^\mu x_\mu} +
a_{\alpha_1}^\dagger(\bm{q}_1) e^{+i q_1^\mu x_\mu} \right]
F_{\alpha_1} (\bm{q}_1)\\
&=& 
\sum_{\alpha_1,\alpha_2=\pi,\eta,\sigma,\delta,\rho,\omega}
g_{\alpha_1} g_{\alpha_2}  2M 
\frac{\overline{u}(\bm{k}_f,\lambda_f) \Gamma_{\alpha_2}
\gamma^+
\Gamma_{\alpha_1}
u (\bm{k}_i,\lambda_i)}{2(2\pi)^3 \sqrt{k_f^+ k_i^+}}
\nonumber\\&&
\left[ \chi^\dagger_{\tau_f} T_{\alpha_2} T_{\alpha_1} \chi_{\tau_i} \right]
\int \frac{d^2k_{m\perp} dk_m^+ 
\theta(k_m^+)}{2k_m^+\sqrt{q_1^+ q_2^+}}
\nonumber\\&& \left[
a_{\alpha_2}        (k_f-k_m) \theta(k_f^+-k_m^+) +
a_{\alpha_2}^\dagger(k_m-k_f) \theta(k_m^+-k_f^+) \right] 
F_{\alpha_2} (\bm{q}_2)
\nonumber\\&&\left[
a_{\alpha_1}        (k_m-k_i) \theta(k_m^+-k_i^+) +
a_{\alpha_1}^\dagger(k_i-k_m) \theta(k_i^+-k_m^+) \right] 
F_{\alpha_1} (\bm{q}_1) \label{eq:instantham}
\end{eqnarray}
Note that the momenta $q_1$ and $q_2$ are implicitly functions of the
momenta $k_f$, $k_m$, and $k_i$.

We could also write out the contact interaction given by
Eq.~(\ref{eq:mediumtme2}) in momentum space, but it is related to
Eq.~(\ref{eq:instantham}) by replacing $\frac{\theta(p^+)\gamma^+}{2p^+}$
with $\frac{s_\alpha}{M}$ and restricting the allowed values of the
$\alpha$'s.

\subsection{Feynman Rules for Nucleon-Nucleon Potentials}

Now that we have the one-meson- and two-meson-exchange expressed in
momentum space, we are now ready to write out the Feynman rules for
diagrams in our model. For simplicity, the only diagrams considered are
those where a meson emitted by one nucleon is absorbed by the other
nucleon. Since we are only interested in the two-nucleon to two-nucleon
potentials, we follow the same approach as outlined in
Ref.~\cite{Cooke:2000ef} to derive the rules. We denote a ``normal''
nucleon propagator by a solid line with an arrow, an instantaneous
nucleon propagator by a solid line with an stroke across it, mesons
of all type by a dashed line, and energy denominator terms by a
vertical, light, dotted line.
\begin{enumerate}
\item Overall factor of 
$\frac{4M^2\delta^{\perp,+}(P_f-P_i)}{2(2\pi)^3\sqrt{k^+_{1f}k^+_{2f}k^+_{1i}k^+_{2i}}}$.
\item Usual light-front rules for $p_\perp$ and $p^+$ momentum
conservation.
\item Factor of $\frac{\theta(q_i)}{q_i}$ for each internal line,
including any instantaneous nucleon lines.
\item
A factor of $\frac{1}{P^--\sum_iq_i^-}$ for each energy denominator.
\item Each meson connects the two nucleons, and each end of the meson
line has a factor of $g_\alpha\Gamma_\alpha T_\alpha F_\alpha(q)$. The
indices of the isospin factors on each end of the meson are contracted
together. The Lorentz indices of the gamma matrices are contracted with
$-g^{\mu\nu}$ for the vector mesons.
\item For each contact vertex, multiply by a factor of $\frac{1}{M}$. If
the vertex is a $\pi-\pi$ vertex, symmetrize the $T_\pi
T_\pi=\tau_i\tau_j$ by replacing it with $\delta_{i,j}$.
\item Factor of
$\frac{\NEG{k}+M}{2M}=\sum_{\lambda}u(k,\lambda)\overline{u}(k,\lambda)$
for each propagating nucleon and $\frac{\gamma^+}{2}$ for an
instantaneous nucleon.
\item Integrate with $4M^2 \int \frac{d^2k_\perp dk^+}{2(2\pi)^3}$ over
any internal momentum loops.
\item Put the spinor factors for nucleon 1 and 2 between
$\overline{u}u$'s and the isospin factors between the initial and final
state isospin.
\end{enumerate}

From this list, it is useful to summarize what needs to be done to
convert a graph with an instantaneous nucleon to one with a contact
interaction: 
\begin{enumerate}
\item Replace $\frac{\theta(k^+)\gamma^+}{k^+}$ with $\frac{1}{M}$.
\item If both mesons are pions, replace $T_i T_j$ with $\delta_{i,j}$.
\end{enumerate}

These rules make it easy to write down what various potentials are.

\subsection{Nucleon-Nucleon Potentials}

The meson exchange potentials have the same basic form as in
Refs.~\cite{Cooke:1999yi,Cooke:2000ef}, however we must include the
contact interaction
and instantaneous nucleon propagators for the nuclear model used
here. First, we discuss how to include the contact diagrams from the
standpoint of the Bethe-Salpeter equation, then we begin to calculate
the light-front potentials.

\subsubsection{The Bethe-Salpeter Equation and Chiral Symmetry}
\label{nt:bseequiv}

\begin{figure}[!p]
\begin{center}
\epsfig{angle=0,width=5.0in,height=0.8in,file=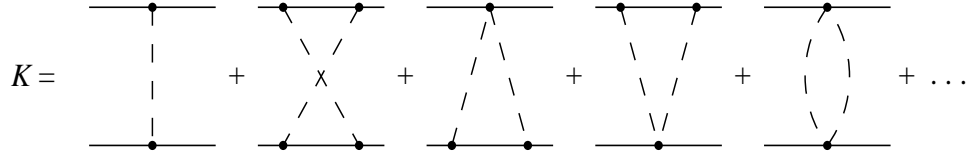}
\caption{The first several terms of the full kernel for the
Bethe-Salpeter equation of the nuclear model with chiral symmetry.
\label{fig:nt.fullbseker}}
\end{center}
\end{figure}

The kernel of the Bethe-Salpeter equation
\cite{Nambu:1950vt,Schwinger:1951ex,%  
Schwinger:1951hq,Gell-Mann:1951rw,Salpeter:1951sz} for this nuclear
model is richer than the one presented in
Refs.~\cite{Cooke:2000ef} for
the Wick-Cutkosky model. This is due mainly to the presence of the
contact interactions which are generated by the chirally invariant
coupling of the pion to the nucleon. Several of the lowest-order pieces
of the full kernel $K$ are shown in Fig.~\ref{fig:nt.fullbseker}.
(Note that for Feynman diagrams, it is useful to combine the ``normal''
nucleon propagator with the instantaneous nucleon propagator
\cite{Schoonderwoerd:1998qj,Schoonderwoerd:1998iy,Schoonderwoerd:1998jm},
and denote the combination with a solid line.)

As discussed in Refs.~\cite{Cooke:1999yi,Cooke:2000ef}, each of these
Feynman
diagrams is covariant. This means that we may choose any of the diagrams
from $K$ to construct a new kernel $K'$, and the infinite series of
potential diagrams physically equivalent to
$K'$ will also be covariant. 

In practice, this means that when deciding which two-meson-exchange
potentials to include for calculating the deuteron wave function, we
may neglect the crossed diagrams. Although including only the box and
contact two-meson-exchange diagrams may affect the exact binding energy
calculated, we should find a partial restoration of rotational
invariance. We reiterate that the focus of this work is on understanding
the effects of the breaking of rotational invariance and how to restore
it; our goal is not precise agreement with experimental results.

We also want to keep the potentials chirally symmetric as well. Whereas
Lorentz symmetry is maintained by using a kernel with any Feynman
diagrams (with potentially arbitrary coefficients), chiral symmetry
relates the strength of the $\pi\pi$ contact interaction to the strength
of the pion-nucleon coupling. 

\begin{figure}[!p]
\begin{center}
\epsfig{angle=0,width=3.0in,height=0.75in,file=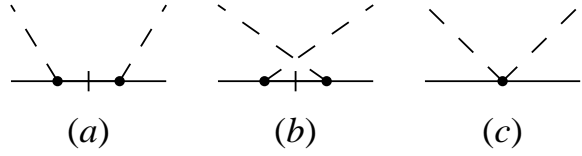}
\caption{The non-vanishing diagrams for pion-nucleon scattering at
threshold: (a) ${\mathcal M}_U$, (b) ${\mathcal M}_X$, and (c)
${\mathcal M}_C$. The mesons here are pions.
\label{fig:nt.chiralcan}}
\end{center}
\end{figure}

Chiral symmetry tells us that for pion-nucleon scattering at threshold,
the sum of the time-ordered graphs approximately cancels
\cite{Miller:1997cr}. Furthermore, upon closer examination, we find that
all the light-front time-ordered graphs for the scattering amplitude
vanish except for the two graphs with instantaneous nucleons and the
contact graph. These graphs are shown in
Fig.~\ref{fig:nt.chiralcan}. Using the Feynman rules, and denoting the
nucleon momentum by $k$ and the pion momentum by $q$, we find that
\begin{eqnarray}
{\mathcal M}_U &=& C \frac{\tau_i\tau_j}{2(k^++q^+)}u(k')\gamma^+u(k), \\
{\mathcal M}_X &=& C \frac{\tau_j\tau_i}{2(k^+-q^+)}u(k')\gamma^+u(k), \\
{\mathcal M}_C &=& C \frac{-\delta_{i,j}}{M}        u(k')        u(k),
\end{eqnarray}
where the factors common to all amplitudes are denoted by $C$.

For threshold scattering, we take $k=k'=M$ and $q=q'=m_\pi$. In that
limit, we find
\begin{eqnarray}
{\mathcal M}_U &=& C'\frac{ \delta_{i,j}+i\epsilon_{i,j,k}\tau_k}{2(M+m_\pi)},\\
{\mathcal M}_X &=& C'\frac{ \delta_{i,j}-i\epsilon_{i,j,k}\tau_k}{2(M-m_\pi)},\\
{\mathcal M}_C &=& C'\frac{-\delta_{i,j}}{M},     
\end{eqnarray}
where $C'=C \overline{u}(k')u(k)$. In the limit that $m_\pi\rightarrow
0$, the sum of these three terms vanishes. The term in these equations
proportional to $\tau_k$ is the famous Weinberg-Tomazowa term
\cite{Brandsden:1973,Adler:1968}.

The fact that the amplitudes cancel only when the contact interaction is
included demonstrates that chiral symmetry can have a significant effect
on calculations. In terms of two-pion-exchange potentials, this result
means that the contact potentials cancel strongly with both the iterated
box potentials and the crossed potentials. This serves to reduce the
strength of the total two-pion-exchange potential, which should lead to
more stable results.

However, since we do not use the crossed graphs for the nucleon-nucleon
potential, we must come up with a prescription which divided the contact
interactions into two parts which cancel the box and crossed diagrams
separately. We do this by formally defining two new contact
interactions, so that 
\begin{eqnarray}
{\mathcal M}_{C_U} &\equiv& \frac{M}{2(M+m_\pi)} {\mathcal M}_C, \\
{\mathcal M}_{C_X} &\equiv& {\mathcal M}_C - {\mathcal M}_{C_U}.
\end{eqnarray}
With these definitions, we find that at threshold and in the chiral
limit,
\begin{eqnarray}
{\mathcal M}_U + {\mathcal M}_{C_U} &=& 0, \\
{\mathcal M}_X + {\mathcal M}_{C_X} &=& 0.
\end{eqnarray}
This indicates that we can incorporate approximate chiral symmetry
without including crossed graphs simply by weighting each graph with a
contact interaction by a factor of $\frac{M}{2(M+m_\pi)}$.

\subsubsection{OME Potential}

\begin{figure}[!p]
\begin{center}
\epsfig{angle=0,width=5.0in,height=1.1in,file=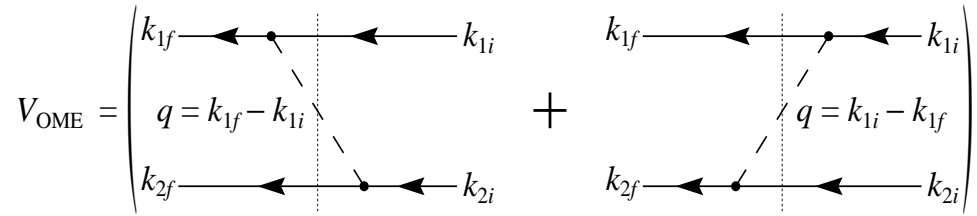}
\caption{The two diagrams which contribute to the OME potential for each
meson.
\label{fig:obepot}}
\end{center}
\end{figure}

The one meson exchange (OME) potential connects an initial state with
two nucleons to a final state with two nucleons, has one meson in
the intermediate state, and has the meson emitted and absorbed by
different nucleons. With these restrictions, along with the fact that in
light-front dynamics the interaction does not allow for particles to be
created from the vacuum, we find that there each meson has only two
diagrams for the OME potential. These diagrams are shown in
Fig.~\ref{fig:obepot}.

The Feynman rules derived in the previous section are used to derive the
potential for these diagrams. We factor out an overall factor of 
$\frac{4M^2\delta(k_{1f}+k_{2f}-k_{1i}-k_{2i})}{2\sqrt{k_{1f}^+k_{1i}^+k_{2f}^+k_{2i}^+}}$
that is common to all the two-nucleon potentials and suppress it from
now on.  Then we get
\begin{eqnarray}
V_{\text{OME},\alpha}
&=& 
\frac{g_\alpha^2 \bm{T}_{\alpha,1}\cdot\bm{T}_{\alpha,2}}{(2\pi)^3}
\overline{u}(\bm{k}_{1f},\lambda_{1f})\Gamma_\alpha u(\bm{k}_{1i},\lambda_{1i})
\overline{u}(\bm{k}_{2f},\lambda_{2f})\Gamma_\alpha u(\bm{k}_{2i},\lambda_{2i})
F_\alpha^2(\bm{q})
\nonumber\\&&
\left[
\frac{\theta(k_{1f}^+-k_{1i}^+)}{
(k_{1f}^+-k_{1i}^+)(P^--k_{1i}^--k_{2f}^-)
 -m_\alpha^2-(k_{1i,\perp}-k_{1f,\perp})^2} \right.\nonumber\\&&\left.
+
\frac{\theta(k_{1i}^+-k_{1f}^+)}{
(k_{1i}^+-k_{1f}^+)(P^--k_{1f}^--k_{2i}^-)
 -m_\alpha^2-(k_{1i,\perp}-k_{1f,\perp})^2}
\right] \label{nuc:eq:obepotraw}.
\end{eqnarray}
In the scattering regime
($P^-=k_{1f}^-+k_{2f}^-=k_{1i}^-+k_{2i}^-$), Eq.~(\ref{nuc:eq:obepotraw})
agrees with Eq.~(4.7) in Ref.~\cite{Miller:1997cr}, after taking into
account the difference in spinor normalization (we use
$\overline{u}u=1$).

To simplify the potential, we define
\begin{eqnarray}
a &=&
\left[ \theta(k_{1f}^+-k_{1i}^+)\times
(k_{1f}^+-k_{1i}^+)(P^--k_{1i}^--k_{2f}^-) \right.+
\nonumber\\&&
\left.
\theta(k_{1i}^+-k_{1f}^+)\times(k_{1i}^+-k_{1f}^+)(P^--k_{1f}^--k_{2i}^-)
\right]
-k_{i,\perp}^2-k_{f,\perp}^2, \\
b &=& 2 k_{i,\perp}k_{f,\perp},
\end{eqnarray}
so
\begin{eqnarray}
V_{\text{OME},\alpha} 
&=& \frac{g_\alpha^2 \bm{T}_{\alpha,1}\cdot\bm{T}_{\alpha,2}}{(2\pi)^3}
F_\alpha(\bm{q})^2 \frac{
\overline{u}(\bm{k}_{1f},\lambda_{1f})\Gamma_\alpha u(\bm{k}_{1i},\lambda_{1i})
\overline{u}(\bm{k}_{2f},\lambda_{2f})\Gamma_\alpha u(\bm{k}_{2i},\lambda_{2i})
}{\left[a -       m_\alpha^2 + b \cos(\phi_f-\phi_i)\right]}.
\end{eqnarray}

Now we consider the precise form to use for the meson-nucleon form
factor. We assume that it has a $n$-pole form \cite{Machleidt:1987hj},
so that the denominator of the meson-nucleon form factor has the
same form as the denominator of the potential in the scattering
regime. In particular, $\Lambda_\alpha$ playing the role of $m_\alpha$
for the form factor. For simplicity, we declare that the
denominator of
the form factor always has the same form of the denominator of the
potential. Thus,
\begin{eqnarray}
F_\alpha(\bm{q}) &=& \left(\frac{\Lambda_\alpha^2 - m_\alpha^2}
{a - \Lambda_\pi^2 + b \cos(\phi_f-\phi_i)} \right)^{n_\alpha}.
\label{eq:formfacexp}
\end{eqnarray}

Inserting the explicit expression of for the meson-nucleon form factor
into the potential results in
\begin{eqnarray}
V_{\text{OME},\alpha}
&=&
\frac{g_\alpha^2 \bm{T}_{\alpha,1}\cdot\bm{T}_{\alpha,2}}{(2\pi)^3}
(\Lambda_\alpha^2-m_\alpha^2)^{2n_\alpha} \nonumber\\&&
\frac{
\overline{u}(\bm{k}_{1f},\lambda_{1f})\Gamma_\alpha u(\bm{k}_{1i},\lambda_{1i})
\overline{u}(\bm{k}_{2f},\lambda_{2f})\Gamma_\alpha u(\bm{k}_{2i},\lambda_{2i})
}{
\left[a -       m_\alpha^2 + b \cos(\phi_f-\phi_i)\right]
\left[a - \Lambda_\alpha^2 + b \cos(\phi_f-\phi_i)\right]^{2n_\alpha}
}.
\end{eqnarray}

Now, since the light-front potentials have exact rotational invariance
about the $z$-axis, they conserve the $J_z$ quantum number $m$. Thus,
these potentials connect only states with the same value of $m$. This
means that, in general, light-front potentials may be written as
\begin{eqnarray}
V(\phi_f,\phi_i) &=& \sum_{m=-\infty}^\infty e^{i m (\phi_i-\phi_f)} V^m,
\end{eqnarray}
where $V^m$ is the potential in the magnetic quantum number basis. This
relation can be inverted to obtain $V^m$ in terms of $V(\phi_f,\phi_i)$,
\begin{eqnarray}
V^m_{\text{OME},\alpha} &=&
\frac{g_\pi^2 \bm{T}_{\alpha,1}\cdot\bm{T}_{\alpha,2}}{(2\pi)^3}
(\Lambda_\alpha^2-m_\alpha^2)^{2n_\alpha}
\nonumber\\&&
\int \frac{d\phi_f}{2\pi} e^{i m \phi_f}
\frac{
\left[
\overline{u}(\bm{k}_{1f},\lambda_{1f})\Gamma_\alpha u(\bm{k}_{1i},\lambda_{1i})
\overline{u}(\bm{k}_{2f},\lambda_{2f})\Gamma_\alpha u(\bm{k}_{2i},\lambda_{2i})
\right]_{\phi_i=0}
}{
\left[a -       m_\alpha^2 + b \cos(\phi_f)\right]
\left[a - \Lambda_\alpha^2 + b \cos(\phi_f)\right]^{2n_\alpha}
}.
\end{eqnarray}

To continue on, we need to get an expression for $\phi_f$ dependence the
$\overline{u}u$ matrix elements.  These are calculated in
Ref.~\cite{CookeThesis}. Summarizing, we can write
\begin{eqnarray}
\overline{u}(\bm{k}_{1f},\lambda_{1f})\Gamma_\alpha u(\bm{k}_{1i},\lambda_{1i})
\overline{u}(\bm{k}_{2f},\lambda_{2f})\Gamma_\alpha u(\bm{k}_{2i},\lambda_{2i})
&=& \sum_j C_\alpha(\Gamma_\alpha,\Gamma_\alpha) e^{ij\phi_f},
\end{eqnarray}
where the $C$ depends implicitly on all the variables on the left-hand
side except $\phi_f$ and $\phi_i$.

Thus,
\begin{eqnarray}
V^m_{\text{OME},\alpha} &=&
g_\alpha^2 (\bm{T}_{\alpha,1}\cdot\bm{T}_{\alpha,2})
\frac{(\Lambda_\alpha^2-m_\alpha^2)^{2n_\alpha}}{(2\pi)^3}
\sum_j C_j(\Gamma_\alpha,\Gamma_\alpha) \nonumber\\&&
\int \frac{d\phi_f}{2\pi}
\frac{e^{i(m+j)\phi_f}}{
\left[a -       m_\alpha^2 + b \cos(\phi_f)\right]
\left[a - \Lambda_\alpha^2 + b \cos(\phi_f)\right]^{2n_\alpha}}.
\end{eqnarray}
The cosine integral is denoted by $I$, and is calculated in
Ref.~\cite{CookeThesis}. Substituting $I$ for the integral, we obtain
\begin{eqnarray}
V^m_{\text{OME},\alpha} &=& g_\alpha^2
(\bm{T}_{\alpha,1}\cdot\bm{T}_{\alpha,2})
\frac{(\Lambda_\alpha^2-m_\alpha^2)^{2n_\alpha}}{(2\pi)^3} \nonumber\\&&
\sum_j C_j(\Gamma_\alpha,\Gamma_\alpha)
I(m+j,a-m_\alpha^2,1,a-\Lambda_\alpha^2,2n_\alpha,b) \label{eq:omefinished}.
\end{eqnarray}
All of the terms in Eq.~(\ref{eq:omefinished}) are known, and the
potential can now be calculated numerically.

\subsubsection{TME Potentials}

\begin{figure}[!p]
\begin{center}
\epsfig{angle=0,width=6in,height=7.0in,file=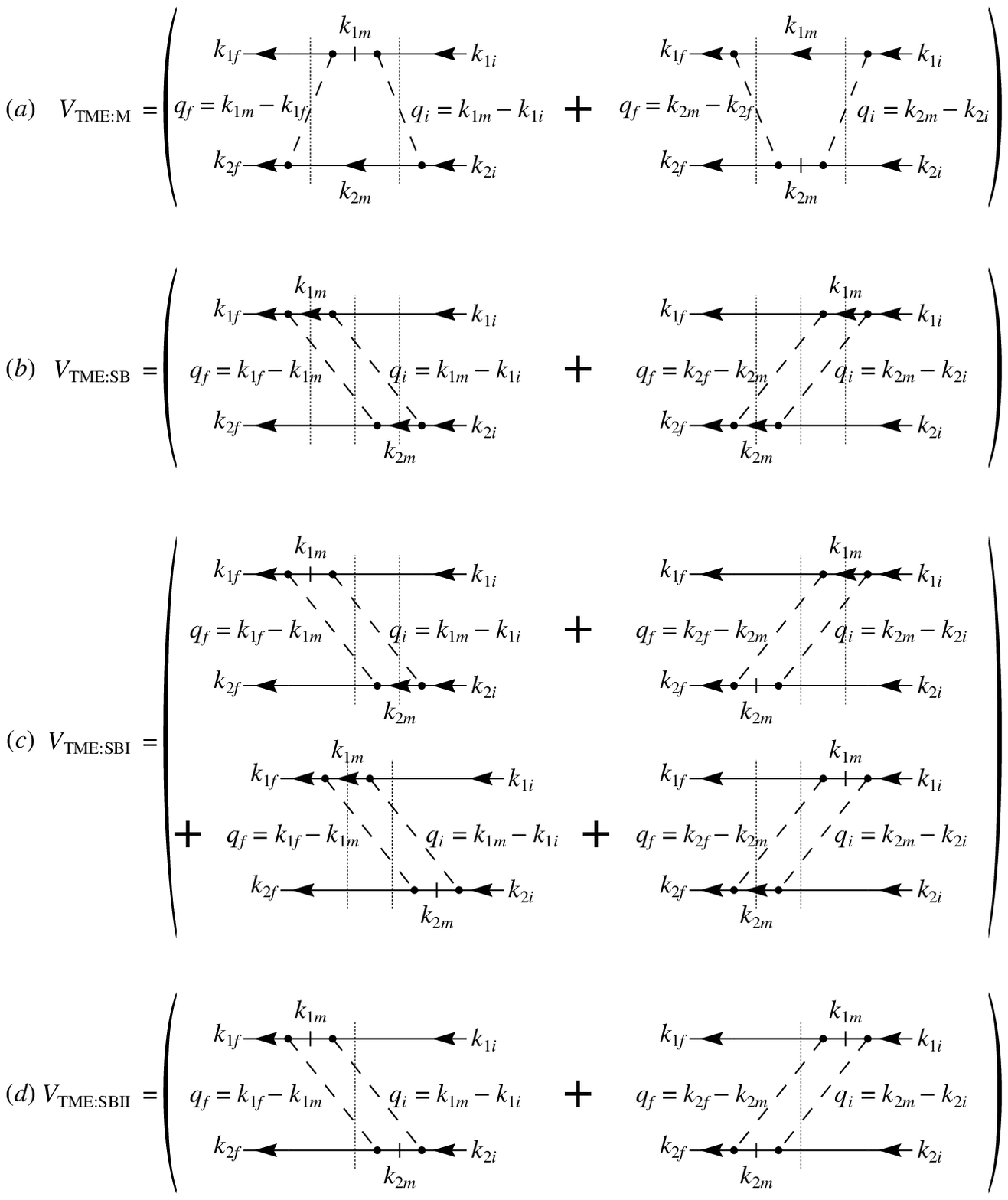}
\caption{The TME potentials for
(a) $V_{\text{TME:M}}$ (the Mesa potential), 
(b) $V_{\text{TME:SB}}$ (the stretched box potential), 
(c) $V_{\text{TME:SBI}}$ (the stretched instantaneous potential), and
(d) $V_{\text{TME:SBII}}$ (the stretched double instantaneous potential).
Note that the graphs on the right side are obtained from the graphs on the
left side by $1\leftrightarrow 2$.
\label{fig:tbepot}}
\end{center}
\end{figure}

\begin{figure}[!p]
\begin{center}
\epsfig{angle=0,width=5in,height=6.75in,file=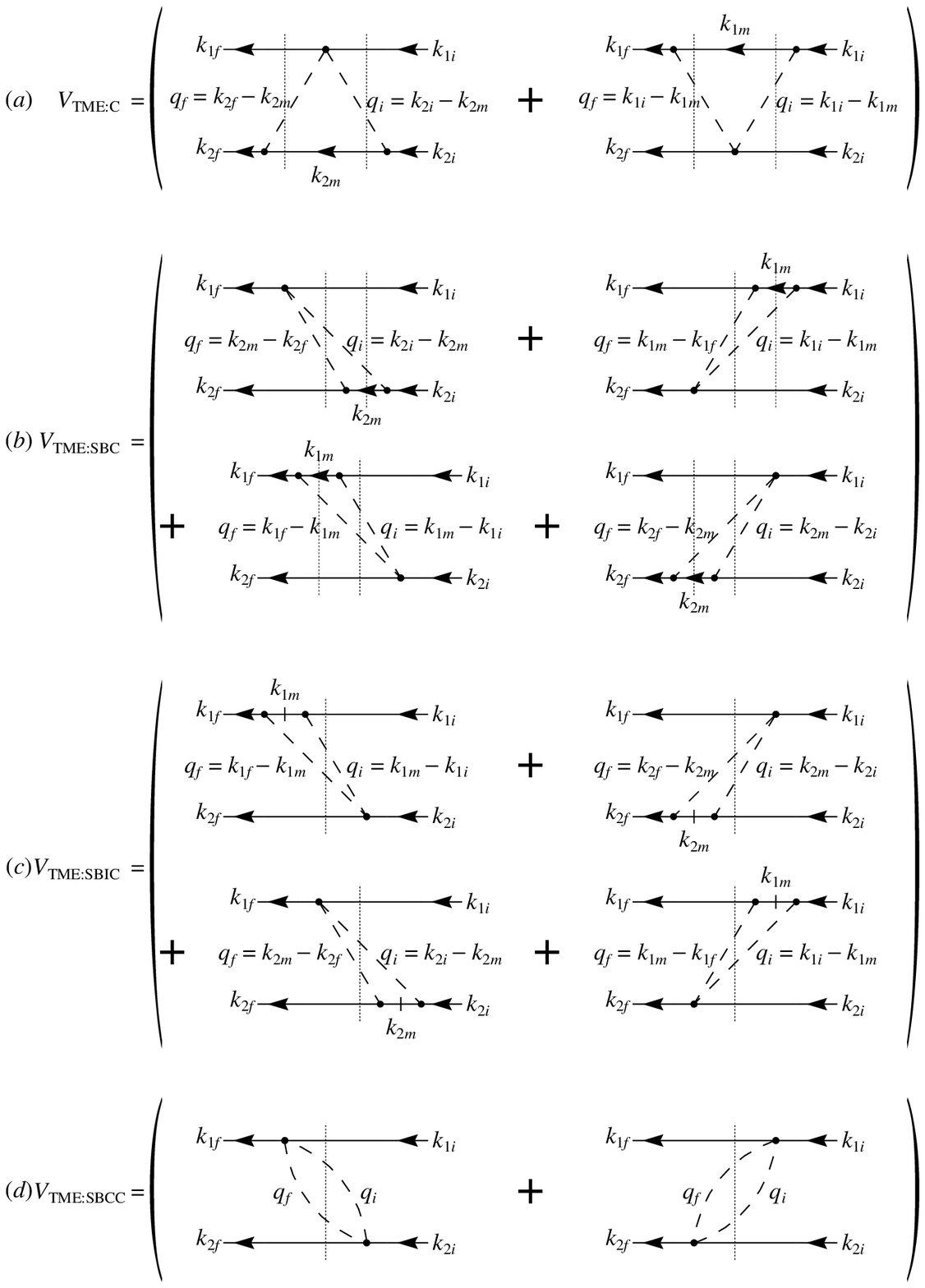}
\caption{The TME potentials that include the contact interaction for
(a) $V_{\text{TME:C}}$ (the contact potential), 
(b) $V_{\text{TME:SBC}}$ (the stretched contact potential), 
(c) $V_{\text{TME:SBIC}}$ (the stretched instantaneous contact potential), and
(d) $V_{\text{TME:SBCC}}$ (the stretched double contact potential).
Note that the graphs on the right side are obtained from the graphs on the
left side by $1\leftrightarrow 2$.
\label{fig:tbecontpot}}
\end{center}
\end{figure}

We two-meson-exchange (TME) potentials consider here are the box
diagrams (see Fig.~\ref{fig:tbepot}) and the contact diagrams (see
Fig.~\ref{fig:tbecontpot}). We do not consider the crossed diagrams
because they are not needed to restore rotational invariance, as shown
in section~\ref{nt:bseequiv}.

\subsubsection{Stretched Box Potential}

We use the Feynman rules to write the stretched-box potential shown in
Fig.~\ref{fig:tbepot}(b),
\begin{eqnarray}
V^{\alpha_f,\alpha_i}_{\text{TME:SB}}
&=& \frac{g^2_{\alpha_f} g^2_{\alpha_i} 4M^2
[\bm{T}_{\alpha_f,1}\cdot\bm{T}_{\alpha_f,2}]
[\bm{T}_{\alpha_i,1}\cdot\bm{T}_{\alpha_i,2}]}{(2\pi)^3}
\int \frac{d^2k_{2m\perp} dk_{2m}^+}{2(2\pi)^3 k_{1m}^+k_{2m}^+}
\nonumber\\&&
{}_{1f}\langle\left( \Gamma_{\alpha_f} \frac{\NEG{k}_{1m}+M}{2M}
\Gamma_{\alpha_i} \right)\rangle_{1m} \times
{}_{2f}\langle\left( \Gamma_{\alpha_f} \frac{\NEG{k}_{2m}+M}{2M}
\Gamma_{\alpha_i} \right)\rangle_{2m}
\nonumber\\&&
\theta(k_{1m}^+)\theta(k_{2m}^+)
\theta(k_{2i}^+-k_{2m}^+)\theta(k_{2m}^+-k_{2f}^+)
\nonumber\\&&
\frac{F_{\alpha_f}(\bm{q}_f)^2}{
(k_{2m}^+-k_{2f}^+)(P^--k_{1m}^--k_{2f}^-) - m_{\alpha_f}^2
- (\bm{k}_{1f\perp}-\bm{k}_{1m\perp})^2}
\nonumber\\&&
\frac{1}{P^- - k_{1i}^- - k_{2f}^- - q_f^- - q_i^-}
\nonumber\\&&
\frac{F_{\alpha_i}(\bm{q}_i)^2}{
(k_{2i}^+-k_{2m}^+)(P^--k_{1i}^--k_{2m}^-) - m_{\alpha_i}^2
- (\bm{k}_{1m\perp}-\bm{k}_{1i\perp})^2 }
\nonumber\\&&
+ \{ 1 \leftrightarrow 2 \}
\label{eqn:strechedfermionbox}.
\end{eqnarray}
To compress notation, we defined
${}_{1f}\langle\equiv\overline{u}(\bm{k}_{1f},\lambda_{1f})$,
$\rangle_{1i}\equiv{}u(\bm{k}_{1i},\lambda_{1i})$, and so on.
The symbol $\{ 1\leftrightarrow 2\}$ means that all labels 1 are
replaced with 2 and vice versa. This is a way of explicitly stating the
potential is invariant under exchange of nucleons 1 and 2.

We also used the following relation to simplify this expression:
\begin{eqnarray}
&&\sum_{\tau_{2m}}
\chi^\dagger_{\tau_{2f}} T_{\alpha_f,j} \chi_{\tau_{2m}}
\chi^\dagger_{\tau_{2m}} T_{\alpha_i,i} \chi_{\tau_{2i}}
\chi^\dagger_{\tau_{1f}} T_{\alpha_f,j} \chi_{\tau_{1m}} 
\chi^\dagger_{\tau_{1m}} T_{\alpha_i,i} \chi_{\tau_{1i}} 
\nonumber\\
&=&
\langle \tau_f | 
[\bm{T}_{\alpha_f,1}\cdot\bm{T}_{\alpha_f,2}]
[\bm{T}_{\alpha_i,1}\cdot\bm{T}_{\alpha_i,2}]
| \tau_i \rangle \nonumber \\
&=&
[\bm{T}_{\alpha_f,1}\cdot\bm{T}_{\alpha_f,2}]
[\bm{T}_{\alpha_i,1}\cdot\bm{T}_{\alpha_i,2}].
\end{eqnarray}

\subsubsection{Mesa Potential}

We consider the Mesa potential next, $V_{\text{TME:M}}$, shown in
Fig.~\ref{fig:tbepot}(a). Using the Feynman rules results in
\begin{eqnarray}
V^{\alpha_f,\alpha_i}_{\text{TME:M}}
&=& \frac{g^2_{\alpha_f} g^2_{\alpha_i} 4M^2
[\bm{T}_{\alpha_f,1}\cdot\bm{T}_{\alpha_f,2}]
[\bm{T}_{\alpha_i,1}\cdot\bm{T}_{\alpha_i,2}]}{(2\pi)^3}
\int \frac{d^2k_{2m\perp} dk_{2m}^+}{2(2\pi)^3 k_{1m}^+k_{2m}^+}
\nonumber\\&&
{}_{1f}\langle\left( \Gamma_{\alpha_f} \frac{\gamma^+}{2}
\Gamma_{\alpha_i} \right)\rangle_{1m} \times
{}_{2f}\langle\left( \Gamma_{\alpha_f} \frac{\NEG{k}_{2m}+M}{2M}
\Gamma_{\alpha_i} \right)\rangle_{2m}
\nonumber\\&&
\theta(k_{1m}^+)\theta(k_{2m}^+)
\theta(k_{2i}^+-k_{2m}^+)\theta(k_{2m}^+-k_{2f}^+)
\nonumber\\&&
\frac{F_{\alpha_f}(\bm{q}_f)^2}{
(k_{2f}^+-k_{2m}^+)(P^--k_{1f}^--k_{2m}^-) - m_{\alpha_f}^2
- (\bm{k}_{1f\perp}-\bm{k}_{1m\perp})^2}
\frac{1}{2M}
\nonumber\\&&
\frac{F_{\alpha_i}(\bm{q}_i)^2}{
(k_{2i}^+-k_{2m}^+)(P^--k_{1i}^--k_{2m}^-) - m_{\alpha_i}^2
- (\bm{k}_{1m\perp}-\bm{k}_{1i\perp})^2 }
\nonumber\\&&
+ \{ 1 \leftrightarrow 2 \}
\label{eqn:mesapot}.
\end{eqnarray}

\subsubsection{Contact Potential}

The last potential we will calculate explicitly is the contact potential 
$V_{\text{TME:C}}$, shown in Fig.~\ref{fig:tbecontpot}(a). The rules
relate this potential to $V_{\text{TME:M}}$ in a simple fashion. If
both the initial and final mesons are not pions, then the potential is
\begin{eqnarray}
V^{\alpha_f,\alpha_i}_{\text{TME:C}}
&=& \frac{g^2_{\alpha_f} g^2_{\alpha_i} 4M^2
[\bm{T}_{\alpha_f,1}\cdot\bm{T}_{\alpha_f,2}]
[\bm{T}_{\alpha_i,1}\cdot\bm{T}_{\alpha_i,2}]}{(2\pi)^3}
\int \frac{d^2k_{2m\perp} dk_{2m}^+}{2(2\pi)^3 k_{2m}^+}
\nonumber\\&&
{}_{1f}\langle\left( \Gamma_{\alpha_f} \frac{1}{2}
\Gamma_{\alpha_i} \right)\rangle_{1m} \times
{}_{2f}\langle\left( \Gamma_{\alpha_f} \frac{\NEG{k}_{2m}+M}{2M}
\Gamma_{\alpha_i} \right)\rangle_{2m}
\nonumber\\&&
\theta(k_{2m}^+)
\theta(k_{2i}^+-k_{2m}^+)\theta(k_{2m}^+-k_{2f}^+)
\nonumber\\&&
\frac{F_{\alpha_f}(\bm{q}_f)^2}{
(k_{2f}^+-k_{2m}^+)(P^--k_{1f}^--k_{2m}^-) - m_{\alpha_f}^2
- (\bm{k}_{1f\perp}-\bm{k}_{1m\perp})^2}
\frac{1}{2M}
\nonumber\\&&
\frac{F_{\alpha_i}(\bm{q}_i)^2}{
(k_{2i}^+-k_{2m}^+)(P^--k_{1i}^--k_{2m}^-) - m_{\alpha_i}^2
- (\bm{k}_{1m\perp}-\bm{k}_{1i\perp})^2 }
\nonumber\\&&
+ \{ 1 \leftrightarrow 2 \}
\label{eqn:contactpot}.
\end{eqnarray}
To get the contact potential for the pions, the following change has to
be made:
\begin{eqnarray}
[\bm{T}_{\alpha_f,1}\cdot\bm{T}_{\alpha_f,2}]
[\bm{T}_{\alpha_i,1}\cdot\bm{T}_{\alpha_i,2}]
&\rightarrow&
\tau_2^2 = 3.
\end{eqnarray}
This is due the additional symmetry that the two pion vertex has.

Before this potential is used in calculations, it must be multiplied by
a factor of $\frac{M}{2(M+m_\pi)}$ (as discussed in
section~\ref{nt:bseequiv}) if the crossed diagrams are not included.

\subsubsection{Common Features and Simplification}

Each of the potentials in Figs.~\ref{fig:tbepot} and
\ref{fig:tbecontpot} can be written schematically in the following
general form:
\begin{eqnarray}
V_{\text{TME}}
&=& \frac{g^4 4M^2 T^4}{(2\pi)^3}
\int \frac{d^2k_{2m\perp} dk_{2m}^+}{2(2\pi)^3}
f(k_{1m}^+,k_{2m}^+,q_f^+,q_i^+)
\nonumber\\&&
{}_{1f}\langle\left( \Gamma_{\alpha_f} {\mathcal M}_1
\Gamma_{\alpha_i} \right)\rangle_{1m} \times
{}_{2f}\langle\left( \Gamma_{\alpha_f} {\mathcal M}_2
\Gamma_{\alpha_i} \right)\rangle_{2m}
\nonumber\\&&
\left[a_f -       m_{\alpha_f}^2 + b_f \cos(\phi_f-\phi_m)\right]^{-n_f}
F_{\alpha_f}(\bm{q}_f)^2
\nonumber\\&&
\left[a_m + b_{mf} \cos(\phi_f-\phi_m)
          + b_{mi} \cos(\phi_m-\phi_i)\right]^{-n_m}
\nonumber\\&&
\left[a_i -       m_{\alpha_i}^2 + b_i \cos(\phi_m-\phi_i)\right]^{-n_i}
F_{\alpha_i}(\bm{q}_i)^2.
\end{eqnarray}
We use the expression for $F_\alpha$ in Eq.~(\ref{eq:formfacexp}) to get
\begin{eqnarray}
V_{\text{TME}}
&=& \frac{g^4 4M^2 T^4(\Lambda^2-m^2)^{4n_\alpha}}{(2\pi)^3}
\int \frac{d^2k_{2m\perp} dk_{2m}^+}{2(2\pi)^3}
f(k_{1m}^+,k_{2m}^+,q_f^+,q_i^+)
\nonumber\\&&
{}_{1f}\langle\left( \Gamma_{\alpha_f} {\mathcal M}_1
\Gamma_{\alpha_i} \right)\rangle_{1m} \times
{}_{2f}\langle\left( \Gamma_{\alpha_f} {\mathcal M}_2
\Gamma_{\alpha_i} \right)\rangle_{2m}
\nonumber\\&&
\left[a_f -       m_{\alpha_f}^2 + b_f \cos(\phi_f-\phi_m) \right]^{-n_f}
\left[a_f - \Lambda_{\alpha_f}^2 + b_f \cos(\phi_f-\phi_m) \right]^{-2n_{\alpha_f}}
\nonumber\\&&
\left[a_m + b_{mf} \cos(\phi_f-\phi_m)
          + b_{mi} \cos(\phi_m-\phi_i) \right]^{-n_m}
\nonumber\\&&
\left[a_i -       m_{\alpha_i}^2 + b_i \cos(\phi_m-\phi_i) \right]^{-n_i}
\left[a_i - \Lambda_{\alpha_i}^2 + b_i \cos(\phi_m-\phi_i) \right]^{-2n_{\alpha_i}}.
\end{eqnarray}

The TME potentials, like the OME potential, have exact rotational
invariance about the $z$-axis, and thus conserve $m$.  We project the
potential onto states of definite $m$ by setting $\phi_i$ to zero and
integrating the potential with $\int d\phi_f e^{im\phi_f}/2\pi$.

Now what we have to obtain the $\overline{u}u$ matrix
elements. In reference \cite{CookeThesis} shows that we can write
\begin{eqnarray}
&&\left[ 
{}_{1f}\langle\left( \Gamma_{\alpha_f} {\mathcal M}_1
\Gamma_{\alpha_i} \right)\rangle_{1m} \times
{}_{2f}\langle\left( \Gamma_{\alpha_f} {\mathcal M}_2
\Gamma_{\alpha_i} \right)\rangle_{2m}
\right]_{\phi_i=0} \nonumber \\
&&\qquad= \sum_{j,k}
C_{j,k}(
\Gamma_{\alpha_f},{\mathcal M}_1,\Gamma_{\alpha_i};
\Gamma_{\alpha_f},{\mathcal M}_2,\Gamma_{\alpha_i})
e^{i j (\phi_f-\phi_m)}
e^{i k \phi_m}.
\end{eqnarray}
Note that the $C$ is implicitly a function of the $p_\perp$, $p^+$,
$\phi$ and $\lambda$ on both sides.

Then, after a change of variables,
$\phi_m\rightarrow\phi'_i$,
$\phi_f-\phi_m\rightarrow\phi'_f$, and
$\phi_f\rightarrow\phi'_f+\phi'_i$, the potential can be written as
\begin{eqnarray}
V_{\text{TME}}
&=& \frac{g^4 4M^2 T^4(\Lambda^2-m^2)^{4n_\alpha}}{(2\pi)^3}
\sum_{j,k}
\int \frac{dk_{2m\perp} \, k_{2m\perp} \, dk_{2m}^+}{2(2\pi)^2}
f(k_{1m}^+,k_{2m}^+,q_f^+,q_i^+)
\nonumber\\&&
C_{j,k}(
\Gamma_{\alpha_f},{\mathcal M}_1,\Gamma_{\alpha_i};
\Gamma_{\alpha_f},{\mathcal M}_2,\Gamma_{\alpha_i})
\nonumber\\&&
\int_0^{2\pi} \frac{d\phi'_f}{2\pi}
\int_0^{2\pi} \frac{d\phi'_i}{2\pi}
e^{i (m+j) \phi'_f}
e^{i (m+k) \phi'_i}
\nonumber\\&&
\left[a_f -       m_{\alpha_f}^2 + b_f    \cos\phi'_f \right]^{-n_f}
\left[a_f - \Lambda_{\alpha_f}^2 + b_f    \cos\phi'_f \right]^{-2n_{\alpha_f}}
\nonumber\\&&
\left[a_m                 + b_{mf} \cos\phi'_f
                          + b_{mi} \cos\phi'_i \right]^{-n_m}
\nonumber\\&&
\left[a_i -       m_{\alpha_i}^2 + b_i    \cos\phi'_i \right]^{-n_i}
\left[a_i - \Lambda_{\alpha_i}^2 + b_i    \cos\phi'_i \right]^{-2n_{\alpha_i}}.
\end{eqnarray}
The azimuthal-angle integrals taken together are denoted as $I$, and the
method for calculating the $\phi'_m$ and $\phi'_f$ integrals is
discussed in Ref.~\cite{CookeThesis}. Summarizing, we can write 
\begin{eqnarray}
V_{\text{TME}}
&=& \frac{g^4 4M^2 T^4(\Lambda^2-m^2)^{4n_\alpha}}{(2\pi)^3}
\sum_{j,k}
\int \frac{dk_{2m\perp} \, k_{2m\perp} \, dk_{2m}^+}{2(2\pi)^2}
f(k_{1m}^+,k_{2m}^+,q_f^+,q_i^+)
\nonumber\\&&
C_{j,k}(
\Gamma_{\alpha_f},{\mathcal M}_1,\Gamma_{\alpha_i};
\Gamma_{\alpha_f},{\mathcal M}_2,\Gamma_{\alpha_i})
\nonumber\\&&
I(
a_f-m_{\alpha_f}^2, a_f-\Lambda_{\alpha_f}^2, b_f, n_f, 2n_{\alpha_f}, m+j;
\nonumber\\&&\phantom{I( }
a_m, b_{mf}, b_{mi}, n_m;
\nonumber\\&&\phantom{I( }
a_i-m_{\alpha_i}^2, a_i-\Lambda_{\alpha_i}^2, b_i, n_i, 2n_{\alpha_i}, m+k).
\end{eqnarray}
This potential is evaluated using the numerical integration techniques
discussed in Ref.~\cite{Cooke:2000ef}.

\subsubsection{Further Development of the Light-front Schr\"odinger
Equation} \label{sec:furtherlfnn}

The potentials derived here possess a high degree of symmetry. To solve
the light-front Schr\"odinger equation efficiently, these symmetries
should be explicitly exploited, as was done in
Refs.~\cite{Cooke:1999yi,Cooke:2000ef}. In addition to the invariance of
the potentials
under parity, there are additional symmetries due to the conservation
of nucleon helicity and invariance under time reversal
\cite{Machleidt:1987hj}. 

We follow Machleidt's approach for taking advantage of the symmetry of
rotationally invariant potentials with helicity
\cite{Machleidt:1989}. However, since the light-front potentials derived
here do not have full rotational invariance, Machleidt's approach must
be modified. The symmetry properties of helicity matrix elements are
rederived in Ref.~\cite{CookeThesis} without assuming full rotational
invariance. These results allow a modified version of Machleidt's
approach to be combined with the exploitation of parity (using the
transformation from light-front coordinates to equal-time coordinates)
discussed in Refs.~\cite{Cooke:1999yi,Cooke:2000ef}.
In particular, the potentials are initially calculated in the
$|p_{\text{ET}},\theta,M,\lambda_1,\lambda_2\rangle$ basis, although the
relations in Ref.~\cite{CookeThesis} are used to transform to the 
$|p_{\text{ET}},J,M,L,S\rangle$ basis to solve for the
wave functions. Note that the potentials connect states with different
$J$ values in general.
Once the symmetries have been explicitly expressed, we can discretize
the Schr\"odinger equation as done in Ref.~\cite{Cooke:2000ef}.

Note that the transformations applied to the potential in order to
simplify the solution of the wave functions. Once the wave functions are
obtained, we may apply the inverse of the transformations to the wave
functions to express them in the helicity basis
($|\bm{p}_\perp,p^+,\lambda_1,\lambda_2\rangle$) or in the Bjorken and
Drell spin basis ($|\bm{p}_\perp,p^+,m_1,m_2\rangle$)
\cite{CookeThesis}.

\section{Deuteron Binding Energies} \label{nnresults}

The next step towards numerically calculating the bound states for these
potentials is to choose the parameters (meson masses, coupling
constants, etc.) for the potentials. 
We consider the full nuclear model where the
nucleon-nucleon interaction is mediated by all the six mesons shown in
Table~\ref{nn:tab:mesparams}.
For numerical work, use the parameters for the light-front
nucleon-nucleon (LFNN) potential from the work of Miller and Machleidt
\cite{Miller:1999ap}. Those parameters were fit for a potential that used
a retarded propagator for the energy in the potentials. Since the
potentials used in this paper have energy dependent denominators, the
parameters must be modified somewhat. We choose to vary the coupling
constant for the $\sigma$ meson. The parameters are given in
table~\ref{nn:tab:mesparams}.

As with all the other deuteron models presented in this paper, 
the light-front one-meson-exchange (OME) potential breaks 
rotational invariance and causes a mass splitting of the deuteron states
with different magnetic quantum numbers. We expect that the
splitting will be removed somewhat by
including higher order potentials.

The first step is to determine which two-meson-exchange potentials
to use. One choice is to use only the two-pion-exchange potentials, TPE
and ncTPE, as defined in the previous section. However, we expect to get
better results using the two-meson-exchange diagrams
generated by all the available mesons, including the contact diagrams
for the pions, which we denote as the two-meson-exchange (TME) potential.
In addition, we can also investigate the effect of leaving out the
contact potentials for the pions, resulting in the non-chiral 
two-meson-exchange (ncTME) potential.

We do not include diagrams with a contact interaction between the nucleon,
a pion, and another meson. This is because, as mentioned in
section~\ref{nt:bseequiv}, the infinite series of the box diagrams is
rotationally invariant and the contact diagrams are not needed to
achieve rotational invariance. Furthermore, they are not required to
control the convergence of the series, since there is no strong
cancellation between the contact diagram and the instantaneous diagrams.

The first step in analyzing the bound states is to determine what range of
$f_\sigma$ gives reasonable results. We iteratively solve for the binding
energy of the deuteron, varying $f_\sigma$ until the binding energy matches
the physical value of the binding energy, for each of the potentials. The
results are shown in Table~\ref{nn:tab:fixbseJ1}. We find that a value of 
$f_\sigma$ in the range 1.2 to 1.3 will give reasonable results for the
binding energy. Note that D-state probability (about $3\%$) is lower in
this model than for the energy-independent light-front used in
Ref.~\cite{Miller:1999ap}, were a value of $4.5\%$ is found. This is
expected since the $f_\sigma$ is greater than 1 in this model, meaning
that the scalar interaction is strengthened relative to the tensor
interaction, leading to a decrease in amount of the D-state present.

We choose two values of $f_\sigma$, one from the low end of the range (1.22)
and one from the high end (1.2815) for our investigations. Using two values
helps ensure that our results are robust.

\begin{figure}[!p]
\begin{center}
\epsfig{angle=0,width=5.0in,height=5.0in,file=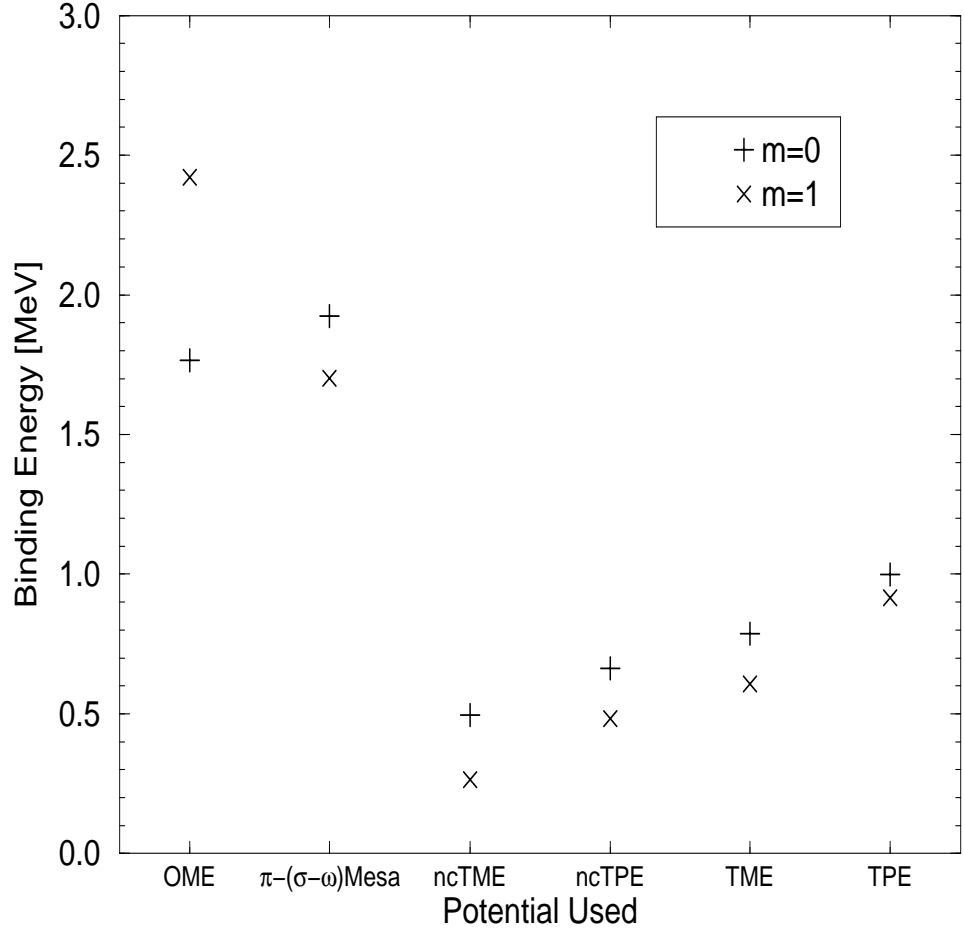}
\caption{The values of the binding energy for the $m=0$ and $m=1$ states
for different nucleon-nucleon light-front potentials. The $\sigma$
coupling constant factor is $f_\sigma=1.22$.
\label{fig:BEforMsAll.1.22}}
\end{center}
\end{figure}

First, we examine the bound states for $f_\sigma=1.22$.
The results for several different choices of the TME potentials 
are shown in Table~\ref{nn:tab:sig122AllJ} and the binding
energies are plotted in Fig.~\ref{fig:BEforMsAll.1.22}. In addition to
the two-meson-exchange potentials mentioned above, we also consider the
$\pi$-$\sigma$ plus $\pi$-$\omega$ Mesa potential.  The reason for
considering this potential is that Carbonell, Desplanques, Karmanov, and
Mathiot \cite{Carbonell:1998rj} have shown that it helps restore
rotational invariance of the deuteron.

In particular, they have used manifestly covariant light-front dynamics 
to analyze the deuteron. They start with a deuteron wave function
calculated in equal-time dynamics, then use a light-front 
one-pion-exchange potential (expanded to lowest order in powers of
$\frac{1}{M}$) to calculate the perturbative corrections to the wave
function. They find that the resulting wave function has an unphysical
dependence on the orientation of the light-front plane, which would
manifest itself as a breaking of rotational invariance in our
formalism. They also use the $\pi$-$\sigma$ and $\pi$-$\omega$ Mesa
potentials (expanded to lowest order in powers of $\frac{1}{M}$) to
calculate the correction to the wave function. When the wave function
corrections are combined, they find that the directional dependence
of the longest range part of the deuteron wave function cancels exactly.

This implies that for our model, using the $\pi$-$\sigma$ plus
$\pi$-$\omega$ Mesa potential (which we denote by
$\pi$-($\sigma$-$\omega$)) should partially restore the rotational
invariance of the deuteron, assuming that the breaking of rotational
invariance is due primarily to the one-pion-exchange potential. Note that
since we solve for the deuteron wave function self-consistently and to
all orders for our potentials, we do not expect to find exactly the same
result as Ref.~\cite{Carbonell:1998rj}. 

The first thing to notice about the data in
Table~\ref{nn:tab:sig122AllJ} is that the results are essentially the
same regardless of if arbitrary angular momentum or is used or if the
potential is restricted to the $J=1$ sector. The same result is also
seen in the pion-only model \cite{CookePiOnly}. It means that the wave
functions are numerically approximate to angular momentum eigenstates.

Next we notice the splittings between masses and D state percentages for
the $m=0$ and $m=1$ states. This implies that the states do not transform
correctly under rotations. All of the two-meson-exchange potentials used
reduce the splittings by similar amounts, by about 60\%
for the binding energy and by about 70\%
for the percent D state. Note also that the mass splittings for the
pion-only model were much larger \cite{CookePiOnly}.

Examining the effects of the individual two-meson-exchange potentials,
we see that $\pi$-($\sigma$-$\omega$)) potential does reduce the mass
splitting, but it does not fully remove it. This is expected since the
OME potential includes more than just the pion potential, and the
potential is relativistic.

Next, we compare the ncTME and ncTPE potentials to the TME and TPE
potentials. The non-chiral potentials reduce the binding energy more
than the chiral potentials, as we expected from our experience from the
pion-only model. However, unlike for the pion-only model, we find that
the chiral and non-chiral potential have fairly similar effects
\cite{CookePiOnly}.

Finally, notice that the mass splitting for the TPE potential is much
smaller than for the other two-meson-exchange potentials. By itself,
this does not imply that the rotational properties of the deuteron
calculated with that potential are significantly better than those from
other two-meson-exchange potentials. The individual potentials that make
up the TME potential are fairly large in magnitude, but vary in
sign. This means that using any subset of those potentials may result in
either a larger or smaller mass splitting. In this case, it is
smaller. To investigate this further, we examine the currents for
the TME and TPE deuteron wave function in section~\ref{ch:ffdeut}.

\begin{figure}[!p]
\begin{center}
\epsfig{angle=0,width=5.0in,height=5.0in,file=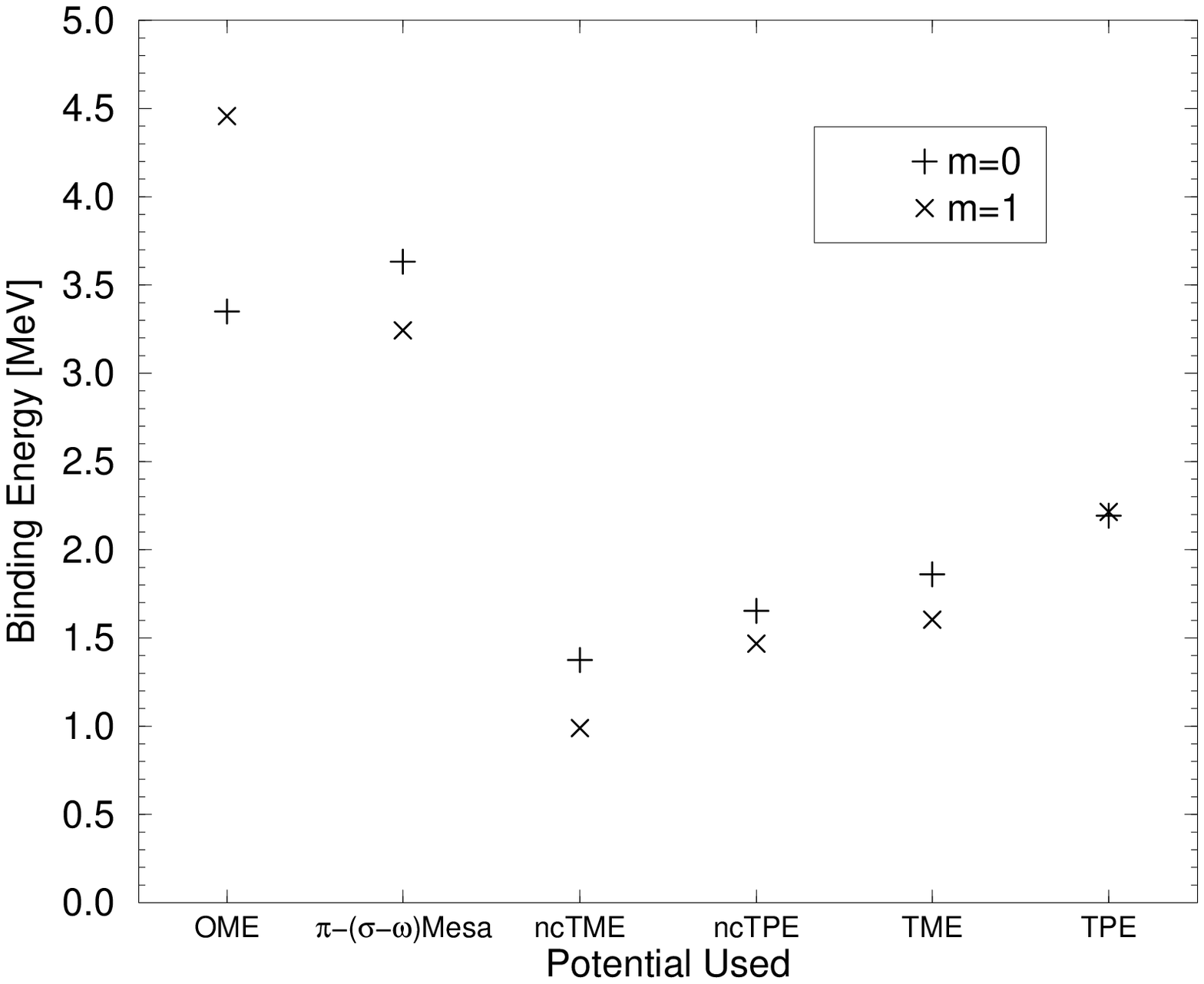}
\caption{The values of the binding energy for the $m=0$ and $m=1$ states
for different nucleon-nucleon light-front potentials. The $\sigma$
coupling constant factor is $f_\sigma=1.2815$.
\label{fig:BEforMsAll.1.28}}
\end{center}
\end{figure}

To verify that our results are independent of the value of $f_\sigma$, we
recalculate the deuteron properties for each of the potentials with
$f_\sigma=1.2815$. The results are summarized in
Table~\ref{nn:tab:sig128AllJ}, and the binding energies are shown in
Fig.~\ref{fig:BEforMsAll.1.28}. The change in $f_\sigma$ increases the
binding of the states, but the rest of the results are qualitatively the
same.

\section{Form Factors of the Deuteron} \label{ch:ffdeut}
In section~\ref{nnresults}, we considered several different
truncations of the light-front nucleon-nucleon (LFNN) potential and used
them to obtain the deuteron wave function. In this section, we use
those wave functions to solve for the matrix elements of the deuteron
current operator, which is used to calculate the deuteron
electromagnetic and axial form factors. 

In this section, we first outline the covariant theory of the
electromagnetic form factors for spin-1 objects, like the deuteron. Then
we recall the features of light-front calculations (including the
breaking of rotational invariance) of the form factors. After that, we
review the covariant and light-front tools for calculating axial form
factors. The formalism is then applied to calculate the electromagnetic
and axial currents and form factors for the light-front deuteron wave
functions.

\subsection{Electromagnetic Form Factors}

\subsubsection{Covariant Theory}

\begin{figure}[!p]
\begin{center}
\epsfig{angle=0,width=3.25in,height=1.8in,file=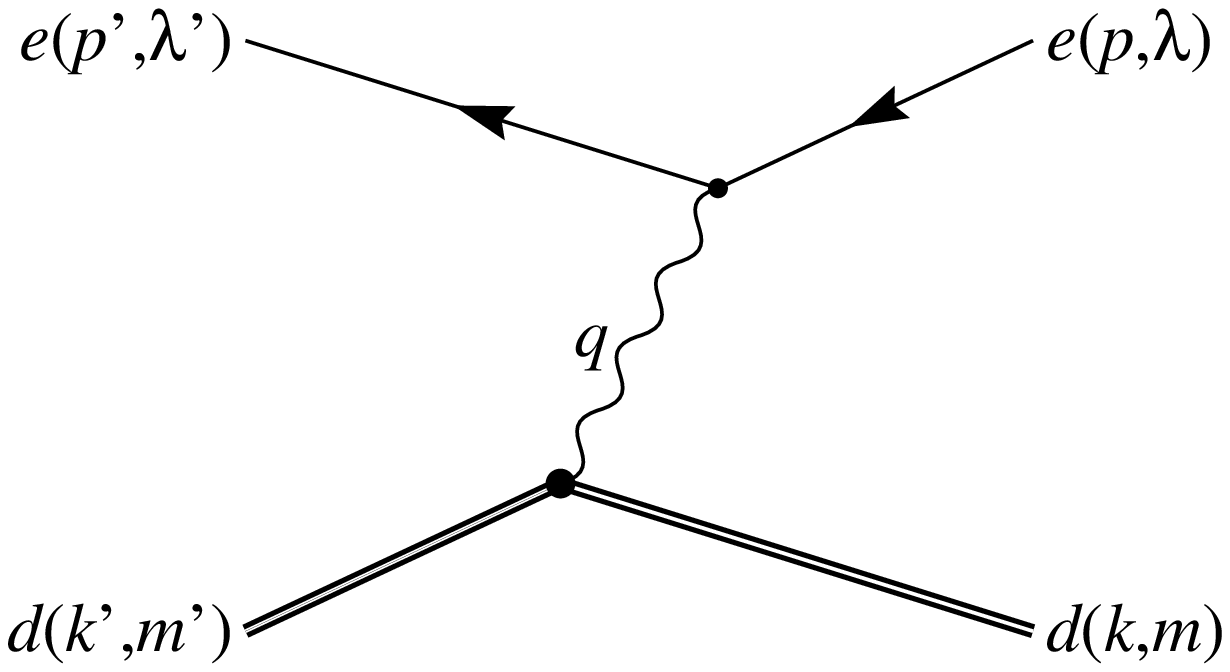}
\caption{The Feynman diagram for one-photon-exchange electron-deuteron
scattering.
\label{form:edscat}}
\end{center}
\end{figure}

In the one-photon-exchange approximation, shown in
Fig.~\ref{form:edscat}, the amplitude of the
scattering process $ed\rightarrow ed$ is just the contraction of the
electron and deuteron currents, multiplied by the photon propagator,
\begin{eqnarray}
\langle p', \lambda' | j_\mu^{e} | p, \lambda \rangle
\frac{1}{q^2}
\langle k', m' | J^\mu_d | k, m \rangle,
\end{eqnarray}
where
\begin{eqnarray}
\langle p', \lambda' | j_\mu^{e} | p, \lambda \rangle
&=&
e \overline{u}(\bm{p}',\lambda') \gamma_\mu u(\bm{p},\lambda).
\end{eqnarray}

From Lorentz covariance, parity invariance, and time reversal invariance
\cite{Hummel:1990zz,Zuilhof:1980ae,Rupp:1990sg,Garcon:2001sz,Arnold:1981zj},
we infer that the deuteron form factor can be written as
\begin{eqnarray}
\langle k', m' | J^\mu_d | k, m \rangle
&=&
- \frac{e}{2 M_d}
(e^*)^\rho(\bm{k}',m')
J^\mu_{\rho \sigma}
e^\sigma(\bm{k},m), \label{eq:deutcurrent}
\end{eqnarray}
where the spin-1 polarization vectors satisfy
\begin{eqnarray}
e_\mu^*(\bm{k},m) e^\mu(\bm{k},m') &=& -\delta_{m,m'}, \\
\sum_{m} e^*_\mu(\bm{k},m) e_\nu(\bm{k},m) &=& - g_{\mu \nu} +
\frac{k_\mu k_\nu}{M_d^2}, \\
k_\mu e^\mu(\bm{k},m) &=& 0,
\end{eqnarray}
and the operator $J^\mu_{\rho\sigma}$ is given by
\begin{eqnarray}
J^\mu_{\rho\sigma} &=& (k'_\mu + k_\mu) \left[
g_{\rho \sigma} F_1(q^2) - \frac{q_\rho q_\sigma}{2 M_d^2} F_2(q^2)
\right] - I^{\mu \nu}_{\rho \sigma} q_\nu G_1(q^2),
\end{eqnarray}
where
$I_{\rho\sigma}^{\mu\nu}=g_\rho^\mu{}g_\sigma^\nu-g_\rho^\nu{}g_\sigma^\mu$
is the generator of infinitesimal Lorentz transformations, $q=k'-k$, and
$F_1$, $F_2$, and $G_1$ are functions of $q^2$, the invariant mass of
the photon. Some conventions \cite{Chung:1988my,Frankfurt:1993ut} denote
$G_1$ as $-F_3$.

The $F_1$, $F_2$, and $G_1$ form factors
are related to the deuteron charge, magnetic, and quadrapole form
factors, denoted by $F_C$, $F_M$, and $F_Q$ respectively, 
by \cite{Coester:1975hj,Hummel:1990zz,Rupp:1990sg}
\begin{eqnarray}
F_C &=& F_1 + \frac{2}{3} \eta \left[ F_1 + (1+\eta) F_2 - G_1 \right],
\label{ff:eq:fc} \\
F_M &=& G_1, \\
F_Q &=& F_1 + (1+\eta) F_2 - G_1, \label{ff:eq:fq}
\end{eqnarray}
where
\begin{eqnarray}
\eta &=& \frac{-q^2}{4 M^2_d}. \label{eq:defeta}
\end{eqnarray}
Note that since both the initial and final electron and deuteron are
on-shell, $-q^2>0$. We define $Q^2=-q^2$.

At zero-momentum transfer, the deuteron charge, magnetic, and
quadrapole form factors are simply
\begin{eqnarray}
F_C(0) &=& 1, \\
F_M(0) &=& \frac{M_d}{m_p} \mu_d = 1.71293,\\
F_Q(0) &=& M_d^2 Q_d = 25.8525,
\end{eqnarray}
where $m_p$ is the mass of the proton,
$\mu_d$ is the magnetic moment of the deuteron (in nuclear magnetons
$\mu_N=\frac{e}{2m_p}$), and $Q_d$ is the electric quadrapole moment of
the deuteron (in units of the deuteron mass, although typically it is
measured in $e\cdot$barns). The numerical values of the form factors are
determined by using experimentally measured values
\cite{Garcon:2001sz,Ericson:1983ei,Mohr:2000}.

Another set of form factors are $G_0$, $G_1$, and $G_2$, commonly used
in light-front dynamics
\cite{Cardarelli:1995yq,Chung:1988my,Frankfurt:1993ut}. They are simply
related to the charge, magnetic, and quadrapole form factors,
\begin{eqnarray}
G_0 &=& F_C, \\
G_1 &=& -F_M, \\
G_2 &=& \frac{\sqrt{8}}{3}\eta F_Q.
\end{eqnarray}

The electron-deuteron cross-section is measured to determine the
deuteron form factors for large momentum transfers. The
unpolarized cross-section has the form 
\cite{Phillips:1998uk,Garcon:2001sz}
\begin{eqnarray}
\frac{d\sigma}{d\Omega_e} &=&
\frac{\sigma_{\text{Mott}}}{ 1+\frac{2E_e}{M_d}\sin^2\frac{\theta_e}{2}}
\left(
A(Q^2) + B(Q^2) \tan^2 \left( \frac{\theta_e}{2} \right) \right),
\end{eqnarray}
where 
\begin{eqnarray}
\sigma_{\text{Mott}} &=&
\left( 
\frac{\alpha \cos\frac{\theta_e}{2}}
{2 E_e \sin^2\frac{\theta_e}{2}} \right)^2,
\end{eqnarray}
and $E_e$ is the electron beam energy, $\theta_e$ is the angle by
which the electron scatters, and $\alpha$ is the fine-structure
constant. The structure functions $A$ and $B$ can be expressed in terms
of the charge, magnetic, and quadrapole form factors,
\begin{eqnarray}
A &=& F_c^2 + \frac{8}{9} \eta^2 F_Q^2 + \frac{2}{3}\eta F_M^2, \\
B &=& \frac{4}{3} \eta (1+\eta) F_M^2.
\end{eqnarray}

The three form factors $F_1$, $F_2$, and $G_1$ cannot be determined
uniquely from the unpolarized scattering data since that measurement of
provides only two structure functions. More information is needed, which
can be obtained
from experiments where the polarization of the electron and/or deuteron
is measured \cite{Garcon:2001sz,Arnold:1981zj}. The most commonly
measured quantity is the tensor polarization observable, $T_{20}$,
\begin{eqnarray}
T_{20} &=& -
\frac{\frac{8}{3}\eta F_C F_Q + \frac{8}{9}\eta^2 F_Q^2 + 
\frac{1}{3}\eta F_M^2 \left(1+2(1+\eta)\tan^2\frac{\theta_e}{2} \right)}
{\sqrt{2}\left(A+B \tan^2\frac{\theta_e}{2}\right)}.
\end{eqnarray}
Note that $T_{20}$ depends on the angle $\theta_e$. For ease of
comparison, data for $T_{20}$ is usually presented for
$\theta_e=70^\circ$. Alternatively, one can eliminate the angular
dependence by defining $\widetilde{T}_{20}$, which is $T_{20}$ with
$F_M$ set equal to zero,
\begin{eqnarray}
\widetilde{T}_{20} &=& -
\frac{\frac{8}{3}\eta F_C F_Q + \frac{8}{9}\eta^2 F_Q^2}
{\sqrt{2}\left(F_C^2 + \frac{8}{9} \eta^2 F_Q^2\right)}.
\end{eqnarray}

The extraction of the structure functions from the data is
difficult. In practice, each cross-section measurement is performed at
a different momentum transfer and angle, making it impossible to exactly
disambiguate, for example, $A$ and $B$. One needs to interpolate between
values of $B$ to calculate values for $A$, and vice versa.
The Jefferson Lab t$_{20}$ Collaboration used a self-consistent method
for obtaining the structure functions from the scattering data
\cite{Abbott:2000ak}. They consider a
variety of theoretical models for the deuteron form factors, then fit
the parameters of the models to the measured cross-sections. After
each model is if, it is used as the the interpolation function 
to disentangle the structure functions. This two-step process helps
minimize model-dependent effects in the values of $A$ and $B$.

\subsubsection{Light Front Calculation} \label{sec:nplfcalcem}

Light-front dynamics is particularly well suited to calculating form
factors. One reason is that the generators of boosts in the one, two,
and plus directions are kinematic, so that wave functions calculated
with a truncated potential transform correctly under boosts. This
feature is especially important for form factors at high momentum
transfer since the wave functions must undergo a large boost.

\begin{figure}[!p]
\begin{center}
\epsfig{angle=0,width=5.75in,height=1.25in,file=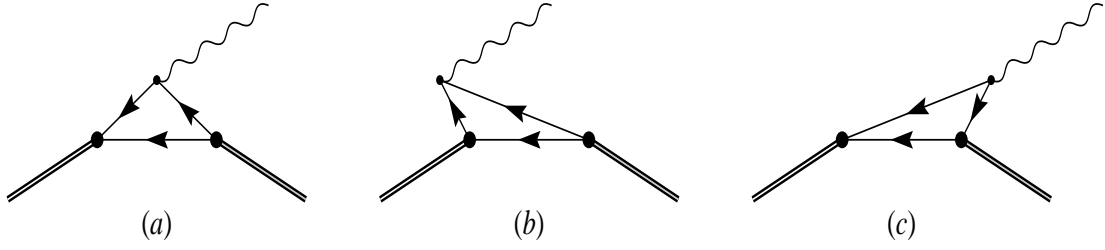}
\caption{The lowest-order graphs which contribute to the deuteron
current matrix element. 
\label{form:deut.cur}}
\end{center}
\end{figure}

Another, more subtle, reason for using the light front is that many of
the graphs which contribute to the current vanish identically. For
example, the three lowest-order graphs for the current are shown in
Fig.~\ref{form:deut.cur}. The double line denotes the deuteron, and the
vertex of the deuteron lines and the nucleon lines represents the
deuteron wave function. The graph labeled (a) does not vanish and is
calculated in section~\ref{ff:impapp}.

Fig.~\ref{form:deut.cur}(b) vanishes in light-front dynamics. To see
why, we first note three facts: the plus component of each particle in
light-front dynamics is non-negative (for massive particles it must be
positive), the plus component of the momentum is conserved, and the plus
momentum of the vacuum is zero. Combining these facts, we find that any
vertex which has particles on one side and vacuum on the other must
vanish. In other words, the vacuum is trivially empty, and no graphs
couple to it.

For Fig.~\ref{form:deut.cur}(c), the coupling of the photon to
the nucleon goes like
$\overline{u}_{\text{LF}}\gamma^\mu{}v_{\text{LF}}$, where the
light-front spinors are defined in the Appendix. For
$\mu=+$, this matrix element is suppressed maximally, and thus $J^+$ is
the ``good'' component of the current
\cite{Frankfurt:1993ut,Dashen:1966,Kondratyuk:1984kq}.
We calculate only $J^+$, since it is the most stable.

We do not consider the contribution of higher-order graphs to the deuteron current, such as graphs where the photon couples to a meson or a nucleon while a meson is present. The omission of the meson-exchange currents does not affect the rotational properties of the current, although it does affect the overall values of the deuteron form factors. This is acceptable since we are only concerned with the rotational properties in this work, not in the detailed results.

The neglect of the graphs where the photon couples to a nucleon while a meson is present may affect the rotational properties of the current. This is because the deuteron current shown in Fig.~\ref{form:deut.cur}(a) is not formally conserved \cite{Phillips:1998uk}, but current conservation is necessary (although not sufficient) for the current to have the correct properties under rotation. To construct the conserved current operator associated with a given wave function, the current must include diagrams that are related to the potential used to calculate the wave function. We expect that these diagrams are small since they contain meson propagators, and that neglecting them should not significantly affect the conservation of the deuteron current or the rotational properties of the current.

\subsubsection{Symmetries of the Electromagnetic Current} 

Now, we use symmetries to relate the components of
$\langle k', m'|J^+(q)| k,m\rangle$. Although in light-front dynamics
some generators of Lorentz transformations are dynamic, such as the
generators of rotations from the $x$-$y$ plane into the $x$ direction
are dynamic, we are free to boost in the plane 
perpendicular to the $z$-axis, boost in the plus direction, and rotate
about the $z$-axis, since the generators of those transformations are
kinematic. In addition, we will find how the states and the
current operator transforms under parity and time-reversal.

The kinematic generators allow us to choose which frame to evaluate the
current in. We choose the Breit frame \cite{Frankfurt:1993ut}, where
$q^+=q^-=q_\perp^y=0$ and $q_\perp^x=Q$. We also choose the plus
momentum of the deuteron to be $M_d$, since $M_d$ is value of the plus
momentum in deuteron's rest frame. To simplify notation, we define the
matrix elements of $J^+$ as 
\begin{eqnarray}
I^+_{m',m}(Q)
&=& 
\left\langle
 \frac{\bm{q}_\perp}{2}, m' \right| J^+(Q) \left|
-\frac{\bm{q}_\perp}{2}, m 
\right\rangle. \label{eq:ff:defimm}
\end{eqnarray}
This quantity is represented with $I_{m'm}$ instead of $J_{m'm}$ because
$J_{m'm}$ is used to represent the matrix elements of $J$ using the
instant-form spin basis. This distinction is discussed later in this
section.

Next, we consider rotation about the $z$-axis by $\pi$ ($R_z(\pi)$),
parity ($\Pi$), and time-reversal ($\Theta$). First note that
the current operator transforms as
\begin{eqnarray}
R_z(\pi) J^\pm R_z(\pi) &=& J^\pm, \\
\Pi J^\pm \Pi &=& J^\mp, \\
\Theta J^\pm \Theta &=& J^\mp.
\end{eqnarray}
Obviously, we must use both parity and time reversal to retain
the plus component of the current. Now look at how the deuteron states
transform:
\begin{eqnarray}
R_z(\pi)  | -\bm{k}_\perp, m \rangle
      &=& | +\bm{k}_\perp, m \rangle (-1)^m, \\
\Theta\Pi | -\bm{k}_\perp, m \rangle
      &=& | -\bm{k}_\perp, -m \rangle (-1)^m.
\end{eqnarray}
Although the symmetry operators can introduce extra, constant phase
factors, they are omitted here since they cancel when calculating matrix
elements. These results make it easy to find what each of these
symmetry operators does to the matrix element. Under rotations a
rotation of $\pi$ about the $z$-axis, the current matrix element
transforms to
\begin{eqnarray}
I^+_{m',m}(Q) &\rightarrow& (-1)^{m-m'} \left\langle
-\frac{\bm{q}_\perp}{2}, m' \right| J^+(Q) \left|
+\frac{\bm{q}_\perp}{2}, m  \right\rangle. \label{eq:ff:currot}
\end{eqnarray}
Under parity followed by time reversal, the matrix element becomes
\begin{eqnarray}
I^+_{m',m}(Q) &\rightarrow& (-1)^{m-m'} \left\langle
-\frac{\bm{q}_\perp}{2}, -m \right| J^+(Q) \left|
+\frac{\bm{q}_\perp}{2}, -m' \right\rangle. \label{eq:ff:curpith} \\
\end{eqnarray}
Since $J^\mu$ is Hermitian, we can also take the complex conjugate of
the matrix element to get
\begin{eqnarray}
I^+_{m',m}(Q) &\rightarrow&  \left\langle
-\frac{\bm{q}_\perp}{2}, m \right| J^+(Q) \left|
+\frac{\bm{q}_\perp}{2}, m' \right\rangle. \label{eq:ff:curcc}
\end{eqnarray}

The appropriate combinations of
Eqs.~(\ref{eq:ff:currot}-\ref{eq:ff:curcc}) give \cite{Chung:1988my}
\begin{eqnarray}
I^+_{m',m}(Q)
&=& (-1)^{m-m'} I^+_{-m',-m }(Q), \label{eq:ff:blah1} \\
&=& (-1)^{m-m'} I^+_{ m , m'}(Q). \label{eq:ff:blah2}
\end{eqnarray}
The same relations also apply for $J^+_{m'm}$ matrix elements.
Eqs.~(\ref{eq:ff:blah1}) and (\ref{eq:ff:blah2})
imply that of the nine possible matrix elements of
$J^+$, there are only four independent components. We choose those
components to be $I^+_{11}$, $I^+_{10}$, $I^+_{1-1}$, and $I^+_{00}$. It
is helpful to express the matrix elements in a matrix to see the
symmetry properties explicitly.
\begin{eqnarray}
I^+_{m',m} &=&
\left(\begin{array}{ccc}
 I^+_{1 1} &  I^+_{10} & I^+_{1-1} \\
-I^+_{1 0} &  I^+_{00} & I^+_{1 0} \\
 I^+_{1-1} & -I^+_{10} & I^+_{1 1} \\
\end{array}\right). \label{eq:deutcurlf}
\end{eqnarray}

\subsubsection{Rotational Invariance and the Angular Condition}
\label{sec:rotinv}

This is as far as we can go with light-front dynamics, but there
should be an additional redundancy in our matrix elements. We have
derived four independent components, whereas in a fully covariant
framework there are only three form factors. The resolution of this
conflict is that full rotational invariance imposes an
{\em angular condition} on the light-front matrix elements.

The angular condition can be found by using the deuteron polarization
vectors
for the Breit frame to calculate the current given in
Eq.~(\ref{eq:deutcurrent}) in terms of $F_1$, $F_2$, and $G_1$.
By comparing that result with Eq.~(\ref{eq:deutcurlf}), we obtain
linear relations between the current matrix elements ($I^+_{11}$,
$I^+_{10}$, $I^+_{1-1}$, and $I^+_{00}$) and the form factors ($F_1$,
$F_2$, and $G_1$) and also find that in general there is a deviation
from the angular condition, which we denote with $\Delta$, given by
\cite{Cardarelli:1995yq}
\begin{eqnarray}
\Delta &=& -I^+_{00} + (1+2\eta) I^+_{11} + I^+_{1-1} - 2\sqrt{2\eta}
I^+_{10},
\end{eqnarray}
where $\eta$ is defined by Eq.~(\ref{eq:defeta}). Since $\Delta$
vanishes when the deuteron current transforms correctly
under rotations, we interpret $\Delta$ as a measure of the extent to
which the current transforms incorrectly.

The form factors are overdetermined by the current matrix elements,
which means there are many different relations between them are
possible. When $\Delta$ is zero, the angular condition can be used to
show that all the relations are equivalent, while a non-zero $\Delta$
means that the form factors depend also on which relation is chosen.

Since $\Delta$ is non-zero in general, it is important to choose the
best relation to obtain the form factor. To do this, we classify the
current matrix elements as
either ``good'' or ``bad''. This classification is similar to the one
made for choosing which component of the current to use. 

In this work we are interested in the overall rotational invariance
properties; the choice of how to relate the form factors is simply
useful for comparing with other approaches. In that spirit, we consider
four different choices \cite{Cardarelli:1995yq}. First, Grach and
Kondratyuk (GK) derive a relation in which $I^+_{00}$ is considered
bad. Using the angular condition to eliminate the $I^+_{00}$ for the
equations, they obtain
\cite{Grach:1984hd}
\begin{eqnarray}
G_{0,GK} &=& \frac{1}{3} \left[ (3-2\eta) I^+_{11}
+ 2\sqrt{2\eta} I^+_{10} + I^+_{1-1} \right], \\
G_{1,GK} &=& 2 \left( I^+_{11} - \frac{1}{\sqrt{2\eta}} I^+_{10} \right), \\
G_{2,GK} &=& \frac{2\sqrt{2}}{3} \left( -\eta I^+_{11}
+ \frac{2\eta} I^+_{10} - I^+_{1-1} \right).
\end{eqnarray}
Brodsky and Hiller (BH) use a prescription where $I^+_{11}$ is bad,
which results in \cite{Brodsky:1992px}
\begin{eqnarray}
G_{0,BH} &=& \frac{1}{3(1+2\eta)} \left[ (3-2\eta) I^+_{00}
+ 8\sqrt{2\eta} I^+_{10} + 2(2\eta-1) I^+_{1-1} \right], \\
G_{1,BH} &=& \frac{2}{1+2\eta} \left[ I^+_{00} - I^+_{1-1} 
+ (2\eta-1) \frac{I^+_{10}}{\sqrt{2\eta}} \right], \\
G_{2,BH} &=& \frac{2\sqrt{2}}{3(1+2\eta)} \left[ \sqrt{2\eta} I^+_{10}
- \eta I^+_{00} - (1+\eta) I^+_{1-1} \right].
\end{eqnarray}

Frankfurt, Frederico, and Strikman (FFS) start by analyzing the deuteron
current using an equal-time spin basis, where the current matrix
elements
are $J^+_{m'm}$ \cite{Frankfurt:1993ut}. Their prescription uses the
Cartesian basis to find that $J^+_{zz}$ is the bad current. Upon
transformation to the spherical basis and use of the Melosh
transformation \cite{Kondratyuk:1984kq,Melosh:1974cu} to relate
$J^+_{m'm}$ to $I^+_{m'm}$, they find 
\begin{eqnarray}
G_{0,FFS} &=& \frac{1}{3(1+\eta)} \left[
(2\eta+3) I^+_{11} + 2 \sqrt{2\eta} I^+_{10} - \eta I^+_{00}
+ (2\eta + 1) I^+_{1-1} \right], \\
G_{1,FFS} &=& \frac{1}{1+\eta} \left[ I^+_{11} + I^+_{00} - I^+_{1-1}
-\frac{2(1-\eta)}{\sqrt{2\eta}} I^+_{10} \right], \\
G_{2,FFS} &=& \frac{\sqrt{2}}{3(1+\eta)} \left[ -\eta I^+_{11}
+ 2\sqrt{2\eta} I^+_{10} - \eta I^+_{00} - (\eta+2) I^+_{1-1} \right].
\end{eqnarray}
Chung, Polyzou, Coester, and Keister (CCKP) choose the canonical
expressions for the form factors $G_0$, $G_1$, and $G_2$ in terms of the
equal-time current as the starting point
\cite{Chung:1988my,Coester:1975hj}. They use rotations and the Melosh
transformation to express the equal-time current matrix elements in
terms of the light-front current matrix elements, and find that
\begin{eqnarray}
G_{0,CCKP} &=& \frac{1}{6(1+\eta)} \left[
(3-2\eta) (I^+_{11} + I^+_{00} ) 
+ 10 \sqrt{2\eta} I^+_{10} + (4\eta-1) I^+_{1-1} \right], \\
G_{1,CCKP} &=& G_{1,FFS}, \label{eq:cckpffs1} \\
G_{2,CCKP} &=& G_{2,FFS}. \label{eq:cckpffs2}
\end{eqnarray}

\subsubsection{Impulse Approximation on the Light Front} \label{ff:impapp}

We now are ready to relate the deuteron wave function to the current
expressed in Eq.~(\ref{eq:ff:defimm}). The deuteron wave functions
solved in section~\ref{ch:pionly} are used. We use the representation of
the wave function in the spin $|\bm{p}_\perp,p^+,m_1,m_2\rangle$
basis, as discussed in Sec.~\ref{sec:furtherlfnn}. Note that these spins
are expressed in the usual Bjorken and Drell representation
\cite{Bjorken:1964}.
By inserting a complete set of states into Eq.~(\ref{eq:ff:defimm})
and making the momentum of
particle 1 explicit, we get
\begin{eqnarray}
J^+_{m',m}(q) &=& \int d^2p_\perp dp^+
\sum_{m'_1,m_1,m_2}
\left\langle \frac{\bm{q}_\perp}{2},m' \left.|
p^+,\bm{p}_\perp+\frac{\bm{q}_\perp}{2}, m'_1, m_2
\right. \right\rangle
\nonumber \\ && \qquad 2
\left\langle p^+,\bm{p}_\perp+\frac{\bm{q}_\perp}{2}, m'_1 \left|
 J_{(S)}^+(\bm{q}_\perp)
\left| p^+,\bm{p}_\perp-\frac{\bm{q}_\perp}{2}, m_1 
\right.
\right.
\right\rangle
\nonumber \\ && \qquad
\left\langle p^+,\bm{p}_\perp-\frac{\bm{q}_\perp}{2}, m_1,
m_2 \left| -\frac{\bm{q}_\perp}{2}, m
\right.\right\rangle, \label{eq:ff:curwf}
\end{eqnarray}
where the spin directions of particles 1 and 2 are labeled $m_1$ and
$m_2$, respectively. Also, since the deuteron is an isoscalar
combination of nucleons, the isovector component of the nucleon current
does not contribute and the isoscalar nucleon current is the same for
both nucleons. This allows us to simply double the isoscalar current of
particle 1 instead of using the isoscalar currents of both particle 1
and 2.

We want to boost the deuteron wave functions to the rest frame, since
that is where the wave functions are calculated. Note that the boosts in
the $x$-$y$ plane 
transforms a general vector $(\bm{k}_\perp,k^+)$ by
$\bm{k}_\perp\rightarrow\bm{k}_\perp-\bm{q}_\perp\frac{k^+}{q^+}$,
leaving $k^+$ unchanged. This means that the boost that puts the
deuteron in its rest frame transforms the individual nucleon momentum by
\begin{eqnarray}
\bm{p}_\perp-\frac{\bm{q}_\perp}{2} \rightarrow 
\bm{p}_\perp-(1-x)\frac{\bm{q}_\perp}{2},
\end{eqnarray}
where we use the Bjorken $x$-variable $x=\frac{p^+}{M_d}$.  Note that
the spin labels do not change.  This is because the boost is in the
perpendicular direction, so the spin labels (defined to point in the $z$
direction) are not affected.

This allows Eq.~(\ref{eq:ff:curwf}) to be written as \cite{Arndt:1999wx}
\begin{eqnarray}
J^+_{m',m}(q) &=& \int d^2p_\perp dp^+
\sum_{m'_1,m_1,m_2}
\left\langle m' \left| p^+,\bm{p}_\perp+(1-x)\frac{\bm{q}_\perp}{2},
m'_1, m_2 \right.\right\rangle
\nonumber \\ && \qquad
2
\left\langle p^+,\bm{p}_\perp+\frac{\bm{q}_\perp}{2}, m'_1 \left|
 J_{(S)}^+(\bm{q}_\perp)
\left| p^+,\bm{p}_\perp-\frac{\bm{q}_\perp}{2}, m_1 
\right.\right.\right\rangle
\nonumber \\ && \qquad
\left\langle\left. p^+,\bm{p}_\perp-(1-x)\frac{\bm{q}_\perp}{2}, m_1,
m_2 \right| m \right\rangle. \label{eq:ff:deutcurwf}
\end{eqnarray}
We have dropped explicit mention of the deuteron's momentum.

Before we continue, we note that we can write the nucleon current matrix
elements using the $u$ spinors,
\begin{eqnarray}
&&\left\langle p^+,\bm{p}_\perp+\frac{\bm{q}_\perp}{2}, m'_1 \left|
 J_{(S)}^+(\bm{q}_\perp)
\left| p^+,\bm{p}_\perp-\frac{\bm{q}_\perp}{2}, m_1 
\right.\right.\right\rangle
\nonumber\\&&\qquad=
\overline{u}_{\text{BD}}\left(p^+,\bm{p}_\perp+\frac{\bm{q}_\perp}{2},
m'_1 \right)
 J_{(S)}^+(\bm{q}_\perp)
u_{\text{BD}}\left(p^+,\bm{p}_\perp-\frac{\bm{q}_\perp}{2},
m_1\right),
\end{eqnarray}
where we have used the ``BD'' subscript to denote these as Bjorken and
Drell spinors. Since the nucleons are on-shell, as they must be in a
Hamiltonian theory, we interpret 
$u_{\text{BD}}\left(p^+,\bm{p}_\perp,m\right)$ as 
$u_{\text{BD}}\left(
p_x=p_{\perp,x},
p_y=p_{\perp,y},
p_z=\frac{p^+-p^-}{2}
\right)$, where $p^-=\frac{M^2+p_\perp^2}{p^+}$.

However, the matrix elements can be calculated easily if we convert the
Bjorken and Drell spinors to light-front spinors. This transformation is
formally accomplished with the Melosh transformation $R_M$, given by
\cite{Chung:1988my,Frankfurt:1993ut},
\begin{eqnarray}
R_M &=& \frac{p^+ + M - i
\bm{\sigma}\cdot(\bm{n}\times\bm{p}_\perp)}{\sqrt{(p^++M)^2+p_\perp^2}},
\end{eqnarray}
where $\bm{n}$ points in the direction of the light front. Note that
the Melosh transformation is unitary.

This transformation converts a Bjorken and Drell spinor to a light-front
spinor in the following manner,
\begin{eqnarray}
u_{\text{LF}}(p^+,\bm{p}_\perp,m) &=& \sum_{m'}(R_M^\dagger)_{m,m'}
u_{\text{BD}}(p^+,\bm{p}_\perp,m').
\end{eqnarray}
Using this transformation, we rewrite Eq.~(\ref{eq:ff:deutcurwf}) as
\begin{eqnarray}
J^+_{m',m}(q) &=& \int d^2p_\perp dp^+
\sum_{m'_1,m_1,m_2}
\left[
\sum_{m^{\prime\prime}_1}
\left\langle m' \left| p^+,\bm{p}_\perp+(1-x)\frac{\bm{q}_\perp}{2},
m''_1, m_2 \right.\right\rangle
(R_M)_{m^{\prime\prime}_1,m'_1}
\right]
\nonumber \\ && \qquad
2
\overline{u}_{\text{LF}}\left( p^+,
\bm{p}_\perp+\frac{\bm{q}_\perp}{2}, m'_1 \right)
 J_{(S)}^+(\bm{q}_\perp)
u_{\text{LF}}\left(p^+,\bm{p}_\perp-\frac{\bm{q}_\perp}{2}, m_1 \right)
\nonumber \\ && \qquad
\left[
\sum_{m^{\prime\prime\prime}_1}
(R_M^\dagger)_{m^{\prime\prime\prime}_1,m_1}
\left\langle\left. p^+,\bm{p}_\perp-(1-x)\frac{\bm{q}_\perp}{2},
m^{\prime\prime\prime}_1, m_2 \right| m \right\rangle
\right]
\label{eq:ff:deutcurwf2}.
\end{eqnarray}

\subsubsection{The Nucleon Form Factors}

From Lorentz covariance, parity invariance, and time reversal invariance
\cite{Zuilhof:1980ae,Rupp:1990sg}, the isoscalar part of the nucleon
current can be expressed as
\begin{eqnarray}
J_{(S)}^\mu(q) &=& \gamma^\mu F^{(S)}_1(q)
+ i \frac{\sigma^{\mu\nu} q_\nu}{2M}  F^{(S)}_2(q).
\end{eqnarray}

Taking the matrix elements of $J^+$ with the light-front spinors gives
\begin{eqnarray}
\overline{u}(k', \lambda') J_{(S)}^+(\bm{q}_\perp) u(k, \lambda)
&=& F^{(S)}_1(q)
\overline{u}_{\text{LF}}(k', \lambda') \gamma^+ u_{\text{LF}}(k, \lambda)
\nonumber\\&&
+ F^{(S)}_2(q)
\frac{q_\perp^i}{2M}
\overline{u}_{\text{LF}}(k',\lambda') \gamma^+ \gamma_\perp^i
u_{\text{LF}}(k,\lambda),
\label{fermcurrent}
\end{eqnarray}
where we have parameterized
\begin{eqnarray}
{k'}^+ &=& k^+, \\
\bm{k}_\perp  &=& \bm{p}_\perp + \frac{\bm{q}_\perp}{2},\\
\bm{k}'_\perp &=& \bm{p}_\perp - \frac{\bm{q}_\perp}{2}.
\end{eqnarray}
We use the representation of the spinors given in the Appendix to
simplify the spinor matrix elements appearing in
Eq.~\ref{fermcurrent}. First consider
\begin{eqnarray}
\overline{u}_{\text{LF}}(k', \lambda') \gamma^+ u_{\text{LF}}(k, \lambda)
&=&
\frac{1}{M k^+}
\chi^\dagger_{\text{LF},\lambda'}
\left[ ( M - \bm{\alpha}^\perp\cdot{\bm{k}'}^\perp ) \Lambda_+ 
+ k^+ \Lambda_- \right]
\gamma^+
\nonumber\\&& \qquad\qquad\times
\left[ \Lambda_- 
( M + \bm{\alpha}^\perp\cdot\bm{k}^\perp )
+ \Lambda_+ k^+ \right] \chi_{\text{LF},\lambda}, \\
&=& \frac{k^+}{M} \delta_{\lambda'\lambda}.
\end{eqnarray}
Next,
\begin{eqnarray}
\overline{u}_{\text{LF}}(k', \lambda') \gamma^+ \gamma_\perp^-
u_{\text{LF}}(k, \lambda)
&=&
\frac{1}{M k^+}
\chi^\dagger_{\text{LF},\lambda'}
\left[ ( M - \bm{\alpha}^\perp\cdot{\bm{k}'}^\perp ) \Lambda_+ 
+ k^+ \Lambda_- \right]
\gamma^+ \gamma^i
\nonumber\\&& \qquad\qquad\times
\left[ \Lambda_- 
( M + \bm{\alpha}^\perp\cdot\bm{k}^\perp )
+ \Lambda_+ k^+ \right] \chi_{\text{LF},\lambda}, \\
&=& \frac{k^+}{M}
\chi^\dagger_{\text{LF},\lambda'} \gamma^+ \gamma_\perp^i
\chi_{\text{LF},\lambda}, \\
&=& - i \epsilon^{ij3} 
\frac{k^+}{M} \chi^\dagger_{\lambda'} \sigma^j \chi_\lambda.
\end{eqnarray}
Combining these, we find \cite{Frankfurt:1993ut,Kondratyuk:1984kq}
\begin{eqnarray}
\langle k', \lambda' | J_{(S)}^+(\bm{q}_\perp) | k, \lambda \rangle
&=& \frac{k^+}{M}
\chi_{\lambda'}^\dagger \left[
F^{(S)}_1(q) - \frac{1}{2M} F^{(S)}_2(q)
\left( i q^i_\perp \epsilon^{ij3} \sigma^j_\perp \right)
\right]
\chi_\lambda \label{eq:ff:nuccurrent}
\end{eqnarray}

We can rewrite Eq.~(\ref{eq:ff:nuccurrent}) as
\begin{eqnarray}
\langle k', \lambda' | J_{(S)}^+(\bm{q}_\perp) | k, \lambda \rangle
&=& 
F^{(S)}_1(q)
\langle k', \lambda' | J_{(1S)}^+(\bm{q}_\perp) | k, \lambda \rangle
+ \nonumber\\&&
F^{(S)}_2(q)
\langle k', \lambda' | J_{(2S)}^+(\bm{q}_\perp) | k, \lambda \rangle,
\end{eqnarray}
which, when inserted into Eq.~(\ref{eq:ff:deutcurwf2}), gives
\begin{eqnarray}
I^+_{m',m}(q) &=&
F^{(S)}_1(q) I^+_{(1)m',m}(q) +
F^{(S)}_2(q) I^+_{(2)m',m}(q),
\end{eqnarray}
where we define
\begin{eqnarray}
I^+_{(i)m',m}(q) &=& \int d^2p_\perp dp^+
\sum_{m'_1,m_1,m_2}
\langle m' | p^+,\bm{p}_\perp+(1-x)\frac{\bm{q}_\perp}{2},
m'_1, m_2 \rangle
\nonumber \\ && \qquad
2
\langle p^+,\bm{p}_\perp+\frac{\bm{q}_\perp}{2}, m'_1 |
 J_{(iS)}^+(\bm{q}_\perp)
| p^+,\bm{p}_\perp-\frac{\bm{q}_\perp}{2}, m_1 \rangle
\nonumber \\ && \qquad
\langle p^+,\bm{p}_\perp-(1-x)\frac{\bm{q}_\perp}{2}, m_1,
m_2 | m \rangle, \label{eq:ff:i12}
\end{eqnarray}
for $i=1,2$. Note that both $J^+_{(1)m',m}(q)$ and $J^+_{(2)m',m}(q)$
must satisfy the same equations as $J^+_{m',m}(q)$ does. In particular,
this means that the angular condition should apply to $J^+_{(1)m',m}(q)$
and $J^+_{(2)m',m}(q)$ independently, so we consider the deviation from
the angular condition for each.

\begin{figure}[!p]
\begin{center}
\epsfig{angle=0,width=5.5in,height=5.5in,file=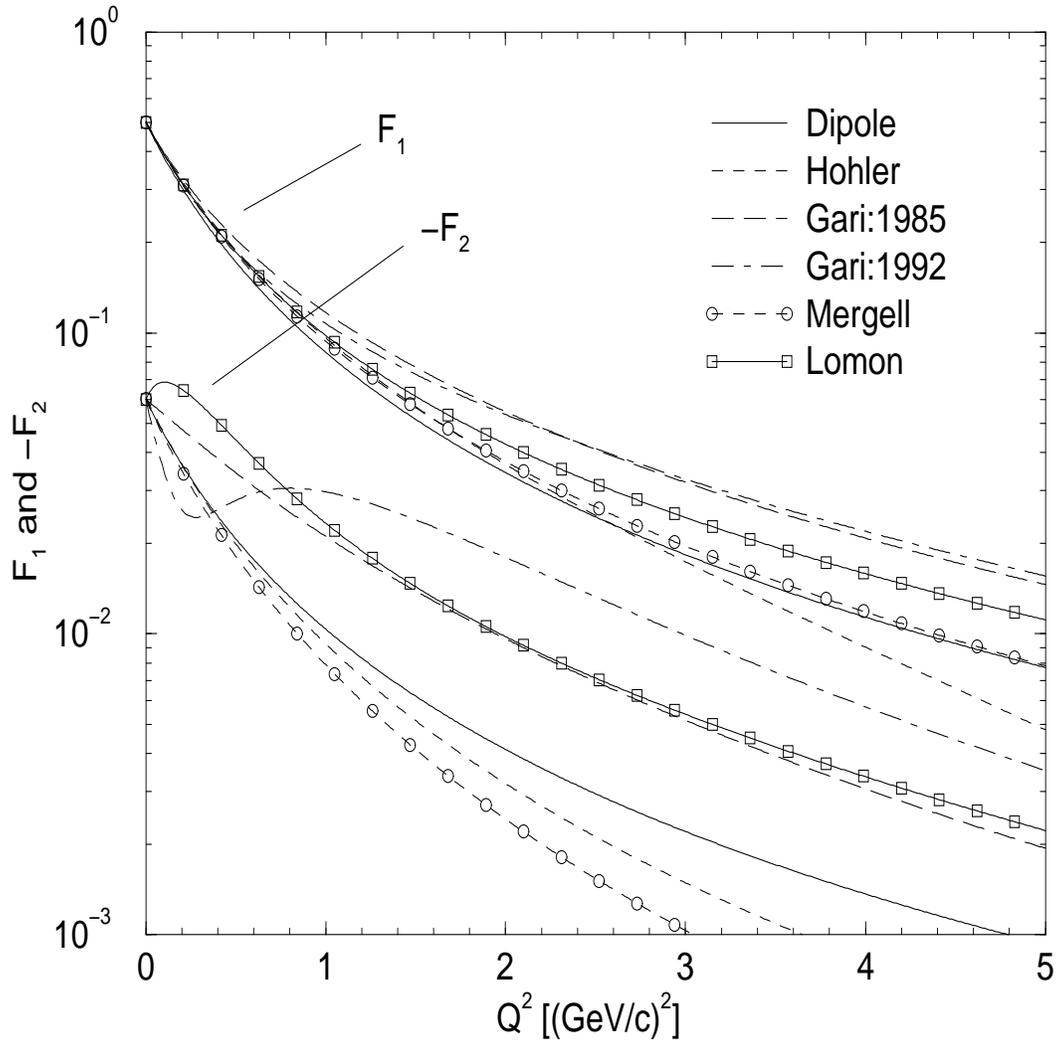}
\caption{The $F_1$ and $-F_2$ isoscalar nucleon form factors for six
different models: the dipole model, Hohler, Gari:1985, Gari:1992,
Mergell, and Lomon. 
\label{form:diag:diffNucFF}}
\end{center}
\end{figure}

There are many parameterizations of the isoscalar nucleon form factors
$F^{(S)}_1(q)$ and $F^{(S)}_2(q)$. Since the measurement of the
electron-nucleon cross section is difficult, the data have large errors
and are consistent with several different models of the nucleon form
factors. Some of the models representative of those proposed in the
literature are: the dipole
model, fit 8.2 of Hohler \cite{Hohler:1976ax}, Gari:1985 \cite{Gari:1985ia},
model 3 of Gari:1992 \cite{Gari:1992qw,Gari:1992}, best fit for the
multiplicative parameterization of Mergell \cite{Mergell:1996bf}, and
model DR-GK(1) of Lomon \cite{Lomon:2001ga}. The $F^{(S)}_1(q)$ and
$F^{(S)}_2(q)$ form factors for each of these models are shown in
Fig.~\ref{form:diag:diffNucFF}.

We can relate the isovector form factors to $G_{Ep}$, $G_{Mp}$,
$G_{En}$, and $G_{Mn}$, the proton electric, proton
magnetic, neutron electric, and neutron magnetic form factors,
respectively, with \cite{deJager:1999ce}
\begin{eqnarray}
F^{(S)}_1 &=& 
\frac{ G_{Ep}+G_{En}+ \tau \left(G_{Mp} + G_{Mn} \right)}{2(1+\tau)}, \\
F^{(S)}_2 &=& 
\frac{-G_{Ep}-G_{En}+ G_{Mp} + G_{Mn}}{2(1+\tau)},
\end{eqnarray}
where $\tau\equiv\frac{Q^2}{4M}$. The value of $\tau$ is approximately 1
at a momentum transfer of about 5~GeV$^2$, the upper range of momentum
transfers that we consider. Since the overall magnitudes of the form
factors are similar at this momentum transfer, it is important to
measure each of the form factors with the same accuracy and cover the
same range of momentum transfers. Currently, the most poorly known form
factor is $G_{En}$, both in terms of the magnitude of the error and in
the number of data points \cite{Lomon:2001ga}.

In section~\ref{ch:ff:results}, we will find that for momentum transfers
greater than about 2~GeV$^2$, the spread in the values of the deuteron
form factors due to the breaking of rotational invariance on the light
front is smaller than the spread in values due to using the various
nucleon form factors. It is uncertainty of the nucleon form factors, not
the use of the light front, that limits the accuracy of the deuteron
form factors at large momentum transfers. Only more accurate
measurements of the nucleon form factors, especially $G_{En}$, will
allow for more accurate deuteron form factor calculations. 

\subsection{Axial Form Factors} \label{sec:axialstuff}

The formalism used for the axial current and form factor is very similar
to that used for the electromagnetic current and form factor. Thus, most
of the discussion from the previous section carries over here with only
slight modifications. We highlight only the differences.

The derivation of the symmetries of the axial current matrix elements is
almost the same as in section~\ref{sec:nplfcalcem}, with the
exception that under parity, the axial current picks up a negative sign
\begin{eqnarray}
\Pi J_5^\pm \Pi &=& -J_5^\mp.
\end{eqnarray}
When this is propagated through the algebra, we find that
\begin{eqnarray}
I^+_{(5)m',m}(Q)
&=& -(-1)^{m-m'} I^+_{(5)-m',-m }(Q), \label{eq:ff:blah1axial} \\
&=& (-1)^{m-m'} I^+_{(5) m , m'}(Q). \label{eq:ff:blah2axial}
\end{eqnarray}
Eqs.~(\ref{eq:ff:blah1axial}) and (\ref{eq:ff:blah2axial})
imply that of the nine possible matrix elements of
$J_5^+$, there are only two independent components. We choose those
components to be $I^+_{(5)11}$, and $I^+_{(5)10}$. It
is helpful to express the matrix elements in a matrix to see the
symmetry properties explicitly.
\begin{eqnarray}
I^+_{(5)m',m} &=&
\left(\begin{array}{ccc}
 I^+_{(5)1 1} &  I^+_{(5)10} &  0            \\
-I^+_{(5)1 0} &  0           & -I^+_{(5)1 0} \\
 0            &  I^+_{(5)10} & -I^+_{(5)1 1} \\
\end{array}\right). \label{eq:deutcurlfaxial}
\end{eqnarray}

We have derived two independent components, but an analysis of the
covariant theory shows that only one deuteron form factor ($F_A$)
contributes for the plus component of the axial current 
\cite{Frederico:1991vb}. This implies that the requirement of full
rotational invariance imposes an angular condition on the light-front
axial current matrix elements. 
The deviation from the angular condition, denoted by $\Delta$, given by
\cite{Frederico:1991vb},
\begin{eqnarray}
\Delta &=& \frac{\sqrt{2\eta}}{2} I^+_{(5)11} - I^+_{(5)10}.
\end{eqnarray}

Since the deuteron axial form factor is overdetermined by the current
matrix elements, we need to classify the current matrix elements as
either ``good'' or ``bad'' to eliminate ambiguity. We consider two such
choices.

Frankfurt, Frederico, and Strikman (FFS) find that the $J^+_{(5)zz}$ is
the bad matrix element \cite{Frankfurt:1993ut}. After transforming to
the spherical basis and using the Melosh transformation, they find that
\begin{eqnarray}
F_A &=& \frac{1}{2(1+\eta)} \left( I^+_{(5)11} + \sqrt{2\eta}
I^+_{(5)10} \right).
\end{eqnarray}
Frederico, Henley, and Miller (FHM) use the behavior of the matrix
elements in the non-relativistic limit to determine that the bad element
is $I^+_{(5)10}$ \cite{Frederico:1991vb}. This means that
\begin{eqnarray}
F_A &=& \frac{1}{2} I^+_{(5)11}.
\end{eqnarray}

The current matrix elements are calculated using the nucleon
axial current. The general form of nucleon axial current is given by
\cite{Frederico:1991vb},
\begin{eqnarray}
J_{(5)n}^\mu &=& \gamma^\mu \gamma^5 F_A^n + q^\mu \gamma^5 F_P^n.
\label{eq:nucaxialff}
\end{eqnarray}
Since we choose to $\mu=+$ and work in the Breit frame, where $q^+=0$,
Eq.~(\ref{eq:nucaxialff}) reduces to
\begin{eqnarray}
J^+_{(5)n} &=& \gamma^+ \gamma^5 F_A^n. \label{eq:nucaxialffplus}
\end{eqnarray}
The light-front spinor matrix elements of Eq.~(\ref{eq:nucaxialffplus})
can be computed using expressions given in the Appendix. The result is
\begin{eqnarray}
\langle k', \lambda' | J_{(5)n}^+(\bm{q}_\perp) | k, \lambda \rangle
&=& \frac{k^+}{M} F^n_A(q) 
\chi_{\lambda'} \sigma^3 \chi_\lambda \label{eq:ff:nucaxialcurrent}.
\end{eqnarray}

We consider only one model for the nucleon axial form factor since the
deuteron axial current has such a simple dependence on it. We choose to
use the dipole model
\begin{eqnarray}
F_A(Q^2) &=& \frac{F_A(0)}{\left(1+\frac{Q^2}{M_A^2}\right)^2},
\end{eqnarray}
where $M_A$ is the axial mass. For our calculations, we use the value
for the axial mass determined by Liesenfeld {\it et
al}.~\cite{Liesenfeld:1999mv}.

\section{Results for the Form Factors} \label{ch:ff:results}

We use the deuteron wave functions obtained for the light-front
nucleon-nucleon potential in section~\ref{ch:pionly} to calculate the
deuteron currents and form factors. This gives a solution where
light-front dynamics is used consistently throughout. For the
potential, we choose the light-front nucleon-nucleon potential with
$f_\sigma=1.2815$. We have verified that the results do not change
significantly when $f_\sigma=1.22$ is used. 

\begin{figure}[!p]
\begin{center}
\epsfig{angle=0,width=5.5in,height=5.5in,file=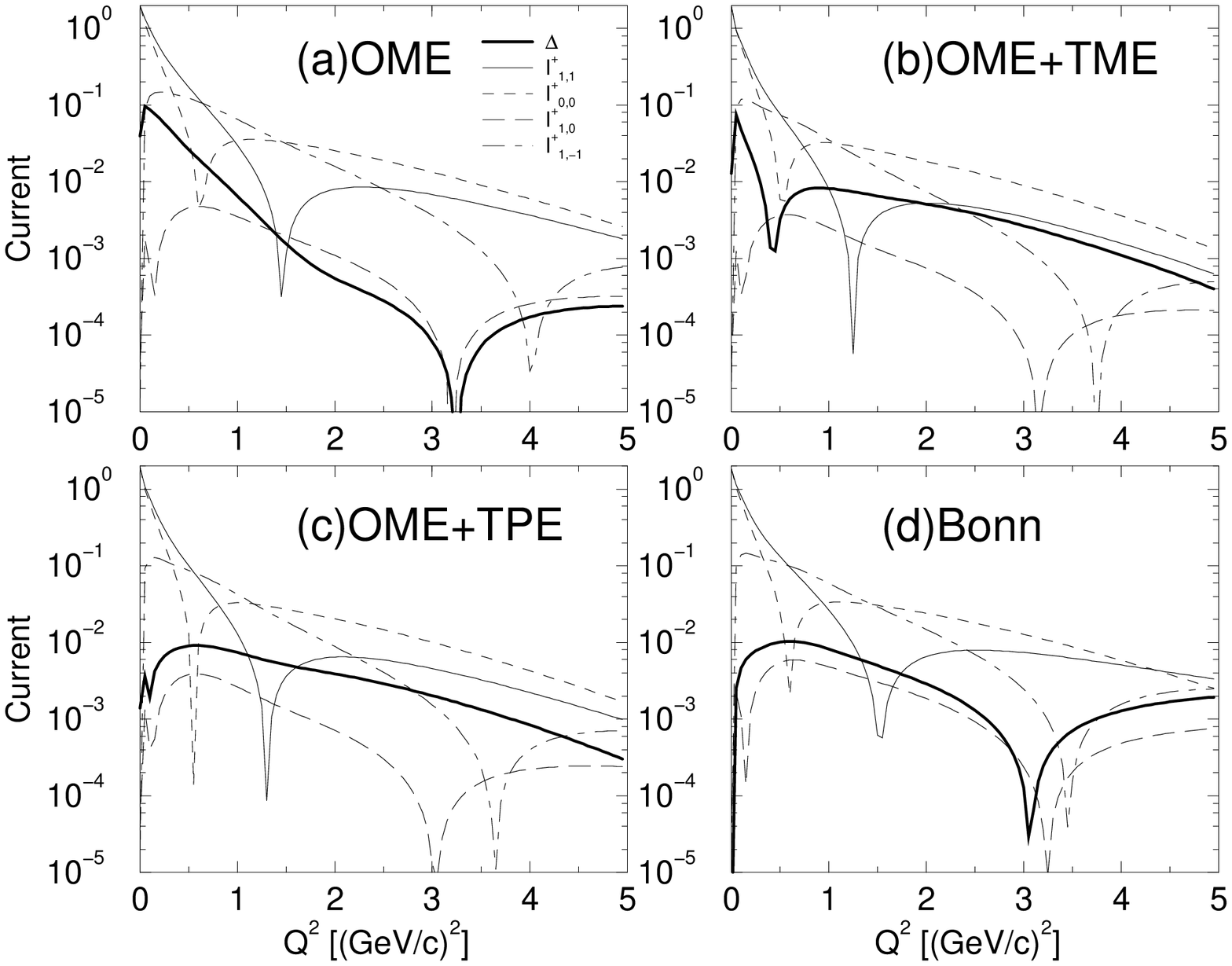}
\caption{The matrix elements of $I^+_{(1)m'm}$, the component of the
electromagnetic current which multiplies the nucleon $F_1$ form factor,
calculated with the wave function from the (a) OME, (b) OME+TME, (c)
OME+TPE, and (d) Bonn potentials.
\label{form:diag:j1}}
\end{center}
\end{figure}

Figure~\ref{form:diag:j1} show the currents and the associated angular
condition for $I^+_{(1)}$, given by Eq.~(\ref{eq:ff:i12}), for several
different deuteron wave functions. Results are shown for the wave
function from the OME, OME+TME, and OME+TPE potentials (calculated in 
section~\ref{ch:pionly}), and the parameterization of the deuteron wave
function for the energy independent Bonn potential
\cite{Machleidt:1987hj}. The currents matrix elements (but not $\Delta$)
are approximately the same regardless of which wave function is used. This
consistency is important, since it verifies that the gross features of
all the models are the same.

We find that $\Delta$ for $I^+_{(1)}$ is much smaller than the largest
matrix elements when using the OME wave function. This means that the
$I^+_{(1)}$ current transforms very well under rotations. This is
somewhat surprising, since we found earlier that the binding energies
for the OME wave functions have a large splitting, indicating that OME
wave functions transforms poorly under rotations. 

Comparing the current calculated with the OME wave function to those
calculated with other potentials, we find that for momentum transfers of
more than 1~GeV$^2$ the OME  $I^+_{(1)}$ current has the best
transformation properties under rotation of all the $I^+_{(1)}$
currents shown.

For smaller momentum transfers, the transformation properties of the Bonn
and OME+TME wave functions are the best. This is expected, since in the
limit of no momentum transfer, the current $I^+_{(1)m'm}$ is simply the
overlap of deuteron wave functions, $\langle m'|m\rangle$. If the
initial and final states have the same mass, the matrix element is
simply $\delta_{m'm}$, which satisfies the angular condition. However,
if the states do not have the same mass (which implies that $m'\neq m$),
there will be a non-zero overlap between the two states, which violates
the angular condition. Since the masses of the deuteron states are
exactly the same for the Bonn wave function, and approximately the same
for the OME+TPE wave function, they have a small $\Delta$ at low
momentum transfer. However, the OME wave functions, having the
largest mass splitting, have the largest $\Delta$ at low momentum
transfers.

\begin{figure}[!p]
\begin{center}
\epsfig{angle=0,width=5.5in,height=5.5in,file=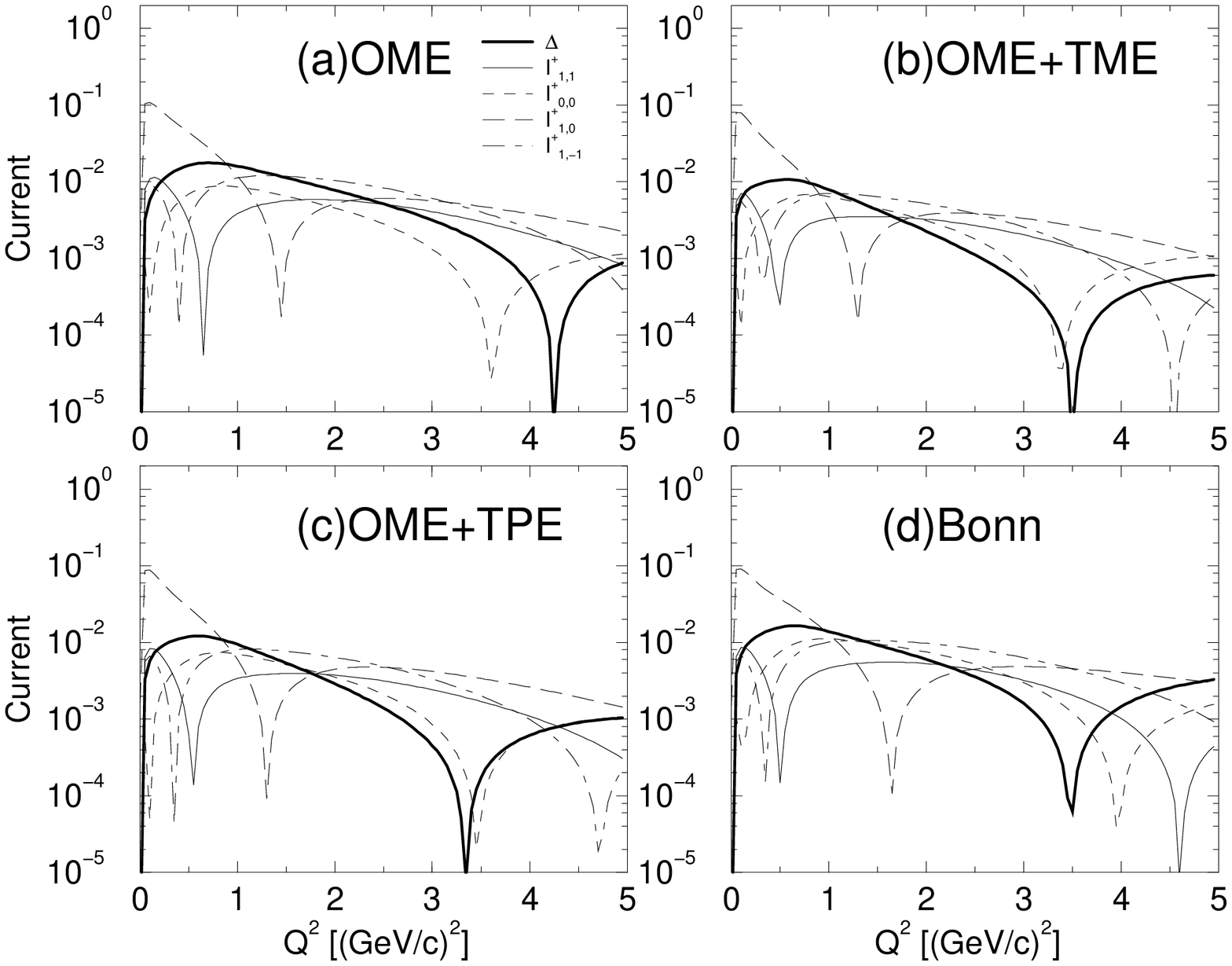}
\caption{The matrix elements of $I^+_{(2)m'm}$, the component of the
electromagnetic current which multiplies the nucleon $F_2$ form factor,
calculated with the wave function from the (a) OME, (b) OME+TME, (c)
OME+TPE, and (d) Bonn potentials.
\label{form:diag:j2}}
\end{center}
\end{figure}

\begin{figure}[!p]
\begin{center}
\epsfig{angle=0,width=5.5in,height=5.5in,file=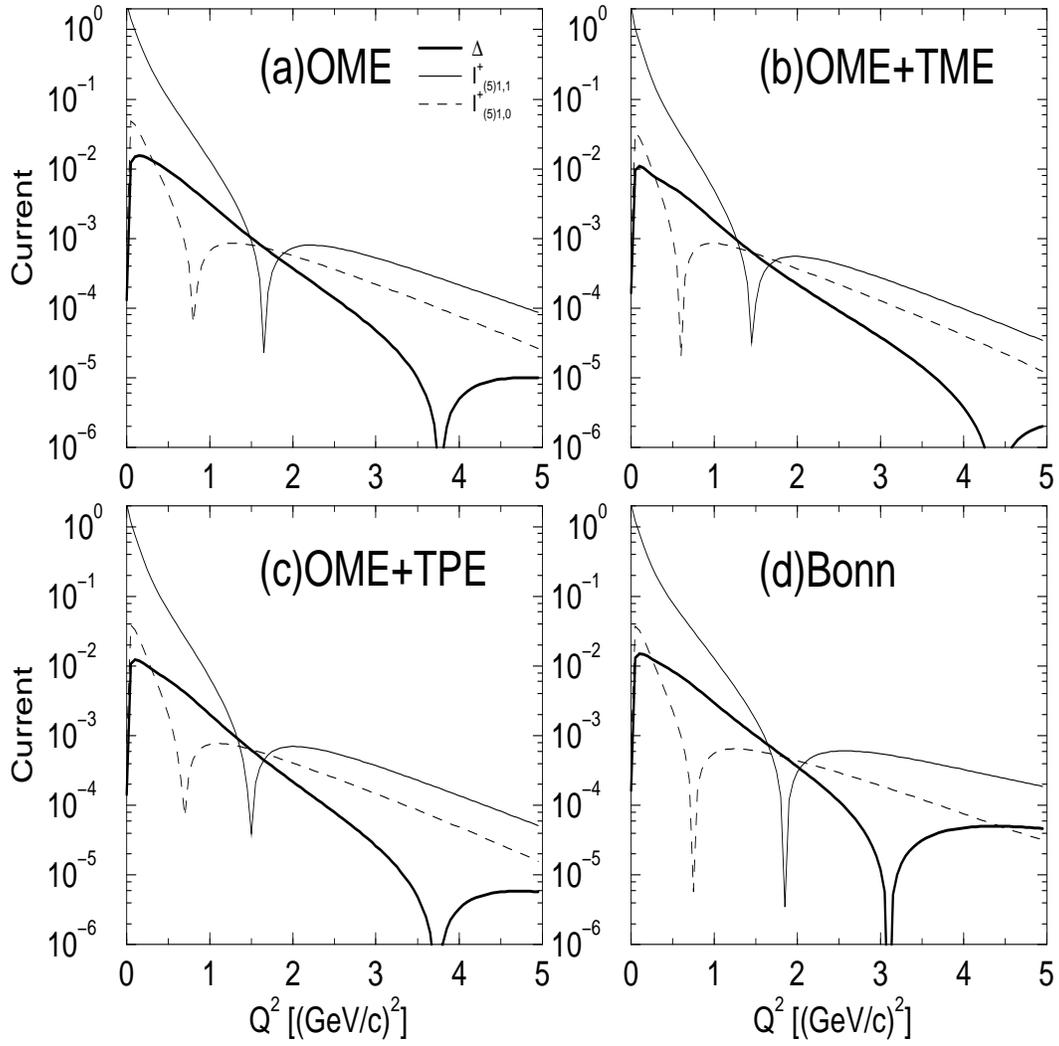}
\caption{The matrix elements of $I^+_{(5)m'm}$, the deuteron axial
current including the nucleon axial form factor, calculated with the
wave function from the (a) OME, (b) OME+TME, (c) OME+TPE, and (d) Bonn
potentials.
\label{form:diag:j5}}
\end{center}
\end{figure}

Figures~\ref{form:diag:j2} and \ref{form:diag:j5} show the current
matrix elements and the angular condition for $I^+_{(2)}$ and
$I^+_{(5)}$, respectively. The general features of these figures are the
same as in Figure~\ref{form:diag:j1}, with one important exception. In
both figures, the $\Delta$ for the
OME wave function has about the same magnitude as the $\Delta$'s for the
other wave functions. This means that the rotational properties
of $I^+_{(2)}$ and $I^+_{(5)}$ currents are approximately the same
regardless of which wave function is used. This is a surprising result
since it indicates that the rotational properties of the current depend
more on how the current is constructed than on which wave function is
used.

In Fig.~\ref{form:diag:j2}, we find that the magnitude of $\Delta$ is
almost the same as the magnitude of the largest matrix element of
$I^+_{(1)}$. This means there is a large deviation from the angular
condition, and that form factors calculated with this current may depend
strongly on which matrix element is chosen as ``bad''. We show below
that this is not the case for the electromagnetic form factors.

We find that $\Delta$ is much smaller than the largest matrix element of
the axial currents shown in Fig.~\ref{form:diag:j5} for most values of
momentum transfer. This means that the deuteron axial form factor will
be essentially independent of which matrix element is chosen as ``bad'',
except for within the range of 1.5 to 2~GeV$^2$.

\begin{figure}[!p]
\begin{center}
\epsfig{angle=0,width=5.5in,height=5.5in,file=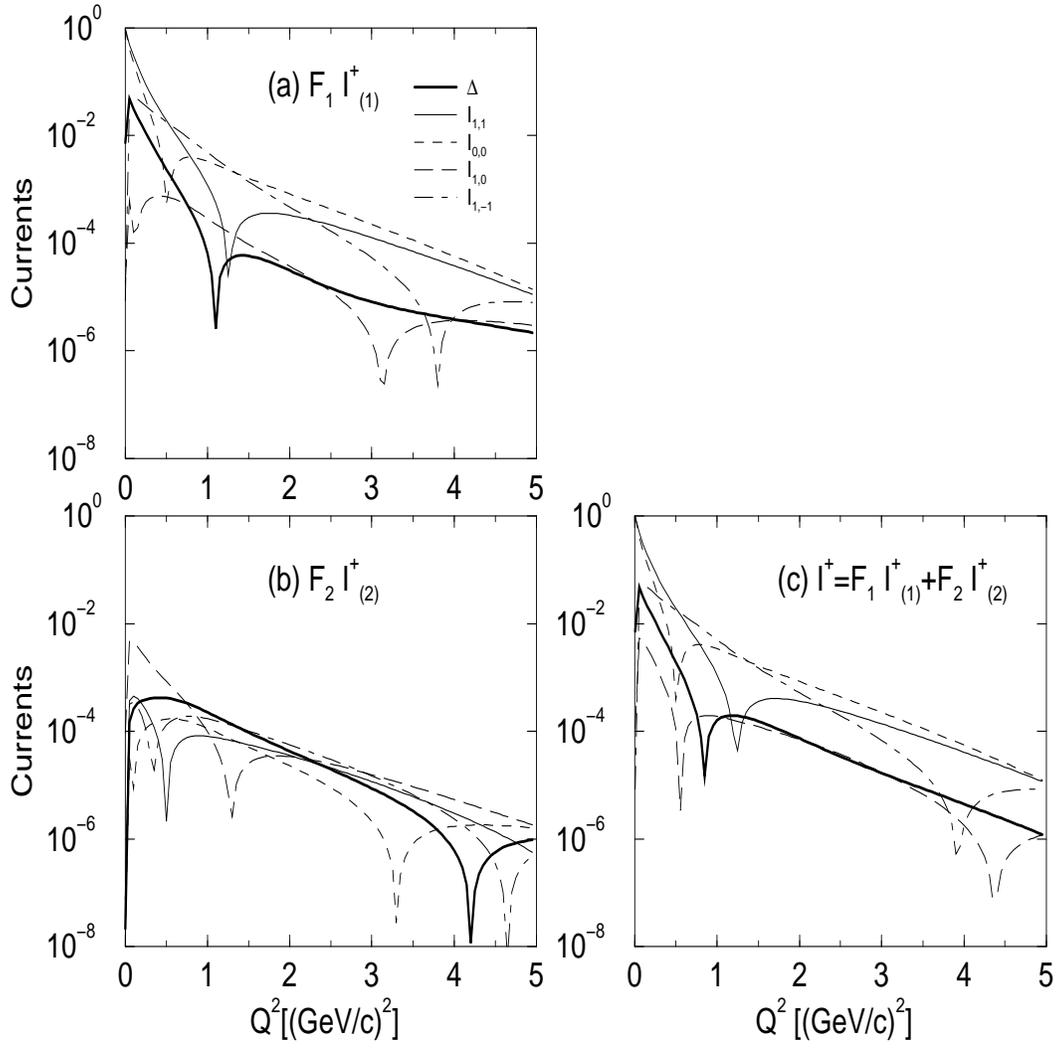}
\caption{The matrix elements for $F_1I^+_{(1)m'm}$, $F_2I^+_{(2)m'm}$, and
$I^+_{m'm}$ calculated with the OME wave functions. The Gari:1985
nucleon isoscalar form factors are used for $F_1$ and $F_2$.
\label{form:diag:gk1985f1f2both}}
\end{center}
\end{figure}

Now we combine the two parts of the electromagnetic current, 
$I^+_{(1)}$ and $I^+_{(2)}$, with the nucleon form factors 
$F_1$ and $F_2$ to get the total current.
Figure \ref{form:diag:gk1985f1f2both} shows the currents for
$F_1I^+_{(1)}$ and $F_2I^+_{(2)}$, as well as the sum, $I^+$.
The Gari:1985 nucleon form factors are used \cite{Gari:1985ia}.
We find that $F_1I^+_{(1)}$ gives the largest contribution to the total
current, and because $\Delta$ is small for $I^+_{(1)}$, $\Delta$ is also
small for the total current, meaning that the total current transforms well
under rotations. Thus, in spite of the fact that $\Delta$ is
approximately the same size as the current matrix elements for
$I^+_{(2)}$, the deuteron electromagnetic form factors should not
depend too strongly on the choice of the ``bad'' matrix element. This is
especially true for the form factors calculated with the OME wave
function. 

\begin{figure}[!p]
\begin{center}
\epsfig{angle=0,width=5.5in,height=5.5in,file=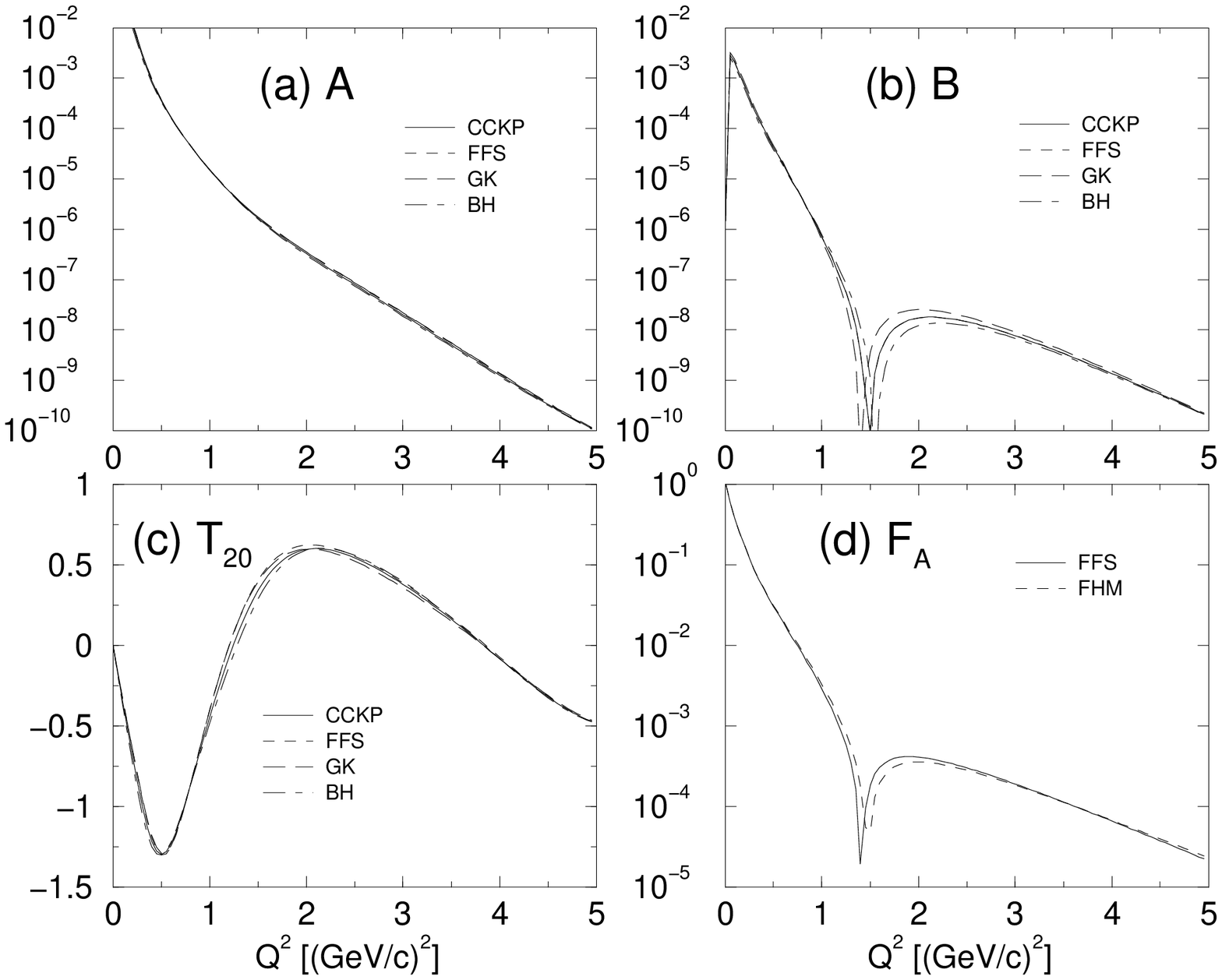}
\caption{The form factors $A$, $B$, $T_{20}$, and $F_A$ calculated using
the various choices of the ``bad'' matrix element. The OME wave function
is used, along with the Lomon nucleon form factors for the
electromagnetic form factors, and the Liesenfeld nucleon form factor for
the axial form factor.
\label{form:diag:ome.allbad}}
\end{center}
\end{figure}

We calculate the form factors $A$, $B$, $T_{20}$, and $F_A$ using the
OME wave function, and show the results in
Fig.~\ref{form:diag:ome.allbad}. In general, the form factors do not
depend strongly on which matrix element is chosen as ``bad'', in
agreement what what we predicted in the previous paragraph. 
The definitions of the ``bad'' matrix elements are given in
sections~ \ref{sec:rotinv} and \ref{sec:axialstuff}.
The only
exception is for the $B$ form factor, and to a lesser extent the $F_A$
form factor, near where they crosses zero. This is not too surprising,
since a small constant shift in any function near a zero-crossing has a
large effect in a logarithmic plot. Also, we note that the FFS and CCKP
choices of the ``bad'' matrix element give the same value for $B$. This
is a consequence of Eqs.~(\ref{eq:cckpffs1}) and (\ref{eq:cckpffs2}).

\begin{figure}[!p]
\begin{center}
\epsfig{angle=0,width=5.5in,height=5.5in,file=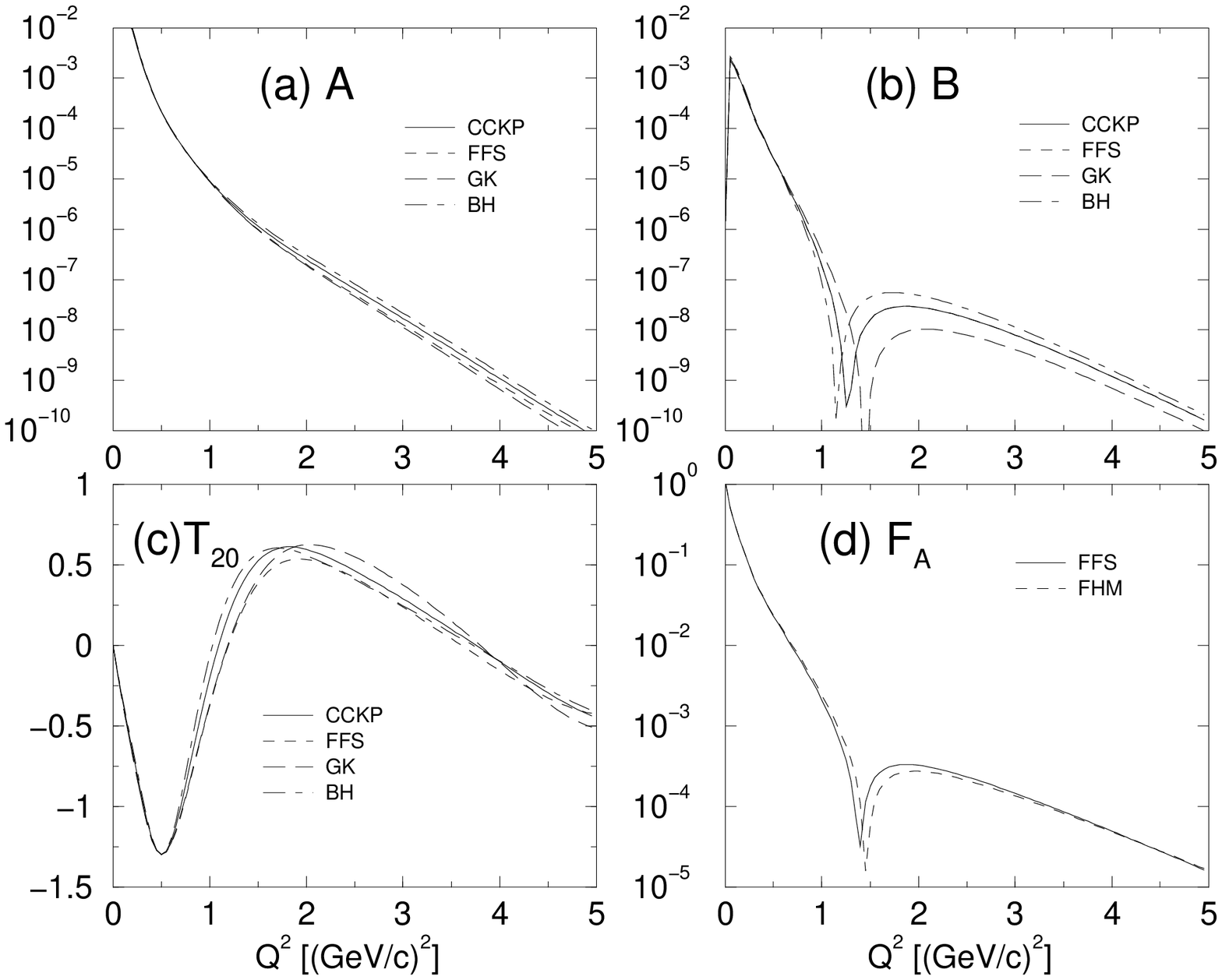}
\caption{The form factors $A$, $B$, $T_{20}$, and $F_A$ calculated using
the various choices of the ``bad'' matrix element. The definitions of
the ``bad'' matrix elements are given in sections~\ref{sec:rotinv} and
\ref{sec:axialstuff}. The OME+TME wave function is used, along with the
Lomon nucleon form factors for the electromagnetic form factors, and the
Liesenfeld nucleon form factor for the axial form factor.
\label{form:diag:tme.allbad}}
\end{center}
\end{figure}

We also use the OME+TME wave function to calculate the form factors $A$,
$B$, $T_{20}$, and $F_A$, which we show in
Fig.~\ref{form:diag:tme.allbad}. We argued earlier that the these
electromagnetic form factors depend more strongly on which matrix
element is chosen as ``bad'' that those calculated with the OME wave
function, and that dependence is clear in this figure. At low momentum
transfers, the dependence on the change is fairly small, but as the
momentum transfer increases, so does the dependence. The axial form
factor is not affected as strongly, primarily because each wave function
generates an axial current which  violates the angular condition by
approximately the same amount. 

\begin{figure}[!p]
\begin{center}
\epsfig{angle=0,width=5.5in,height=5.5in,file=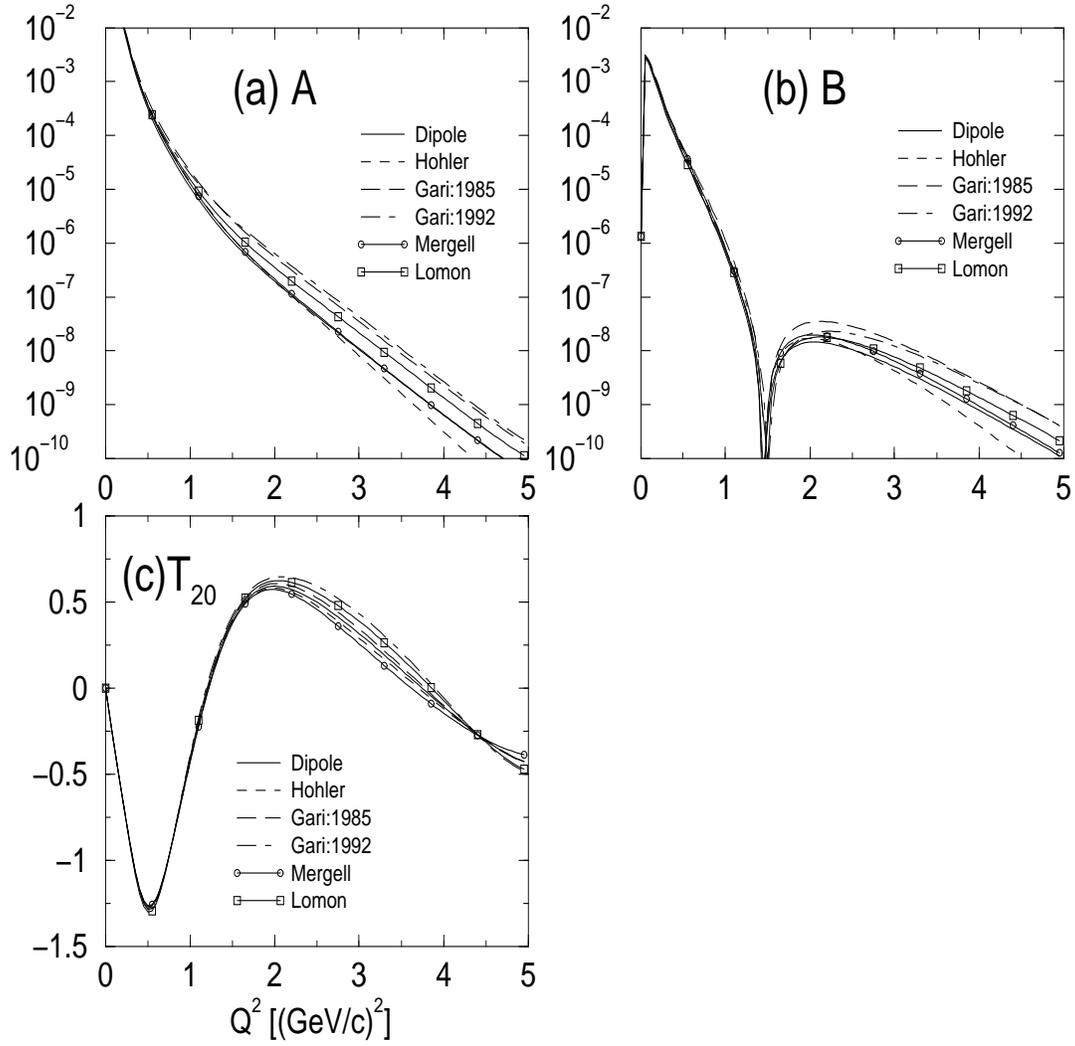}
\caption{The electromagnetic form factors $A$, $B$, and $T_{20}$
calculated using
the various choices of the nucleon isoscalar form factors. The
OME wave function is used, along with the FFS choice of the ``bad''
deuteron current matrix. The axial form factor is not shown since its
dependence on different form factors is trivial.
\label{form:diag:ome.allff}}
\end{center}
\end{figure}

Since there are many different models of the nucleon electromagnetic
form factors, we calculate the deuteron electromagnetic form factors
using each of them to see what effect the differences have. The results
are shown in Fig.~\ref{form:diag:ome.allff}. At low momentum transfers,
all the nucleon form factors give close to the same results. However,
when the momentum transfers is large, we find a large spread in the
values due to nucleon form factors. In fact, this spread is larger than
the spread of values obtained from using different ``bad'' matrix
elements with the OME+TME wave functions. In other words, in order to
obtain accurate results for momentum transfers over 2~GeV$^2$, it is
more important to determine which nucleon form factor to use than when
``bad'' matrix to use.

\begin{figure}[!p]
\begin{center}
\epsfig{angle=0,width=5.5in,height=5.5in,file=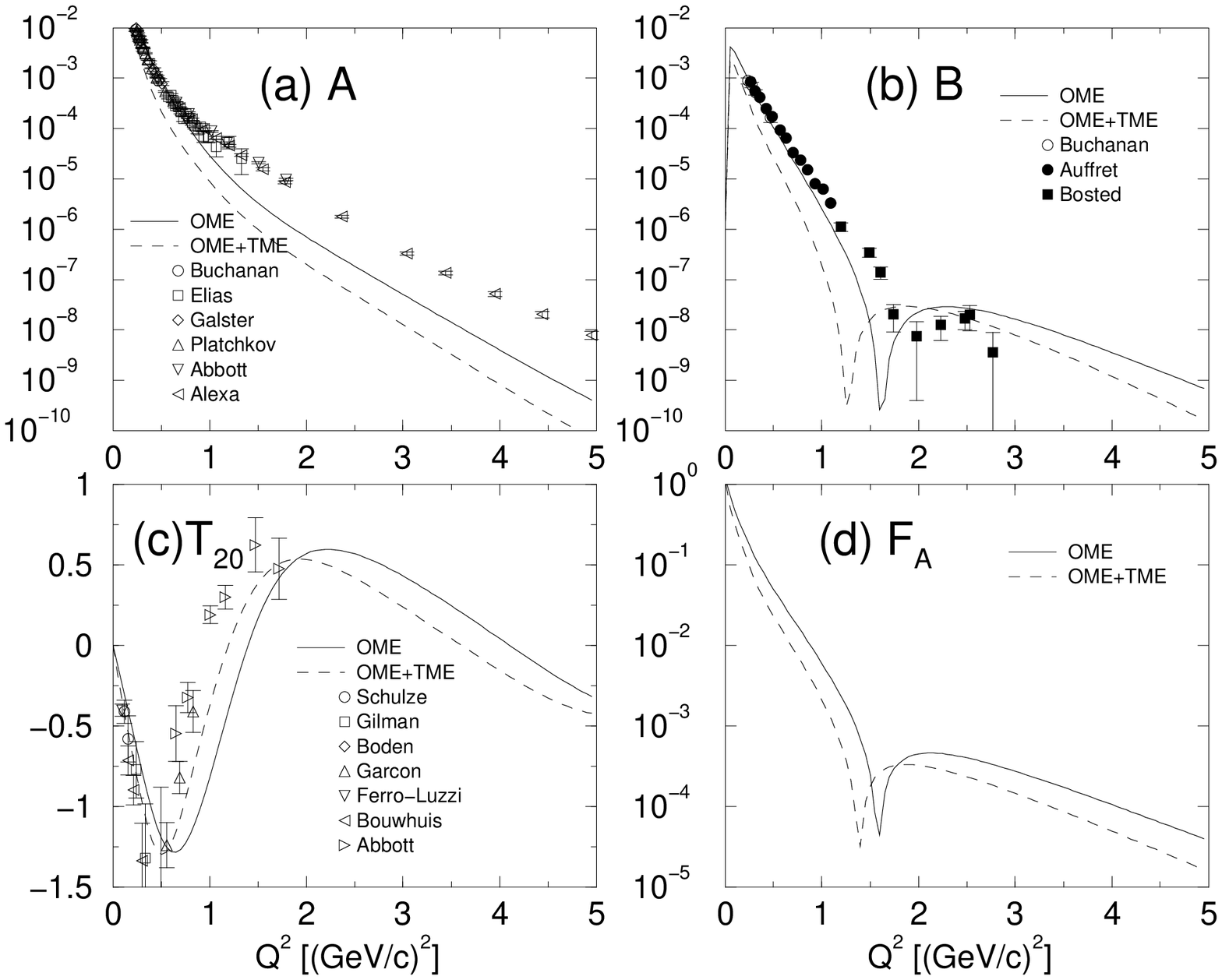}
\caption{The $A$, $B$, $T_{20}$, and $F_A$ form factors for the OME and
OME+TME potentials, along with data. See the accompanying text for an
explanation of the data.
\label{form:diag:OME.expt2}}
\end{center}
\end{figure}

Finally, in Fig.~\ref{form:diag:OME.expt2}, we compare the $A$, $B$,
$T_{20}$, and $F_A$ form factors for the OME and OME+TME wave functions
to experimental data. The ``bad'' component was chosen according to FFS,
and the nucleon form factors of Lomon were used for $A$, $B$, and $T_{20}$,
while the Liesenfeld axial nucleon form factor was used for $F_A$. The
data for $A$ is from: Buchanan {\it et al.} \cite{Buchanan:1965},
Elias {\it et al.} \cite{Elias:1969},
Galster {\it et al.} \cite{Galster:1971kv},
Platchkov {\it et al.} \cite{Platchkov:1990ch},
Abbott {\it et al.} \cite{Abbott:1998sp}, and
Alexa {\it et al.} \cite{Alexa:1999fe};
the data for $B$ is from: Buchanan {\it et al.} \cite{Buchanan:1965},
Auffret {\it et al.} \cite{Auffret:1985}, and
Bosted {\it et al.} \cite{Bosted:1990hy};
and the data for $T_{20}$ is from:
Schulze {\it et al.} \cite{Schulze:1984ms}, 
Gilman {\it et al.} \cite{Gilman:1990vg}, 
Boden {\it et al.} \cite{Boden:1991un}, 
Garcon {\it et al.} \cite{Garcon:1994vm}, 
Ferro-Luzzi {\it et al.} \cite{Ferro-Luzzi:1996dg}, 
Bouwhuis {\it et al.} \cite{Bouwhuis:1998jj}, and
Abbott {\it et al.} \cite{Abbott:2000fg}.

There is a rather large difference between the form factors calculated
with the OME and OME+TME wave functions. This difference is due
primarily to the fact that the OME wave functions are more deeply bound
than the OME+TME wave functions, and it can be reduced by choosing a
different sigma coupling constant $f_\sigma$ for the OME and OME+TME
potentials. However, for our analysis of rotational invariance, it is
important to keep $f_\sigma$ fixed.

The difference between the calculated form factors and the data is also
quite large. This is not unexpected, since in our model of the current,
meson exchange currents are not included. It is known that these have a
large effect on the form factors at large momentum transfers
\cite{Phillips:1998uk,Wiringa:1995wb,Schiavilla:1991ug}. Including these
effects would bring the form factors into better agreement with the
data. However, we emphasize again that agreement with the data is not a
priority of this work. Our goal is to gain a better understanding of the
breaking of rotational invariance by the light front, and how to restore
that invariance. Only after we have that understanding can we pursue
accurate calculation of the form factors with light-front dynamics.

\section{Conclusions} \label{ch:conclusions}

The issue of rotational invariance in light-front dynamics must be
addressed before one attempt to use light-front dynamics for
high-precision calculations. In this paper, we have sought to
find ways to quantify the level to which rotational invariance is
broken. In addition, we have investigated potentials of different
orders, since we know that if all orders of the potential were included,
rotational invariance would be restored. We have used light-front
dynamics to obtain new light-front nucleon-nucleon one-meson-exchange
(OME) and two-meson-exchange (TME) potentials.

In section~\ref{ch:pionly}, we derive OME and TME potentials for a
model Lagrangian for nuclear physics which includes chiral
symmetry. These potentials are then used in section~\ref{nnresults} to
calculate the binding energy and wave function for the $m=0$ and $m=1$
states of the deuteron. We find that the splitting between the $m=0$ and
$m=1$ states was smaller for the OME+TME potential as compared to the
OME potential gives.

In section~\ref{ch:ff:results}, the wave functions obtained in
section~\ref{ch:pionly} are used to calculate the form factors of the
deuteron using only light-front dynamics throughout.
In light-front dynamics, there are four independent components
of the deuteron current. However, the requirement of rotational
invariance introduces an angular condition that the four components must
satisfy, reducing the number of physically independent components to
three. The deviation of the calculated current components from the
angular condition is denoted by $\Delta$. We find that $\Delta$ is very
small for the deuteron wave functions calculated with the OME
potential. This is an important result, since it means that although
{\em in principle} the light-front calculation of the deuteron current
does not transform correctly under rotations, {\em in practice} it does
quite well. The smallness of $\Delta$ means that any reasonable
prescription for eliminating the dependent component of the current gives
essentially the same results; the uncertainty introduced by the various
nucleon form factors is much greater.

We also found that $\Delta$ is significantly larger when the TME
potentials are used. Since the results in
Refs.~\cite{Cooke:1999yi,Cooke:2000ef,CookePiOnly} indicate that the 
rotational properties of the TME wave function are better that for the
OME wave function, we interpret the increase in $\Delta$ as an
indication that extra diagrams need to be included in the current
calculation to restore rotational invariance. That is, having a wave
function with good rotational properties is not sufficient to obtain 
matrix elements of the current operator with good rotational
properties. To verify this, the
contribution to the deuteron form factors due to the photon coupling to
an intermediate two-meson-exchange type of diagram must be calculated.
This may restore the rotational invariance of the deuteron form factor
when the wave function is calculated using the two-meson-exchange
potentials.

\begin{acknowledgments}
This work is supported in part by the U.S.~Dept.~of Energy under Grant
No.~DE-FG03-97ER4014.  We are grateful to Daniel Phillips for many 
many useful discussions about this work.
\end{acknowledgments}

\appendix*

\section{Notation, Conventions, and Useful Relations}

For a general four-vector $a$ with components $(a^0,a^1,a^2,a^3)$ in the
equal-time basis, we define the light-front variables 
\begin{eqnarray}
a^\pm          &=& a^0 \pm a^3, \\
\bm{a}_\perp &=& (a^1,a^2),
\end{eqnarray}
so the 4-vector $a^\mu$ can be denoted in the light-front basis as
\begin{eqnarray}
a&=&(a^+,a^-,\bm{a}_\perp).
\end{eqnarray}
Using this, we find that the scalar product is
\begin{eqnarray}
a \cdot b &=& a^\mu b_\mu
= \frac{1}{2}\left(a^+ b^- + a^- b^+ \right) - \bm{a}_\perp \cdot 
\bm{b}_\perp.
\end{eqnarray}
This defines $g_{\mu\nu}$, with $g_{+-}=g_{-+}=1/2$, $g_{11}=g_{22}=-1$,
and all other elements of $g$ vanish.  The elements of $g^{\mu\nu}$ are
obtained from the condition that $g^{\mu\nu}$ is the inverse of
$g_{\mu\nu}$, so
$g^{\alpha\beta}g_{\beta\lambda}=\delta^\alpha_\lambda$.  Its elements
are the same as those of $g_{\mu\nu}$, except that
$g^{-+}=g^{+-}=2$. Thus,
\begin{eqnarray}
a^\pm &=& 2 a_\mp.
\end{eqnarray}
and the partial derivatives are similarly given by
\begin{eqnarray}
\partial^\pm &=& 2 \partial_\mp = 2 \frac{\partial}{\partial x^\mp}.
\end{eqnarray}

To find the physical consequences of this coordinate system, consider
the commutation relations $[p^\mu,x^\nu] = i g^{\mu\nu}$, which yield
\begin{eqnarray}
\,[       p^\pm     ,      x^\mp      ] &=& 2i, \\
\,[ \bm{p}_\perp^i,\bm{x}_\perp^j ] &=& -i \delta_{i,j},
\end{eqnarray}
with the other commutators equal to zero.  This means that
$\bm{x}_\perp^i$ is
canonically conjugate to $\bm{p}_\perp^i$, and $x^\pm$ is
conjugate to
$p^\mp$.  Since $x^+$ plays the role
of time (the light-front time) in light-front dynamics, and $p^-$ is
canonically conjugate to $x^+$, this means that $p^-$ is the
light-front energy and that the light-front Hamiltonian is given by
$P^-$.

In any Hamiltonian theory, particles have the an energy defined
by the on-shell constraint $k^2=m^2$.  This implies that the light-front
energy of a particle is
\begin{eqnarray}
k^- &=& \frac{m^2 +\bm{k}_\perp^2}{k^+}.
\end{eqnarray}
The independent components of the momentum can be written as a
light-front three-vector $\bm{k}_{\text{LF}}$, denoted by
\begin{eqnarray}
\bm{k}_{\text{LF}} &=& (k^+,\bm{k}_\perp).
\end{eqnarray}

For dealing with spin, we require the Pauli sigma matrices, which are
\begin{eqnarray}
(\sigma^1,\sigma^2,\sigma^3) &=&
\left(
\left(\begin{array}{cc} 0 &  1 \\ 1 &  0 \end{array} \right),
\left(\begin{array}{cc} 0 & -i \\ i &  0 \end{array} \right),
\left(\begin{array}{cc} 1 &  0 \\ 0 & -1 \end{array} \right)
\right) \label{sigmamat}.
\end{eqnarray}

The Bjorken and Drell convention \cite{Bjorken:1964} for the gamma
matrices is used in this work. They specify that 
\begin{eqnarray}
\gamma^0 = \beta &=&
\left(\begin{array}{cc} 1 & 0 \\ 0 & -1 \end{array} \right), \\
\bm{\gamma} = \beta \bm{\alpha} &=&
\left(\begin{array}{cc} 0 & \sigma \\ -\sigma & 0 \end{array} \right), \\
\gamma^5 = i \gamma^0 \gamma^1 \gamma^2 \gamma^3 &=&
\left(\begin{array}{cc} 0 & 1 \\ 1 & 0 \end{array} \right).
\end{eqnarray}

The spin matrices $S^i$ then are
\begin{eqnarray}
S^i &=& \frac{1}{2} \Sigma^i = -\frac{1}{2} \gamma^5 \gamma^i, \\
\Sigma^i &=& 
\left(\begin{array}{cc} \sigma^i & 0 \\ 0 & -\sigma^i
\end{array} \right).
\end{eqnarray}
Using $\bm{\Sigma}$, we can express the helicity operator as
$H=\widehat{p}\cdot\bm{\Sigma}$, which has eigenvalues $\pm 1$. This
is useful since the helicity is invariant under rotations.

It is useful to define the spinor projection operators $\Lambda_\pm$ by
\begin{eqnarray}
\Lambda_\pm &=& \frac{1}{4} \gamma^\mp \gamma^\pm = \frac{1}{2} \gamma^0
\gamma^\pm = \frac{1}{2} (I\pm \alpha^3). \label{app:deflampm}
\end{eqnarray}
These satisfy the requirements for projection operators,
\begin{eqnarray}
\Lambda_+ + \Lambda_- &=& 1, \\
(\Lambda_\pm)^2 &=& \Lambda_\pm, \\
\Lambda_\pm \Lambda_\mp &=& 0.
\end{eqnarray}

We summarize the effect these projection operators have on the gamma
matrices: 
\begin{eqnarray}
\Lambda_\pm \gamma^0 &=& \gamma^0 \Lambda_\mp, \\
\Lambda_\pm \gamma^\pm &=& 0 = \gamma^\pm \Lambda_\mp, \\
\Lambda_\pm \gamma^\mp &=& \gamma^\mp = \gamma^\mp \Lambda_\mp, \\
\Lambda_\pm \gamma^\perp &=& \gamma^\perp \Lambda_\pm,
\end{eqnarray}
and under conjugation,
\begin{eqnarray}
\gamma^0 \Lambda_\pm^\dagger \gamma^0 &=& \Lambda_\mp.
\end{eqnarray}

The spinors used by Bjorken and Drell \cite{Bjorken:1964} are polarized
in the $\widehat{z}$ direction, and so the $\chi$'s are chosen to be
\begin{eqnarray}
\chi_{+1/2} &=& \left(\begin{array}{c} 1 \\ 0 \end{array} \right), \\
\chi_{-1/2} &=& \left(\begin{array}{c} 0 \\ 1 \end{array} \right).
\end{eqnarray}

The light-front spinors are defined to be \cite{Miller:1997cr}
\begin{eqnarray}
u_{\text{LF}}(k,\lambda)
&\equiv& \frac{1}{\sqrt{M k^+}} \left[ M \Lambda_- + (k^+ +
\bm{\alpha}^\perp\cdot\bm{k}^\perp ) \Lambda_+ \right]
\chi_{\text{LF},\lambda} \\
&=& \frac{1}{\sqrt{M k^+}} \left[ \Lambda_- 
( M + \bm{\alpha}^\perp\cdot\bm{k}^\perp )
+ \Lambda_+ k^+ \right] \chi_{\text{LF},\lambda}, \\
\chi_{\text{LF},\lambda} &\equiv&
\left( \begin{array}{c} \chi_\lambda \\ 0 \end{array} \right),
\end{eqnarray}
where $\chi_\lambda$ is the usual Pauli spinor, and the $\Lambda_\pm$
are the spinor projection operators defined in Eq.~(\ref{app:deflampm}).
We find that
\begin{eqnarray}
\overline{u}_{\text{LF}}(k,\lambda)
&=& \frac{1}{\sqrt{M k^+}} \chi^\dagger_{\text{LF},\lambda}
\left[ \Lambda_+ M + \Lambda_- (k^+ +
\bm{\alpha}^\perp\cdot\bm{k}^\perp ) \right] \\
&=& \frac{1}{\sqrt{M k^+}} \chi^\dagger_{\text{LF},\lambda}
\left[ ( M - \bm{\alpha}^\perp\cdot\bm{k}^\perp ) \Lambda_+ 
+ k^+ \Lambda_- \right].
\end{eqnarray}
Note that these spinors are normalized to satisfy
$\overline{u}_{\text{LF}}(k,\lambda')u_{\text{LF}}(k,\lambda)=\delta_{\lambda'\lambda}$.

For helicity spinors, we choose the eigenvectors of the helicity
operator ($\bm{\sigma}\cdot\widehat{\bm{p}}$) as the $\chi$'s. In
particular, 
$(\widehat{\bm{p}}\cdot\bm{\Sigma})u(\bm{p},\lambda)=hu(\bm{p},\lambda)$,
where $h=2\lambda$.  This choice allows us to write
\begin{eqnarray}
u(\bm{p},\lambda) &=& \sqrt{\frac{W}{2M}}
\left(\begin{array}{c} 1 \\ f \end{array} \right)
\chi_\lambda(\widehat{\bm{p}}),
\end{eqnarray}
and
\begin{eqnarray}
\chi_\lambda(\widehat{\bm{p}}) &=&
\left\{\begin{array}{ll}
\left(\begin{array}{r}  c_2e^{-i\phi/2}\\ s_2e^{+i\phi/2} \end{array} \right)
& \mbox{ if $h=+1$} \\
\left(\begin{array}{r} -s_2e^{-i\phi/2}\\ c_2e^{+i\phi/2} \end{array} \right)
& \mbox{ if $h=-1$} \\
\end{array} \right. \phantom{\left. \right\}} 
\end{eqnarray}
where $c_2=\cos\frac{\theta}{2}$, $s_2=\sin\frac{\theta}{2}$,
$f=\frac{hp}{W}$, and $h=2\lambda$.

When there are two fermions in the center-of-momentum frame, 
we can define $\phi\equiv\phi_1$ and $\theta\equiv\theta_1$ and for
particle two $\phi_2=\pi+\phi$ and $\theta_2=\pi-\theta$. This means
that
\begin{eqnarray}
u(\bm{p}_i,\lambda_i) &=& \sqrt{\frac{W}{2M}}
\left(\begin{array}{c} 1 \\ f_i \end{array} \right)
\chi_{i,\lambda_i}(\widehat{\bm{p}}),
\end{eqnarray}
where $i=1,2$ and 
\begin{eqnarray}
\chi_{1,\lambda_1}(\widehat{\bm{p}})
&=&   \chi_{ \lambda_1}(\widehat{\bm{p}}), \\
\chi_{2,\lambda_2}(\widehat{\bm{p}})
&=& i \chi_{-\lambda_2}(\widehat{\bm{p}}).
\end{eqnarray}

\newpage

\begin{table}[!p]
\caption{The parameters for the $\pi$, $\eta$, $\rho$, $\omega$,
$\delta$, and $\sigma$ mesons. 
For the meson type, ``iv'' and ``is'' stand
for isovector and isoscalar, while ``ps'', ``v'', and ``s'' stand for
pseudoscalar, vector, and scalar.
The $\sigma$ meson is also known as
$f_0(400-1200)$, and the $\delta$ meson is the $a_0(980)$.
\label{nn:tab:mesparams}}
\begin{center}
\begin{ruledtabular}\begin{tabular}{ccccccc}
Meson     & type  & mass [MeV]  & $\Lambda$ [GeV] & $n$ & $\frac{g^2}{4\pi}$ &
$\frac{f}{g}$ \\ \hline
$\pi$     & iv,ps & 138.04 & 1.2  & 1 &                 14.0    &     \\
$\eta$    & is,ps & 547.5  & 1.5  & 1 &                  3.0    &     \\
$\rho$    & iv,v  & 769.   & 1.85 & 2 &                  0.9    & 6.1 \\
$\omega$  & is,v  & 782.   & 1.85 & 2 &                 24.5    & 0.0 \\
$\delta$  & iv,s  & 983.   & 2.0  & 1 &                  2.0723 &     \\
$\sigma$  & is,s  & 550.   & 2.0  & 1 & $f_\sigma\times$ 8.9602 &     \\
\end{tabular}\end{ruledtabular} 
\end{center}
\end{table}
%\clearpage

\begin{table}[!p]
\caption{The values of $f_\sigma$ required to give the physical value of
the deuteron mass for a given potential and state with $J_z=m$.
The percent of the wave function in the D-state and in the $J=1$ state
are also shown.
\label{nn:tab:fixbseJ1}}
\begin{center}
\begin{ruledtabular}\begin{tabular}{cccccccc}
Potential &
\multicolumn{3}{c}{$f_\sigma$} &
\multicolumn{2}{c}{\% D state} &
\multicolumn{2}{c}{\% $J=1$} \\ \hline
& m=0 & m=1 & Diff & m=0 & m=1 & m=0 & m=1  \\ \hline
OME &
1.2407  &  1.2125  &   0.0282 & 2.87 & 3.55 & 99.99 & 99.97 \\ \hline
\begin{tabular}{c}OME\\+TPE\end{tabular} &
1.2829  &  1.2819  &   0.0010 & 2.96 & 3.23 & 99.99 & 99.97 \\ \hline
\begin{tabular}{c}OME\\+TME\end{tabular} &
1.2968  &  1.3079  &  -0.0111 & 2.95 & 3.28 & 99.99 & 99.96 \\ \hline
\begin{tabular}{c}OME\\+ncTPE\end{tabular} &
1.3064  &  1.3121  &  -0.0057 & 2.99 & 3.16 & 99.98 & 99.97 \\ \hline
\begin{tabular}{c}OME\\+ncTME\end{tabular} &
1.3198  &  1.3397  &  -0.0199 & 2.98 & 3.21 & 99.98 & 99.96\\
\end{tabular}\end{ruledtabular} 
\end{center}
\end{table}
%\clearpage

\begin{table}[!p]
\caption{The values of the binding energy, percentage of the wave
function in the D state, and the percentage of the wave function in the
$J=1$ state for the $m=0$ and $m=1$ states for different potentials.
The $\sigma$ coupling constant factor is
$f_\sigma=1.22$.
\label{nn:tab:sig122AllJ}}
\begin{center}
\begin{ruledtabular}\begin{tabular}{cccccccc}
Potential &
\multicolumn{3}{c}{Binding Energy (MeV)} &
\multicolumn{2}{c}{D state (\%)} &
\multicolumn{2}{c}{$J=1$ (\%)} \\ \hline
& m=0 & m=1 & Diff & m=0 & m=1 & m=0 & m=1 \\ \hline
OME only & 
-1.7653  &-2.4200  &0.6547   &2.73     &3.61     &99.99    &99.96 \\ \hline
\begin{tabular}{c}OME\\+$\pi$-($\sigma$-$\omega$) Mesa\end{tabular} &
-1.9236  &-1.7021  &-0.2215  &2.80     &3.38     &99.99    &99.96   \\ \hline
\begin{tabular}{c}OME\\+ncTME\end{tabular} &
-0.4948  &-0.2646  &-0.2302  &1.97     &1.76     &99.99    &99.98   \\ \hline
\begin{tabular}{c}OME\\+ncTPE\end{tabular} &
-0.6620  &-0.4825  &-0.1795  &2.16     &2.09     &99.99    &99.98   \\ \hline
\begin{tabular}{c}OME\\+TME\end{tabular} &
-0.7861  &-0.6060  &-0.1801  &2.25     &2.31     &99.99    &99.97   \\ \hline
\begin{tabular}{c}OME\\+TPE\end{tabular} &
-0.9981  &-0.9155  &-0.0826  &2.42     &2.57     &99.99    &99.98   \\
\end{tabular}\end{ruledtabular} 
\end{center}
\end{table}
%\clearpage

\begin{table}[!p]
\caption{The values of the binding energy, percentage of the wave
function in the D state, and the percentage of the wave function in the
$J=1$ state for the $m=0$ and $m=1$ states for different potentials.
The $\sigma$ coupling constant factor is
$f_\sigma=1.2815$, 
distinguishes this table from Table~\ref{nn:tab:sig122AllJ}.
\label{nn:tab:sig128AllJ}}
\begin{center}
\begin{ruledtabular}\begin{tabular}{cccccccc}
Potential &
\multicolumn{3}{c}{Binding Energy (MeV)} &
\multicolumn{2}{c}{D state (\%)} &
\multicolumn{2}{c}{$J=1$ (\%)} \\ \hline
& m=0 & m=1 & Diff & m=0 & m=1 & m=0 & m=1 \\ \hline
OME only & 
-3.3500  &-4.4546  & 1.1046  &3.09     &3.97     &99.99    &99.96   \\ \hline
\begin{tabular}{c}OME\\+$\pi$-($\sigma$-$\omega$) Mesa\end{tabular} &
-3.6331  &-3.2408  &-0.3923  &3.10     &3.85     &99.99    &99.95   \\ \hline
\begin{tabular}{c}OME\\+ncTME\end{tabular} &
-1.3766  &-0.9901  &-0.3865  &2.67     &2.64     &99.99    &99.97   \\ \hline
\begin{tabular}{c}OME\\+ncTPE\end{tabular} &
-1.6532  &-1.4693  &-0.1839  &2.81     &2.88     &99.99    &99.97   \\ \hline
\begin{tabular}{c}OME\\+TME\end{tabular} &
-1.8617  &-1.6032  &-0.2585  &2.85     &3.05     &99.99    &99.96   \\ \hline
\begin{tabular}{c}OME\\+TPE\end{tabular} &
-2.1915  &-2.2137  &0.0222   &2.95     &3.23     &99.99    &99.97   \\
\end{tabular}\end{ruledtabular} 
\end{center}
\end{table}
%\clearpage
\clearpage
\newpage

%%%%%%%%%%%%%%%%%%%%%%%
%%% Now the figures %%%
%%%%%%%%%%%%%%%%%%%%%%%

%\clearpage

%\clearpage

%\clearpage

%\clearpage

%\clearpage

%\clearpage

%\clearpage

%\clearpage

%\clearpage

%\clearpage

%\clearpage

%\clearpage

%\clearpage

%\clearpage

%\clearpage

%\clearpage

%\clearpage

%\clearpage


\begin{thebibliography}{100}
\bibitem{breakpoin}
This problem is not unique to light-front dynamics. In any form of 
Hamiltonian dynamics, at least three of the Poincar\'{e} group
generators are dynamic, and thus complicated.
\end{thebibliography}
\end{document}